\documentstyle[epsfig]{elsart}

\newcommand{\bea}{\begin{equation}}
\newcommand{\eea}{\end{equation}}
\newcommand{\beq}{\begin{eqnarray}}
\newcommand{\eeq}{\end{eqnarray}}
\newcommand{\ba}{\begin{array}}
\newcommand{\ea}{\end{array}}
\newcommand{\dd}{{\rm d}}
\newcommand{\dfrac}{\displaystyle\frac}
\newcommand{\nn}{\nonumber}
\newcommand{\gmu}{\gamma_\mu}
\newcommand{\gmup}{\gamma^\mu}
\newcommand{\gfi}{\gamma_5}

\newcommand{\smunu}{\sigma_{\mu\nu}}
\newcommand{\smunup}{\sigma^{\mu\nu}}
\newcommand{\eps}{\epsilon}
\newcommand{\raw}{\rightarrow}
\def\ijmp#1#2#3{{\it Int. Jour. Mod. Phys. }{\bf #1~}(19#2)~#3}
\def\fp#1#2#3{{\it Fortschr. Phys. }{\bf #1~}(19#2)~#3}
\def\plb#1#2#3{{\it Phys. Lett. }{\bf B#1~}(19#2)~#3}
\def\zpc#1#2#3{{\it Z. Phys. }{\bf C#1~}(19#2)~#3}
\def\prl#1#2#3{{\it Phys. Rev. Lett. }{\bf #1~}(19#2)~#3}
\def\rmp#1#2#3{{\it Rev. Mod. Phys. }{\bf #1~}(19#2)~#3}
\def\prep#1#2#3{{\it Phys. Rep. }{\bf #1~}(19#2)~#3}
\def\prd#1#2#3{{\it Phys. Rev. }{\bf D#1~}(19#2)~#3}
\def\npb#1#2#3{{\it Nucl. Phys. }{\bf B#1~}(19#2)~#3}
\def\mpl#1#2#3{{\it Mod. Phys. Lett. }{\bf #1~}(19#2)~#3}
\def\arnps#1#2#3{{\it Annu. Rev. Nucl. Part. Sci. }{\bf #1~}(19#2)~#3}
\def\sjnp#1#2#3{{\it Sov. J. Nucl. Phys. }{\bf #1~}(19#2)~#3}

\def\ptp#1#2#3{{\it Prog. Theor. Phys. }{\bf #1~}(19#2)~#3}
\def\hepph#1{{\bf hep-ph}/#1}
\newcommand{\lsim}{\raisebox{-0.13cm}{~\shortstack{$<$ \\[-0.07cm] $\sim$}}~}

\def\pplus{{\mathbf{\hat p}_{\mathbf{+}}}}
\def\pminus{{\mathbf{\hat p}_{\mathbf{-}}}}

\def\kplus{{\mathbf{\hat k}_{\mathbf{+}}}}
\def\kminus{{\mathbf{\hat k}_{\mathbf{-}}}}
\def\qplus{{\mathbf{q}_{\mathbf{+}}}}
\def\qminus{{\mathbf{q}_{\mathbf{-}}}}

\def\sone{{\mathbf{s}_{\mathbf{1}}}}
\def\stwo{{\mathbf{s}_{\mathbf{2}}}}
\def\soner{{\mathbf{s}^{\mathbf{*}}_{\mathbf{1}}}}
\def\stwor{{\mathbf{s}^{\mathbf{*}}_{\mathbf{2}}}}
\def\splus{{\mathbf{s}_{\mathbf{+}}}}
\def\sminus{{\mathbf{s}_{\mathbf{-}}}}
\def\spm{{\mathbf{s}_{\mathbf{\pm}}}}
\def\smp{{\mathbf{s}_{\mathbf{\mp}}}}

\def\spmr{{\mathbf{s}^{\mathbf{*}}_{\mathbf{\pm}}}}

\def\pplusn{{\mathbf{p}_{\mathbf{+}}}}
\def\pminusn{{\mathbf{p}_{\mathbf{-}}}}
\def\ppmn{{\mathbf{p}_{\mathbf{\pm}}}}
\def\pmpn{{\mathbf{p}_{\mathbf{\mp}}}}
\def\kplusn{{\mathbf{k}_{\mathbf{+}}}}
\def\kminusn{{\mathbf{k}_{\mathbf{-}}}}
\def\kpmn{{\mathbf{k}_{\mathbf{\pm}}}}
\def\kmpn{{\mathbf{k}_{\mathbf{\mp}}}}
\def\qplusn{{\mathbf{\hat q}_{\mathbf{+}}}}
\def\qminusn{{\mathbf{\hat q}_{\mathbf{-}}}}
\def\qpmn{{\mathbf{q}_{\mathbf{\pm}}}}
\def\qmpn{{\mathbf{q}_{\mathbf{\mp}}}}
\newcommand{\s}{}
\input{epsf.tex}
\begin{document}

\vspace*{-2cm}
\begin{flushright}
KA--TP--10--1998\\
{\tt hep-ph/9808408}\\
October 2001 (rev.)
\end{flushright}

\begin{frontmatter}

\title{Dipole Form Factors and Loop--induced \\
       CP violation in Supersymmetry}

\author{W. Hollik, J.I. Illana, C. Schappacher, 
        D. St{\"o}ckinger,}\footnote{E-mail addresses:
                \{hollik,jillana,cs,ds\}@particle.physik.uni-karlsruhe.de} 
\address{Institut f{\"u}r Theoretische Physik, Universit{\"a}t Karlsruhe, \\
D--76128 Karlsruhe, Germany}

\author{S. Rigolin}\footnote{E-mail address: rigolin@delta.ft.uam.es} 
\address{Departamento de F{\'\i}sica Te{\'o}rica, Universidad Aut{\'o}noma de
Madrid, \\ Cantoblanco, E--28049 Madrid, Spain}

\begin{abstract}

The one--loop Minimal Supersymmetric Standard Model (MSSM) contributions to the 
weak and electromagnetic dipole form factors of heavy fermions are reviewed. 
For the $Z$ boson on shell, the weak--magnetic and weak--electric dipole 
moments of the $\tau$ lepton and the $b$ quark can be defined and directly 
connected to observables. But far from the $Z$ peak, the weak and 
electromagnetic dipole form factors are not enough to account for all the 
new physics effects.
In the context of the calculation of the process $e^+e^-\to t\bar{t}$ 
to one loop in the MSSM, we compare the impact on the phenomenology of the 
CP--violating dipole form factors of the top quark with the contribution from 
CP--violating box graphs. 
Some exemplificative observables are analyzed and the relevance of both the 
contributions is pointed out. The set of tensor integrals employed,
the one--loop expressions for the electromagnetic and weak dipole 
form factors in a general renormalizable theory and the SM and MSSM couplings 
and conventions are also given.

\end{abstract}

\end{frontmatter}

\setcounter{footnote}{0}

\section{Introduction}

The investigation of the electric and magnetic dipole moments of fermions 
provides deep insight in particle theory. The measurement of the 
intrinsic {\em magnetic dipole moment} (MDM) of the electron 
proved the correctness of the hypothesis of half--integer spin particles 
\cite{gelectron} and is one of the most spectacular achievements of 
quantum field theory predictions. More precise studies of electron 
and muon showed afterwards the presence of an {\em anomalous} contribution to 
the MDM (AMDM) and imply very accurate tests of the quantum structure 
of the Standard Model (SM). The measurements of the $(g_e-2)$ and $(g_\mu-2)$ 
available \cite{databook} are in perfect agreement with the  
SM predictions to several orders in the perturbative expansion of the theory
(cf. \cite{smemdm} and references therein). Furthermore, 
with the expected precision at the E821 Brookhaven experiment \cite{brook} it 
will be possible to improve the previous measurement of $(g_\mu-2)$ by a 
factor 20. Therefore the MDMs can be used, together with the precision tests at 
the $Z$ resonance from LEP and SLC and the new results of LEP2 and TEVATRON, to 
set bounds on possible New Physics effects beyond the SM \cite{munpemdm}.  

The importance of the analysis of the {\em electric dipole moment} (EDM) of 
elementary and composite particles is intimately related to the CP violating 
character of the theory. In the electroweak SM there is only one 
possible source of CP violation, the $\delta_{\rm CKM}$ phase of 
the Cabibbo--Kobayashi--Maskawa (CKM) mixing matrix 
for quarks \cite{ckm}. Currently the only place where CP violation has been 
measured, the neutral $K$ system, fixes the value of this phase but does not 
constitute itself a test for the origin of CP violation \cite{cpreview}. 
On the other hand, if the baryon asymmetry of the universe has been 
dynamically generated, CP must be violated. The SM cannot account for the 
size of the observed asymmetry \cite{b-asym}. In extended models (beyond the 
SM) many other possible explanations of CP violation can be given. In 
particular, in supersymmetric (SUSY) models \cite{susycp,dugan} CP violation 
can appear assuming complex soft--SUSY--breaking terms. 
Two physical phases remain in the GUT constrained MSSM 
\cite{dugan,gavela,relax}, enough to provide the correct size of baryon 
asymmetry in some range of parameters \cite{b-susyasym}.
But the most significant effect of the CP violating phases in the phenomenology 
is their contribution to the EDMs \cite{dipoles}. Unlike the SM, where the 
contribution to the EDM of fermions arises beyond two loops \cite{smedm}, the 
MSSM can give a contribution already at the one--loop level \cite{susycp}. 

%
The measurements of the neutron, electron and muon EDMs \cite{edm-n,edm-l} 
constrain the phases and the supersymmetric spectrum in a way that may 
demand fine tuning (supersymmetric CP problem): SUSY particles very heavy 
(several TeV \cite{nath1}) or phases of ${\cal O}(10^{-2})$ \cite{susycp}. 
Very large soft--SUSY--breaking masses are unappealing as it seems natural to 
demand the SUSY spectrum to be at the electroweak scale.\footnote{Moreover
if the SUSY spectrum is in the TeV region this could also give rise to relic 
densities unacceptably large.} 
%
%
On the other side, general universal soft--SUSY--breaking terms can be 
taken only if the genuine SUSY CP phases are vanishing. 
In this case the CP violation is originated via the usual SM CKM mechanism 
and the supersymmetric spectrum affects the observables only through 
renormalization group equations and/or one--loop radiative contributions. 
Of course in this scenario one has to construct models in which the SUSY 
phases naturally vanish \cite{phase0} and at the same time provide some 
other non--standard mechanism for explaining electroweak baryogenesis. 
%
%
There exist also ways of naturally obtaining small non--zero soft--phases 
which leave sufficient CP violation for baryogenesis \cite{relax}. 
But, in general, to get observable effects in most electroweak processes 
one has to relax the assumption of soft--term universality. 
%
%
Several attempts have been made, following this 
direction, to use CP violation from top--squark mixing: a complex 
parameter $A_t$ would yield large CP violating effects in collider 
processes involving top quarks \cite{cptop}.\footnote{ 
Large non--SM CP violating top--quark couplings could be probed at high
energies colliders like the NLC \cite{nlc}.} 
%
%
Besides, due to renormalization--group--induced effects on the other 
low energy phases, the phase of $A_t$ is constrained by the EDM of the neutron 
\cite{garisto}. 
%
%
One can also make the hypothesis that, due to cancellation among the 
different components of the neutron EDM (constituent quarks and gluons), 
the SUSY phases can still be kept of ${\cal O}(1)$ and the SUSY spectrum at 
the electroweak scale satisfying the experimental bounds \cite{nath2}.
%
%
%
Finally, in \cite{falkolive} it is shown that, in the constrained MSSM, large CP
violating phases are compatible with the bounds on the electron and neutron 
EDMs as well as with the cosmological relic densities.
%
%
In view of all these arguments we keep our analysis completely general and 
consider the SUSY CP--phases as free parameters.

Some attention has also been payed to the study of possible CP violating 
effects in the context of R--parity violating models \cite{abel}. In this 
class of models new interactions appear providing extra sources of CP 
violation (still preventing fast proton decay). They can explain the 
CP violation in the $K$ system (with no need of the CKM phase) without 
introducing anomalous Flavor Changing Neutral Current (FCNC) contributions 
\cite{fcnc-cp}. 
In any case we assume in the following R--parity conservation for simplicity.

Recently a number of works have been devoted to the analysis of {\em weak dipole 
moments} (WDM). The WDMs are defined, in analogy to the usual DMs, taking 
the corresponding on--shell chirality--flipping form factors of the $Zff$ 
effective vertex. 
The SM one--loop contribution to 
the {\em anomalous weak magnetic dipole moment} (AWMDM) has been calculated 
for the $\tau$ lepton and the $b$ quark in \cite{ber95,ber97}. 
The CP--violating {\em weak electric dipole moment} (WEDM) is in the SM 
a tiny three--loop effect. The WDMs are gauge invariant and can directly 
connected to physical observables. While for the $\tau$ case, using 
appropriate observables \cite{ber95,ber94,ber95p}, an experimental analysis 
is feasible, for the $b$ case the situation is complicated by hadronization 
effects \cite{mele}.
The SM predictions are far below the sensitivity reachable at LEP 
\cite{ber95} but non--standard interactions can enhance these
expectations (2HDM \cite{ber95b}, MSSM \cite{hirs1}) especially for the 
(CP--violating) WEDMs (2HDM, leptoquark models \cite{bernre97b}, 
MSSM \cite{hirs2}).
The experimental detection of non--zero AWMDM or WEDM of heavy fermions, at 
the current sensitivity, would be a clear evidence of new physics beyond the SM.

In the perspective of the next generation of linear $e^+ e^-$ colliders, 
we extend the previous analyses on the WDMs to consider the $t$ quark
{\em dipole form factors} (DFF) and, in particular, the 
CP--violating ones. In \cite{vienna} an independent analysis of the $t$ quark 
EDFF and WEDFF can be found.
%
Since the $t$ quark is very heavy one expects this fermion to be the best 
candidate to have larger DFFs. Beyond the $Z$ peak 
($s>4m^2_t$) other effects are expected to give contributions to 
the physical observables. In fact the DFFs in a general model are not 
guaranteed to be gauge independent. An exception is the 
one--loop MSSM contribution to the CP--violating {\em electromagnetic} and 
{\em weak} DFFs (EDFF and WEDFF) which do not involve any gauge boson in 
internal lines. Anyway, although this contribution is indeed gauge invariant 
any CP--odd observable will be sensitive to not only the DFFs but to the 
complete set of one--loop CP--violating diagrams involved. In this work we 
present the full calculation of the expectation value of a set of these 
observables in the context of the MSSM and compare the size of the different 
contributions.
 
The paper is organized as follows.
In Section 2 we present the general effective vertex describing the interaction
of on--shell fermions with a neutral vector boson. The definitions and the 
generic expressions of the DFFs for all the contributing topologies at the 
one--loop level are also given.
In Section 3 we evaluate these expressions for on--shell $Z$ and the case
of the $\tau$ lepton and the $b$ quark (their AWMDM and WEDM) in the context 
of the SM and the MSSM. A discussion of the supersymmetric limit is also given.
The evaluation of the $t$ quark EDFFs and WEDFFs at $\sqrt{s}=500$ GeV 
is presented in Section 4. 
In Section 5 we review observables sensitive to the $\tau$ AWMDM at the $Z$
peak 
as well as other CP--odd observables valid in general.
In Section 6 we evaluate specific CP--odd observables for the $t$ quark pair
production in $e^+e^-$ colliders to one loop in the MSSM 
and compare the influence of the $t$ EDFF and WEDFF with
that of the CP--violating box diagrams.
Our conclusions are presented in Section 7.
In the Appendices one can find the definition of the one--loop 
tensor integrals and the SM and MSSM couplings and conventions that have 
been used. 

%
\section{The dipole form factors}

\subsection{The $Vff$ effective vertex}
%

The most general effective Lagrangian describing the interaction 
of a neutral vector boson $V$ with two fermions can be written, using at 
most dimension five operators, as a function of ten independent terms:
\beq
\hspace{-2mm}
{\cal L}_{Vff} 
&=& 
V^\mu (x) \bar{\Psi} (x) \Bigg[ \gmu \left(g_{\rm V} - g_{\rm A} \gfi \right) 
\ + \ {\rm i} \stackrel{\leftrightarrow}{\partial_\mu} 
\left(g_{\rm M} + {\rm i}g_{\rm E} \gfi \right)   \nn \\
& & \qquad \qquad \qquad +
{\rm i}\stackrel{\leftrightarrow}{\partial^\nu} \smunu 
\left( g_{\rm TS} + {\rm i} g_{\rm TP} \gfi \right) \Bigg] \Psi (x) \ \nn \\
&+& 
\left({\rm i} \partial^\mu V^\nu (x) \right) \bar{\Psi} (x) \Bigg[ 
g_{\mu \nu} \left({\rm i}g_{\rm S} + g_{\rm P} \gfi \right) \ + \ 
\smunu \left({\rm i} g_{\rm TM} + g_{\rm TE} \gfi \right) \Bigg] \Psi (x).
\label{efflag}
\eeq
The first two coefficients, i.e. $g_{\rm V}$ and $g_{\rm A}$, are the 
usual vector and axial--vector couplings. They are connected to chirality 
conserving dimension four operators. All the other coefficients in 
Eq.~(\ref{efflag}) multiply chirality flipping dimension five operators 
and can receive a contribution only through radiative corrections in a
renormalizable theory. The operators associated to $g_{\rm V}$, $g_{\rm A}$, 
$g_{\rm M}$, $g_{\rm P}$, $g_{\rm TM}$ and $g_{\rm TP}$ are even under a CP 
transformation. The presence of non vanishing $g_{\rm E}$, $g_{\rm S}$, 
$g_{\rm TE}$ and $g_{\rm TS}$ yields a contribution to CP--violating 
observables. 
In Table~\ref{tab11} we summarize the C, P, T and chirality properties of 
each operator introduced in the effective Lagrangian.
\begin{table}
\vspace{0.3cm}
\caption{\em C, P, T properties of the operators in the effective Lagrangian of 
Eq.~(\ref{efflag}). Their chirality flipping behavior is also displayed.}
\label{tab11}
\vspace*{0.2cm}
\begin{center}
\begin{tabular}{|l|c|c|c|c|c|}  
\hline 
Operator & Coefficient & P & CP & T & Chirality Flip \\
 \hline \hline
$V^\mu \bar{\Psi} \gmu \Psi$ 
         & $g_{\rm V}$ & $+$ & $+$ & $+$ & NO \\
$V^\mu \bar{\Psi} \gmu \gfi \Psi$ 
         & $g_{\rm A}$ & $-$ & $+$ & $+$ & NO \\
$V^\mu \bar{\Psi}{\rm i}\stackrel{\leftrightarrow}{\partial_\mu} \Psi$
         & $g_{\rm M}$ & $+$ & $+$ & $+$ & YES \\
$V^\mu \bar{\Psi} \stackrel{\leftrightarrow}{\partial_\mu} \gfi \Psi$
         & $g_{\rm E}$ & $-$ & $-$ & $-$ & YES \\
\hline 
$V^\mu \bar{\Psi}{\rm i}\stackrel{\leftrightarrow}{\partial^\nu} \smunu \Psi$
         & $g_{\rm TS}$ & $+$ & $-$ & $-$ & YES \\
$V^\mu \bar{\Psi} \stackrel{\leftrightarrow}{\partial^\nu} \smunu \gfi \Psi$
         & $g_{\rm TP}$ & $-$ & $+$ & $+$ & YES \\
$\left(\partial \cdot V \right) \bar{\Psi} \Psi$  
         & $g_{\rm S}$ & $+$ & $-$ & $-$ & YES \\
$\left({\rm i}\partial \cdot V \right) \bar{\Psi} \gfi \Psi$ 
         & $g_{\rm P}$ & $-$ & $+$ & $+$ & YES \\
$\left(\partial^\mu V^\nu \right) \bar{\Psi} \smunu \Psi$ 
         & $g_{\rm TM}$ & $+$ & $+$ & $+$ & YES \\
$\left({\rm i}\partial^\mu V^\nu \right) \bar{\Psi} \smunu \gfi \Psi$ 
         & $g_{\rm TE}$ & $-$ & $-$ & $-$ & YES \\
\hline
\end{tabular}
\end{center}
\vspace*{0.2cm}
\end{table}

By Fourier transform of Eq.~(\ref{efflag}) one obtains the most general 
Lorentz structure for the vertex $Vff$ in the momentum space:
\beq
\Gamma^{Vff}_\mu & = & {\rm i}  \Bigg[
\gmu \left(f_{\rm V} - f_{\rm A} \gfi \right) \ + \ 
(q-\bar{q})_\mu \left(f_{\rm M} + {\rm i}f_{\rm E} \gfi \right) + \ p_\mu 
\left({\rm i}f_{\rm S} + f_{\rm P} \gfi \right) \nn \\ 
& & \qquad  +
(q - \bar{q})^\nu \smunu 
\left( f_{\rm TS} + {\rm i}f_{\rm TP} \gfi \right) \ + \ 
p^\nu \smunu \left({\rm i}f_{\rm TM} + f_{\rm TE} \gfi 
\right) \Bigg],
\label{efflagm}
\eeq
where $q$ and $\bar{q}$ are the outgoing momenta of the fermions and 
$p = (q+\bar{q})$ is the total incoming momentum of the neutral boson $V$. 
The form factors $f_i$ are functions of the kinematical invariants. 
Actually they are more 
general than the coefficients $g_i$. In fact any operator of dimension higher 
than five added to the Lagrangian of Eq.~(\ref{efflag}) is related to a new 
coefficient $g_i$. But every new coefficient can contribute only to the ten 
independent form factors $f_i$ introduced in Eq.~(\ref{efflagm}). The 
parameters $g_i$ and the form factors $f_i$ can be complex in general. 
Their real parts account for dispersive effects (CPT--even) whereas 
their imaginary parts are related to absorptive contributions.

It is possible to lower the number of independent form factors in 
Eq.~(\ref{efflagm}) by imposing on--shell conditions on the 
fermionic and/or bosonic fields. For instance, in the case of on--shell 
fermions, making use of the Gordon identities:
\beq
2 m_f ~\bar{u}~ \gmup ~v & = & \Big\{ \bar{u}~(q - \bar{q})^\mu~v + 
\bar{u}~{\rm i}~(q + \bar{q})_\nu \smunup ~v \Big\}, \nn \\
2 m_f ~\bar{u}~ \gmup \gfi ~v & = & \Big\{ \bar{u}~(q + \bar{q})^\mu
\gfi~v + \bar{u}~{\rm i}~(q - \bar{q})_\nu \smunup \gfi ~v \Big\}, \nn \\
0 & = & \Big\{ \bar{u}~(q + \bar{q})^\mu ~v + 
\bar{u}~{\rm i}~(q - \bar{q})_\nu \smunup v \Big\}, \nn \\
0 & = & \Big\{ \bar{u}~(q - \bar{q})^\mu~\gfi ~v + 
\bar{u}~{\rm i}~(q + \bar{q})_\nu \smunup \gfi ~v \Big\}~,
\label{gordon}
\eeq
one can eliminate $f_{\rm TM}$, $f_{\rm TE}$, $f_{\rm TS}$ and $f_{\rm TP}$ 
from the effective Lagrangian. The number of relevant form factors can be 
further reduced taking also the boson $V$ on its mass shell. In this case the 
condition $p_\mu \eps^\mu=0$ automatically cancels all the contributions 
coming from $f_{\rm S}$ and $f_{\rm P}$. The same situation occurs for 
off--shell vector boson $V$ when, in the process $e^+e^- \to V^* \to f\bar{f}$, 
the electron mass is neglected. Therefore $f_{\rm S}$ and $f_{\rm P}$ will
be ignored in the following. With all these assumptions the $Vff$ effective 
vertex for on--shell fermions is conventionally written as: 
\bea
\hspace{-5mm}
\Gamma^{Vff}_\mu(s) = {\rm i}e
        \Bigg\{\gamma_\mu \left[V^V_f(s)-A^V_f(s) \gamma_5\right] + 
        \sigma_{\mu\nu} (q+\bar{q})^\nu \Bigg[ {\rm i}\frac{a^V_f(s)}{2m_f} - 
                \frac{d^V_f(s)}{e}\gamma_5 \Bigg]\Bigg\},
\label{vertex}
\eea
where $e$ and $m_f$ are respectively the electric unit charge and the mass
of the external fermion. The form factors in Eq.~(\ref{vertex}) depend only on 
$s$. As mentioned above $V^V_f(s)$ and $A^V_f(s)$ parameterize the vector and 
axial--vector current. They are 
connected to the chirality conserving CP--even sector. The form factors 
$a^{V}_f(s)$ and $d^{V}_f(s)$ are known respectively as anomalous magnetic 
dipole form factor (AMDFF) and electric dipole form factor (EDFF). They 
are both connected to chirality flipping operators. In a renormalizable 
theory they can receive contribution exclusively by quantum corrections. 
The EDFFs contribute to the CP--odd sector and constitute a source of 
CP--violation.

The {\em dipole moments} (DM) are defined taking the corresponding vector 
bosons on shell, $s=M^2_V$.
For $V=\gamma$ one gets the usual definitions of the photon {\em anomalous 
magnetic dipole moment} (AMDM) and {\em electric dipole moment} (EDM).
The definitions of electric charge, magnetic and electric 
dipole moments \cite{iz} consistent with our convention for the covariant 
derivative (\ref{covder}) are respectively:
\beq
      \mbox{charge}&\equiv& -e\ V^\gamma_f(0)=e\ Q_f\ , \\
      \mbox{MDM}   &\equiv& \frac{e}{2m_f}(V^\gamma_f(0)+a^\gamma_f(0))\ , \\
      \mbox{EDM}   &\equiv& d^\gamma_f(0)\ .
\eeq
Thus the AMDM of a fermion is $a^\gamma_f(0)=-Q_f(g_f-2)/2$ with 
$g_f$ being the gyromagnetic ratio. The axial-vector 
coupling $A^\gamma_f$ vanishes. For $V=Z$, the quantities $a^w_f\equiv 
a^Z_f(M^2_Z)$ and $d^w_f\equiv d^Z_f(M^2_Z)$ define the {\em anomalous 
weak--magnetic dipole moment} (AWMDM) and the {\em weak--electric dipole 
moment} (WEDM).\footnote{ 
Massless and neutral fermions may have magnetic moments. The usual 
parametrization in (\ref{vertex}) must be generalized by the replacement 
$e\ a^V_f(s)/2m_f\to\mu^V(s)$. The dipole moments of neutrinos are then
given by $\mu^V(M^2_V)$.} 

%
%
\subsection{One--loop generic expressions of the dipole form factors}
\begin{figure}[t]
\begin{center}
\begin{tabular}{c}
\epsfig{figure=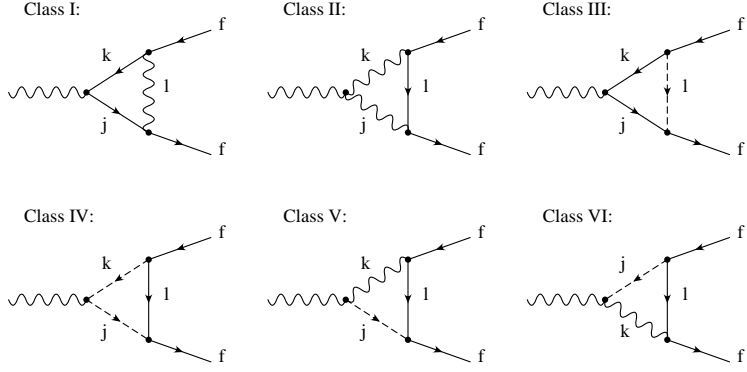,height=5cm}
\end{tabular}
\end{center}
\caption{\em The one-loop $Vff$ diagrams with general couplings.
\label{fig1}}
\vspace*{0.2cm}
\end{figure}

%
%
All the possible one--loop contributions to the $a^V_f(s)$ and $d^V_f(s)$ 
form factors can be classified in terms of the six classes of triangle 
diagrams depicted in Fig.~\ref{fig1}. 
The vertices involved are labelled by generic couplings for vector bosons
$V^{(k)}_\mu=A_\mu,\ Z_\mu,\ W_\mu,\ W^\dagger_\mu$, fermions $\Psi_k$ and 
scalar bosons $\Phi_k$, according to the following interaction Lagrangian:
\beq
{\cal L} 
& = & 
     {\rm i} e J (W^\dagger_{\mu\nu}W^\mu V^\nu - W^{\mu\nu}W^\dagger_\mu V_\nu
             + V^{\mu\nu}W^\dagger_\mu W_\nu)
     + {\rm i} e G_{jk} V^\mu \Phi_j^\dagger\stackrel{\leftrightarrow}
             {\partial}_\mu\Phi_k \nn \\
& & 
     + \Big\{e \bar{\Psi}_f(S_{jk}-P_{jk}\gamma_5)\Psi_k\Phi_j\ 
     + e K_{jk} V^\mu V^{(k)}_\mu\Phi_j + {\rm h.c.} \Big\} \nn \\  
& &
     + e V^{(k)}_\mu\bar{\Psi}_j\gamma^\mu(V^{(k)}_{jl}-A^{(k)}_{jl}\gamma_5)
       \Psi_l \ .
\label{genlag}
\eeq
Every class of diagrams is calculated analytically and expressed in terms 
of the couplings introduced in (\ref{genlag}) and the one--loop three--point 
integrals $\bar{C}_i$ (see App.~\ref{appendix-a}). 
The result is given in the 't Hooft-Feynman gauge.
\begin{itemize}
\item{[Class I]: vector--boson exchange:}
\beq
\frac{a^V_f(s)}{2 m_f}({\rm I}) 
&=& 
        \frac{\alpha}{4\pi}  \Big\{ 4m_f\sum_{jkl}{\rm Re} \Big[
   V^{(V)}_{jk}(V^{(l)}_{fj}V^{(l)*}_{fk} + A^{(l)}_{fj}A^{(l)*}_{fk})  \nn \\
& &
        \hspace{2truecm}+ A^{(V)}_{jk}(V^{(l)}_{fj}A^{(l)*}_{fk} + 
   A^{(l)}_{fj}V^{(l)*}_{fk}) \Big] \Big[2C^+_2 - 3C^+_1 + C_0\Big]_{kjl}
   \nn \\
& &
   \hspace{0.8truecm} + 4\sum_{jkl} m_k {\rm Re}
   \Big[V^{(V)}_{jk}(V^{(l)}_{fj}V^{(l)*}_{fk}-A^{(l)}_{fj}A^{(l)*}_{fk}) \nn \\
& & \hspace{2truecm}
    -A^{(V)}_{jk}(V^{(l)}_{fj}A^{(l)*}_{fk}-A^{(l)}_{fj}V^{(l)*}_{fk})\Big]
    \Big[2C^{+}_1 - C_0\Big]_{kjl} \Big\} 
\label{mdm1} \\
\frac{d^V_f(s)}{e}({\rm I}) 
&=& 
      \frac{\alpha}{4\pi}  \Big\{ 4m_f\sum_{jkl}{\rm Im}
      \Big[V^{(V)}_{jk}(V^{(l)}_{fj}A^{(l)*}_{fk}+A^{(l)}_{fj}V^{(l)*}_{fk})  \nn \\
& &  \hspace{2truecm} 
      +A^{(V)}_{jk}(V^{(l)}_{fj}V^{(l)*}_{fk}+A^{(l)}_{fj}A^{(l)*}_{fk}) \Big]
      \Big[2C^{+-}_2-C^-_1 \Big]_{kjl}  \nn \\
& &
    \hspace{0.8truecm} -4\sum_{jkl}m_k{\rm Im}
    \Big[V^{(V)}_{jk}(V^{(l)}_{fj}A^{(l)*}_{fk}-A^{(l)}_{fj}V^{(l)*}_{fk})\nn \\
& & \hspace{2truecm}
        -A^{(V)}_{jk}(V^{(l)}_{fj}V^{(l)*}_{fk}-A^{(l)}_{fj}A^{(l)*}_{fk}) \Big] 
        \Big[2C^{+}_1-C_0 \Big]_{kjl} \Big\}
\label{edm1}
\eeq
In the case of gluon exchange one has to substitute $\alpha$ for $\alpha_s$
and $V^{(l)}_{fi}$ for $T_l$ (the SU(3) generators) and take
$A^{(l)}_{fj}=0$. The sum over the index $l$ yields a 
color factor $C_F=4/3$. The gluon contribution to the CP--violating form factor 
$d^V_f({\rm I}$) vanishes in general.
\item{[Class II]: fermion exchange and two internal vector bosons:}
\beq
\frac{a^V_f(s)}{2 m_f}({\rm II}) 
&=& 
       \frac{\alpha}{4\pi} \Big\{ 2m_f\sum_{jkl}{\rm Re}
   \Big[J (V^{(j)}_{fl}V^{(k)*}_{fl} + A^{(j)}_{fl}A^{(k)*}_{fl}) \Big] 
             \Big[4C^{+}_2 + C^+_1 \Big]_{kjl} \nn \\
& &
   \hspace{1truecm} - 6\sum_{jkl} m_l{\rm Re} 
   \Big[J(V^{(j)}_{fl}V^{(k)*}_{fl} - A^{(j)}_{fl}A^{(k)*}_{fl}) \Big]
             \Big[C^+_1 \Big]_{kjl} \Big\} \\
\label{mdm2}
\frac{d^V_f(s)}{e}({\rm II}) 
&=&
       -\frac{\alpha}{4\pi} \Big\{ 2m_f\sum_{jkl}{\rm Im}
   \Big[J(V^{(j)}_{fl}A^{(k)*}_{fl}+A^{(j)}_{fl}V^{(k)*}_{fl}) \Big]
            \Big[4C^{+-}_2-C^-_1 \Big]_{kjl} \nn \\
& &
  \hspace{1truecm} + 6\sum_{jkl} m_l{\rm Im}
   \Big[J(V^{(j)}_{fl}A^{(k)*}_{fl}-A^{(j)}_{fl}V^{(k)*}_{fl}) \Big]
           \Big[ C^+_1 \Big]_{kjl} \Big\}
\label{edm2}
\eeq
\item{[Class III]: scalar exchange:}
\beq
\frac{a^V_f(s)}{2 m_f}({\rm III}) &=&
       \frac{\alpha}{4\pi}  \Big\{2m_f\sum_{jkl}{\rm Re}
   \Big[V^{(V)}_{jk}(S_{lj}S^*_{lk} + P_{lj}P^*_{lk}) \nn \\
& &
        \hspace{2truecm} + A^{(V)}_{jk}(S_{lj}P^*_{lk} + P_{lj}S^*_{lk}) \Big] 
    \Big[2C^{+}_2 - C^+_1 \Big]_{kjl} \nn \\
& & \hspace{0.8truecm}
 - 2\sum_{jkl} {m}_k {\rm Re}\Big[V^{(V)}_{jk}(S_{lj}S^*_{lk} - P_{lj}P^*_{lk})
 \nn \\
& & \hspace{2truecm}
    -A^{(V)}_{jk}(S_{lj}P^*_{lk}-P_{lj}S^*_{lk}) \Big] 
    \Big[C^+_1 + C^-_1 \Big]_{kjl} \Big\} 
\label{mdm3}\\
\frac{d^V_f(s)}{e}({\rm III})
&=&  
         \frac{\alpha}{4\pi} \Big\{-2m_f\sum_{jkl}{\rm Im}
   \Big[V^{(V)}_{jk}(P_{lj}S^*_{lk}+S_{lj}P^*_{lk}) \nn\\
& & \hspace{2truecm}
   +A^{(V)}_{jk}(S_{lj}S^*_{lk}+P_{lj}P^*_{lk}) \Big] 
   \Big[2C^{+-}_2-C^-_1 \Big]_{kjl} \nn \\
& & \hspace{0.8truecm}
   +2\sum_{jkl} {m}_k {\rm Im}\Big[V^{(V)}_{jk}(P_{lj}S^*_{lk}-S_{lj}P^*_{lk})
   \nn\\
& & \hspace{2truecm}
   +A^{(V)}_{jk}(S_{lj}S^*_{lk}-P_{lj}P^*_{lk}) \Big]
   \Big[C^+_1+C^-_1 \Big]_{kjl} \Big\}
\label{edm3}
\eeq
\item{[Class IV]: fermion exchange and two internal scalars:}
\beq
\frac{a^V_f(s)}{2 m_f}({\rm IV})
& = & 
       - \frac{\alpha}{4\pi}  \Big\{ 2 m_f\sum_{jkl}{\rm Re}
   \Big[G_{jk}(S_{jl}S^*_{kl} + P_{jl}P^*_{kl}) \Big]
   \Big[2C^{+}_2 - C^+_1 \Big]_{kjl} \nn \\
& & \hspace{0.8truecm}
   +\sum_{jkl} {m}_l {\rm Re}\Big[G_{jk}(S_{jl}S^*_{kl} - P_{jl}P^*_{kl}) \Big] 
   \Big[2C^+_1 - C_0 \Big]_{kjl} \Big\} 
\label{mdm4}\\
\frac{d^V_f(s)}{e}({\rm IV})
&=&
         \frac{\alpha}{4\pi}  \Big\{2m_f\sum_{jkl} {\rm Im}
  \Big[G_{jk}(S_{jl}P^*_{kl}+P_{jl}S^*_{kl})\Big]\Big[2C^{+-}_2-C^-_1\Big]_{kjl} 
   \nn \\  
& & \hspace{0.8truecm}
   -\sum_{jkl} {m}_l {\rm Im}\Big[G_{jk}(S_{jl}P^*_{kl} - P_{jl}S^*_{kl}) \Big]
   \Big[2C^+_1 - C_0 \Big]_{kjl} \Big\}
\label{edm4}
\eeq
\item{[Class V+VI]: fermion exchange, one vector and one
                    scalar internal boson:}
\beq
\frac{a^V_f(s)}{2 m_f}({\rm V+VI})
&=&
         \frac{\alpha}{4\pi} 2\sum_{jkl}{\rm Re} 
   \Big[K_{jk}(V^{(k)}_{fl}S_{jl}^*+A^{(k)}_{fl}P_{jl}^*) \Big]
   \Big[C^+_1+C^-_1 \Big]_{kjl} \\ 
\label{mdm56}
\frac{d^V_f(s)}{e}({\rm V+VI})
&=&
         -\frac{\alpha}{4\pi} 2\sum_{jkl}{\rm Im}
   \Big[K_{jk}(V^{(k)}_{fl}P_{jl}^*+A^{(k)}_{fl}S_{jl}^*) \Big]
   \Big[C^+_1+C^-_1 \Big]_{kjl}
\label{edm56}
\eeq
\end{itemize}
In Eqs.~(\ref{mdm1}--\ref{edm56}) the shorthand notation $[\bar{C}]_{kjl}$ 
stands for the three--point tensor integrals $\bar{C}(-\bar{q},q,M_k,M_j,M_l)$. 
The integrals appearing in the previous Eqs. are UV and IR finite. 
All the expressions are proportional to some positive 
power of a fermion mass, either internal or external, consistently with 
the chirality flipping character of the dipole moments.\footnote{
This is as expected when applying the mass--insertion method: to induce a 
flip in the fermion chirality one introduces a mass in either one of the 
internal fermion lines, picking a mass term from the propagator, or in the 
external fermion lines, using the equations of motion for the free fermion.}
For class V and VI diagrams the mass dependence is hidden in the product 
of the Yukawa couplings $S_{ij}$ ($P_{ij}$) and the dimensionful 
parameter $K_{ij}$. Hence, the heaviest fermions are the most promising 
candidates to have larger DFFs. Eqs.(\ref{mdm1}--\ref{edm56}) also show 
that, in general, the DFFs for massless fermions are not vanishing 
but proportional to masses of fermions running in the loop. The SM 
cancellation of the massless neutrino DFFs is only due to the absence of
right--handed neutrinos. Finally, 
notice that all the contributions to the EDFFs are proportional to the 
imaginary part of certain combinations of couplings. A theory with real 
couplings has manifestly vanishing EDFFs.
%

%
\section{SM and MSSM predictions for the \boldmath{$\tau$} and \boldmath{$b$} 
         WDMs}
%
%
In the previous Section we derived the DFFs for the generic interaction 
Lagrangian of Eq.~(\ref{genlag}). In this Section we focus in 
the calculation of the WDMs in two specific models: the SM and the MSSM. 
These quantities, defined at the $Z$ peak, are gauge invariant and can be
directly related to observables.
The $\tau$ lepton and the $b$ quark are the heaviest fermions to which
a $Z$ boson can decay and hence the optimal candidates to have larger
weak dipole moments.
%
\subsection{The $\tau$ and $b$ magnetic and electric WDMs in the SM}  
\begin{figure}[t]
\begin{center}
\begin{tabular}{cc}
\epsfig{figure=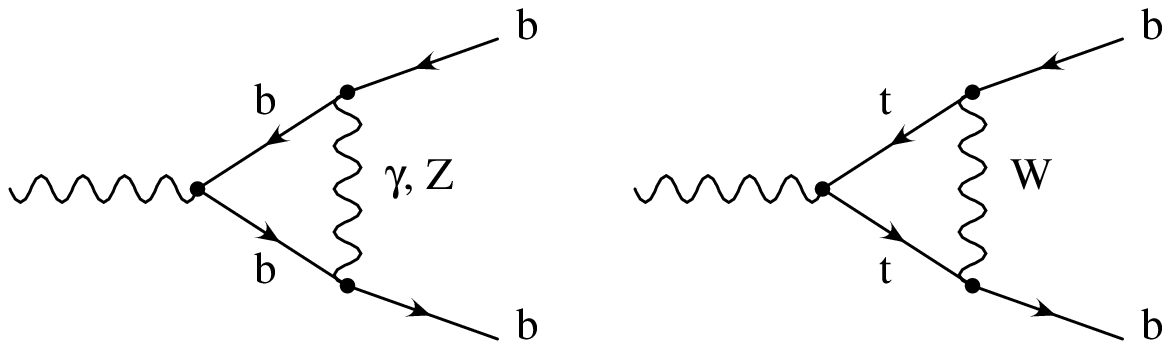,height=2cm,angle=0} &
\epsfig{figure=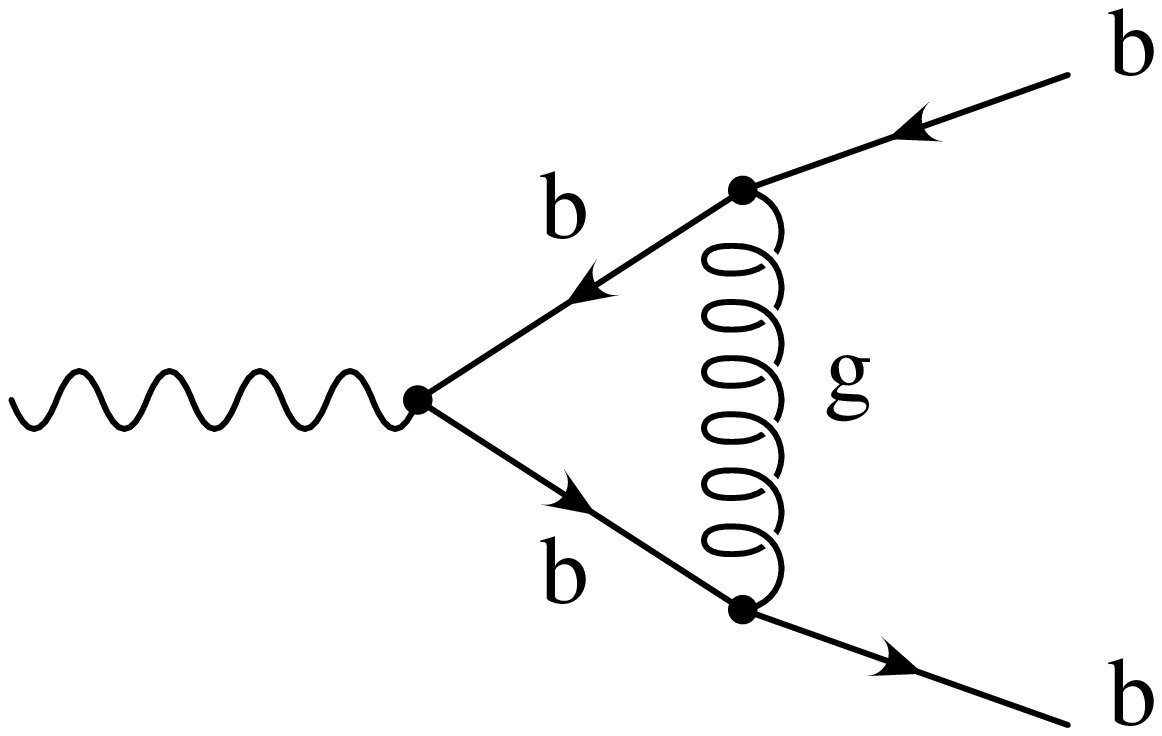,height=2cm,angle=0}
\epsfig{figure=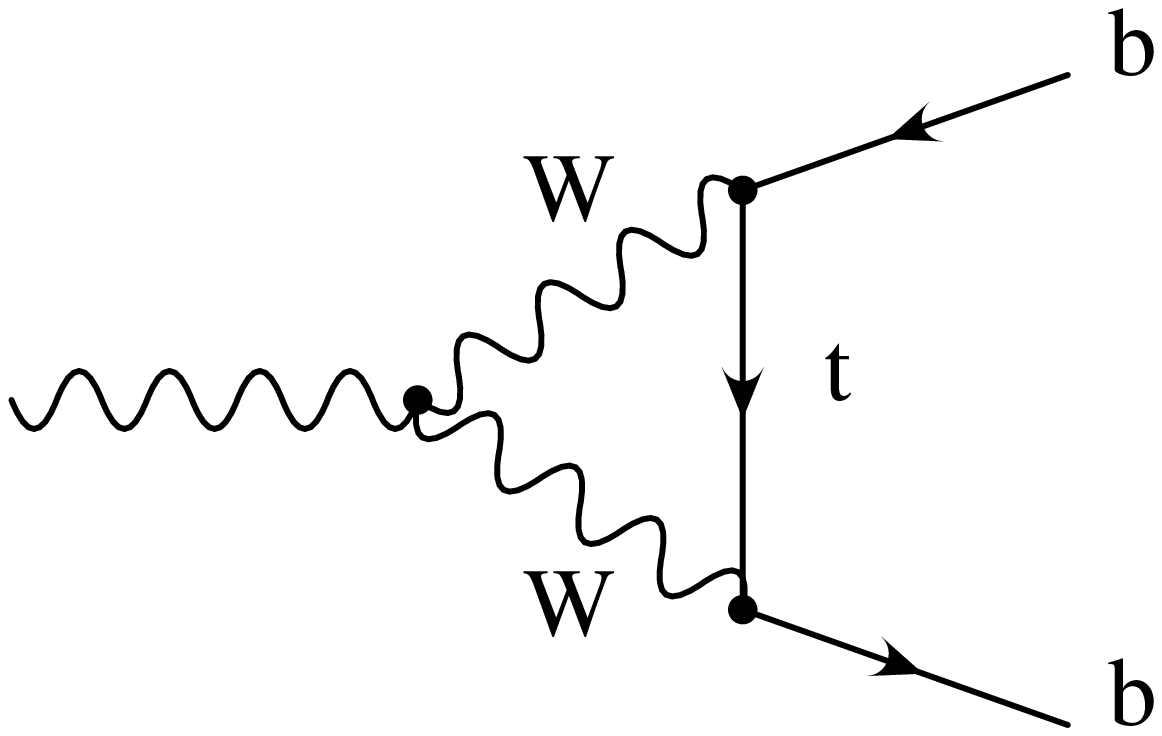,height=2cm,angle=0}
\end{tabular}
\end{center}
\caption{\em SM diagrams for the $Zbb$ vertex (Higgs sector excluded).}
\label{fig31a}
\vspace*{0.3cm}
\end{figure}

\begin{figure}[t]
\begin{center}
\begin{tabular}{cc}
\epsfig{figure=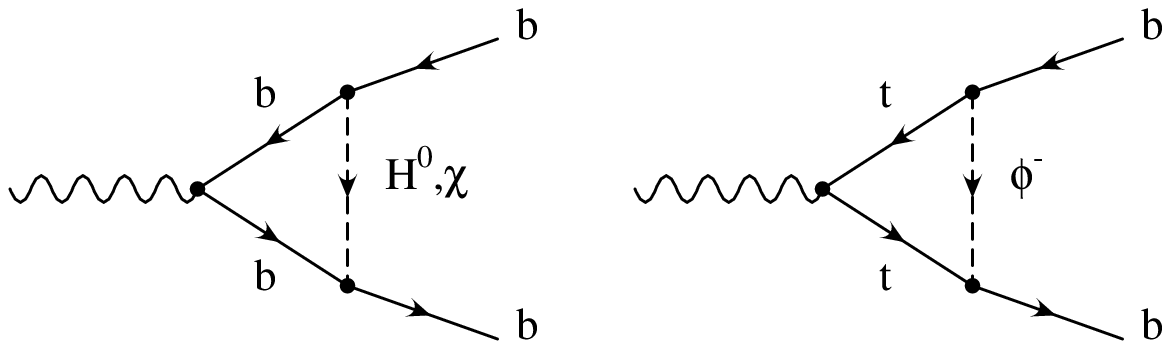,height=2cm,angle=0}&
\epsfig{figure=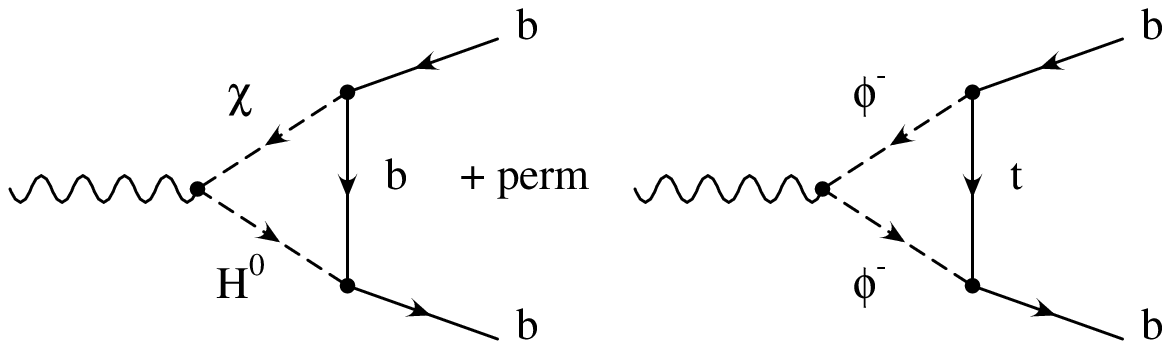,height=2cm,angle=0}\\
\epsfig{figure=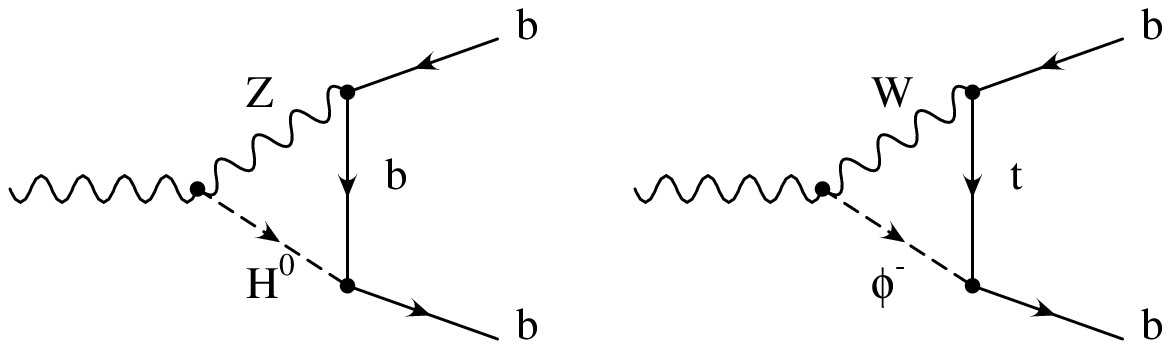,height=2cm,angle=0} &
\epsfig{figure=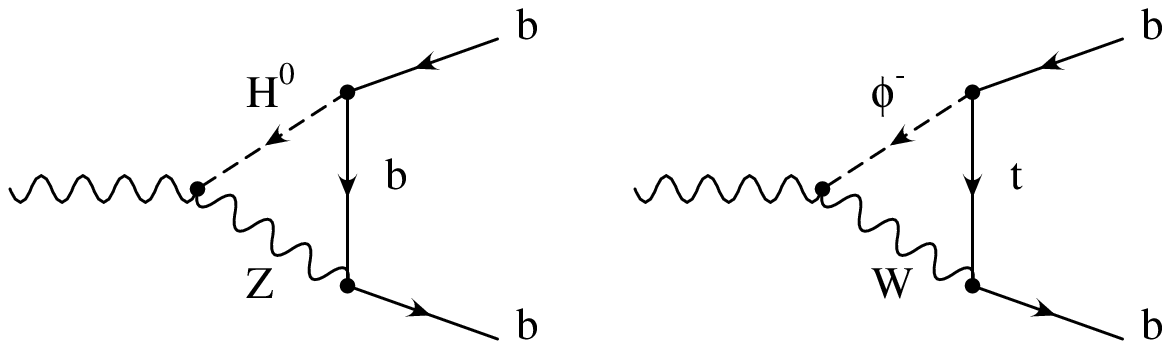,height=2cm,angle=0}
\end{tabular}
\end{center}
\caption{\em SM Higgs diagrams for the $Zbb$ vertex.}
\label{fig31b}
\vspace*{0.3cm}
\end{figure}

%
In the electroweak sector of the SM, adopting the 't Hooft-Feynman gauge, 
there are 14 diagrams contributing at one loop to the $Zff$ boson 
interaction. One more diagram has to be considered (the gluon exchange) if also 
QCD is included. All the possible SM diagrams for the $Zbb$ vertex 
are given in Figs.~\ref{fig31a} and \ref{fig31b}.
The calculation in Ref.~\cite{hirs1} is in perfect agreement with the results 
of Refs.~\cite{ber95,ber97} 
for both the $\tau$ lepton and the $b$ quark. Taking as input parameters 
$m_{\tau}=1.777$ GeV, $m_b=4.5$ GeV, $m_t=175$ GeV, $M_Z=91.19$ GeV, 
$s^2_W=0.232$, $\alpha=1/128$ and $\alpha_s=0.118$, the pure electroweak 
contributions are:
\beq  
a^w_{\tau}[{\rm EW}] &=& (2.10+0.61~{\rm i}) \times 10^{-6}, 
\label{tausmwm}\\ 
a^w_b[{\rm EW}] &=& [(1.1;\ 2.0;\ 2.4)-0.2~{\rm i}]\times10^{-6}, 
\label{bewwm}
\eeq
for three different values of the Higgs mass (respectively $M_{H^0}=M_Z$, 
$2M_Z$ and $3M_Z$). Note that for 
the $\tau$ case the Higgs boson contribution has a very small impact, 
changing the final result by less than $1 \%$ in the range of the Higgs boson 
masses considered. A much stronger dependence is observed for the $b$ case. 
The CKM matrix element $V_{tb}$ is taken  
equal to one and all the off-diagonal entries vanishing. The whole SM 
contribution (EW+QCD) for the $b$ quark is dominated by the gluon exchange:
\bea
a^w_b[{\rm SM}]=(-2.98 + 1.56~{\rm i}) \times 10^{-4}. \nn
\label{bsmwm}
\eea
This large enhancement is as expected by the replacement $\alpha\raw \alpha_s$. 

The only phase present in the SM, the $\delta_{\rm CKM}$, is not sufficient 
for generating one-loop contributions to the (W)EDM.\footnote{We implicitly 
assume here a vanishing $\theta_{\rm QCD}$ phase.} One needs to go beyond the 
two--loop level \cite{smedm}. A very crude estimate 
of the tiny three--loop SM contribution to the (W)EDM can be done using simple 
power counting arguments. For a general fermion with mass $m_f$ 
$d^w_f[{\rm SM}]$ is of the order of $e G_F m_f\alpha^2\alpha_sJ/(4\pi)^5$ with 
$J=c_1c_2c_3s_1^2s_2s_3s_{\delta}$ being an invariant under reparameterizations
of the CKM matrix \cite{dipoles}. For the $\tau$ lepton and the $b$ quark we 
have: 
\bea
d^w_{\tau}[{\rm SM}] \lsim 1.5\times 10^{-19}\ \mu_\tau \qquad {\rm and} \qquad 
d^w_b[{\rm SM}] \lsim 1.5\times 10^{-18}\ \mu_b .
\label{smwedm}
\eea 
We use the scale unit $\mu_f \equiv e/2 m_f$ (``magneton") with
$\mu_\tau\ (\mu_b) = 5.55\ (2.19)\times 10^{-15}\ e$cm, respectively.
%
\subsection{The $\tau$ and $b$ AWMDMs in the MSSM}  
\begin{figure}[t]
\begin{center}
\begin{tabular}{cc}
\epsfig{figure=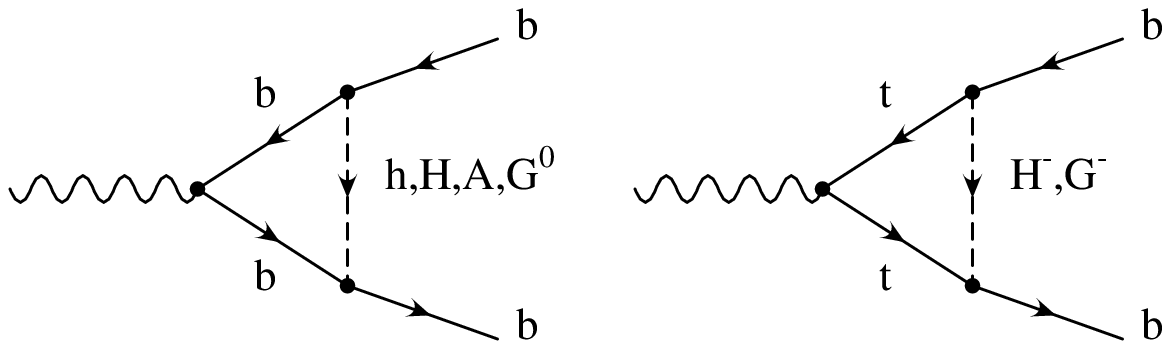,height=2cm,angle=0} &
\epsfig{figure=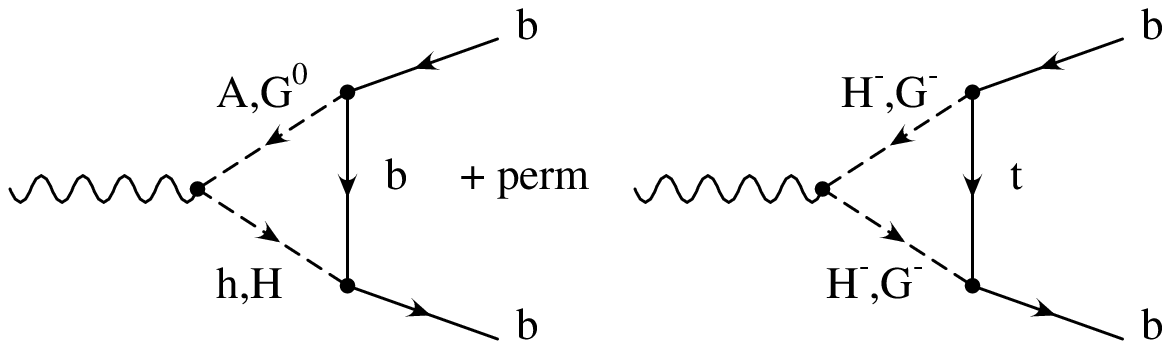,height=2cm,angle=0} \\
\epsfig{figure=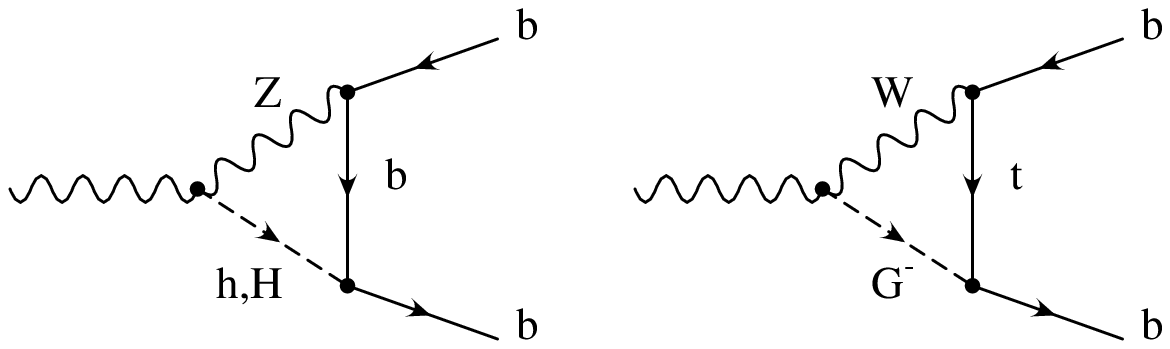,height=2cm,angle=0} &
\epsfig{figure=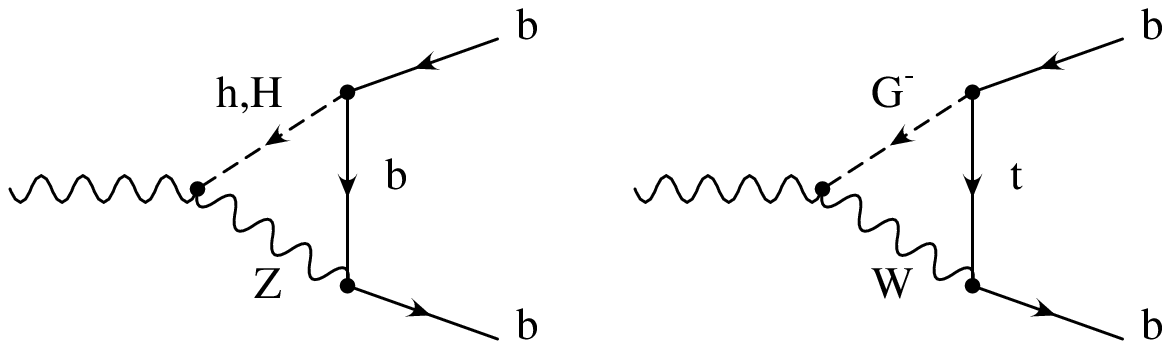,height=2cm,angle=0} 
\end{tabular}
\end{center}
\caption{\em MSSM Higgs diagrams for $Zbb$.}
\label{fig32a}
\vspace*{0.3cm}
\end{figure}

\begin{figure}[t]
\begin{center}
\begin{tabular}{cc}
\epsfig{figure=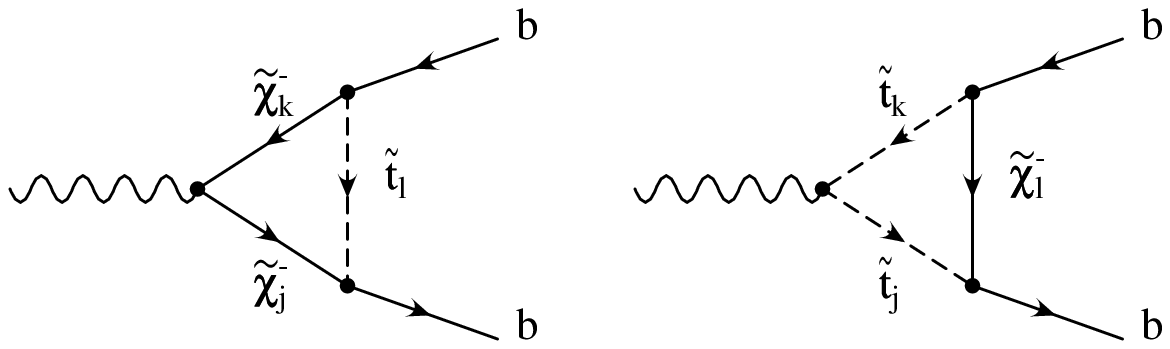,height=2cm,angle=0} & \\
\epsfig{figure=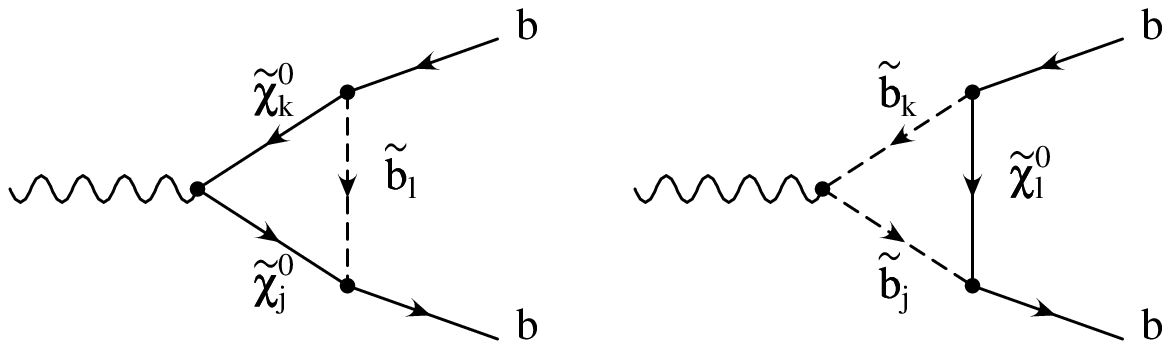,height=2cm,angle=0} &
\raisebox{1cm}[0cm][0cm]{\epsfig{figure=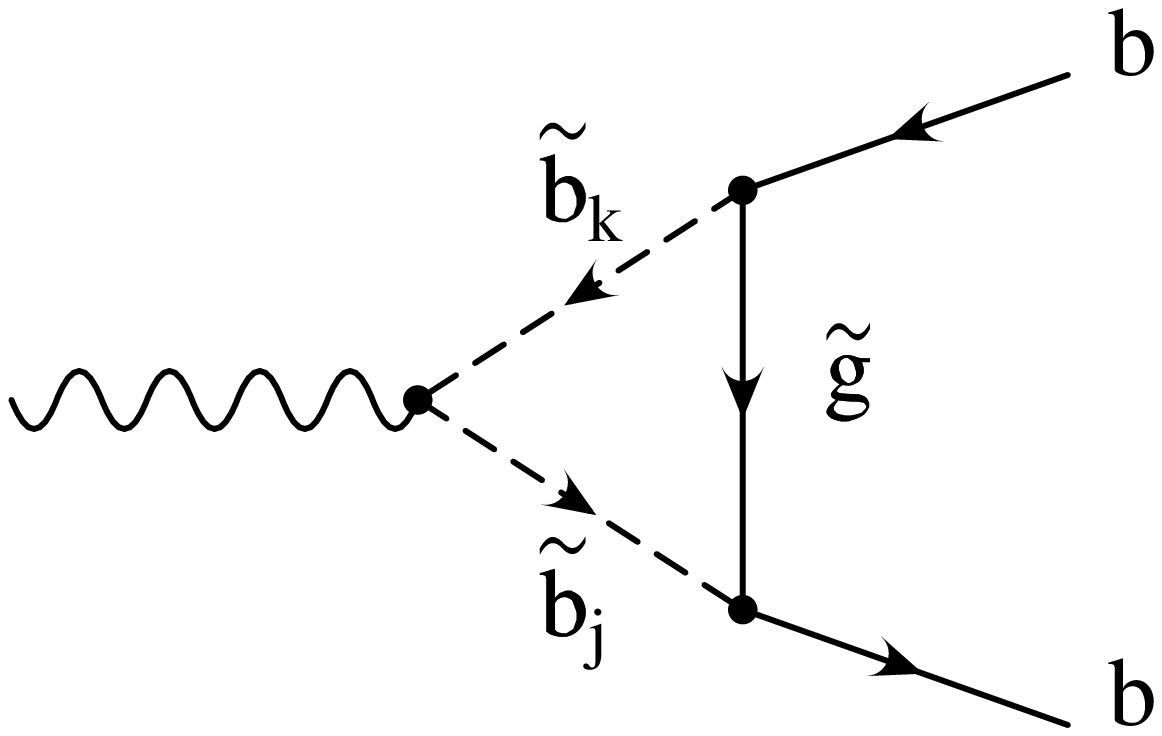,height=2cm,angle=0}} 
\end{tabular}
\end{center}
\caption{\em MSSM diagrams for $Zbb$.}
\label{fig32b}
\vspace*{0.3cm}
\end{figure}

%
The particle content of the MSSM includes the SM spectrum, the SUSY
partners and an extended Higgs sector consisting of two doublets.
We assume implicitly R--parity conservation.
The new non SM diagrams are depicted in Figs.~\ref{fig32a} and \ref{fig32b}.
They will be grouped into sets of diagrams including Higgs bosons, 
neutralinos, charginos and gluinos, as in Ref.~\cite{hirs1}. All of them 
belong to the topologies of classes III, IV, V and VI. 

The MSSM involves a large set of new parameters currently unknown but 
constrained by present experimental data. For sake of simplicity we 
assume the usual GUT relations for the soft--breaking gaugino mass terms,
\bea
\quad M_1=\frac{5}{3}\tan^2\theta_W M_2\ , 
\quad M_3=\frac{\alpha_s}{\alpha} s^2_W M_2
\label{gut}
\eea
and a common scalar fermion mass parameter. Moreover we restrict ourselves to 
two typical scenarios, low and high $\tan\beta$, set respectively to 
$\tan\beta = 1.6$ and $\tan\beta = 50$.\footnote{Currently our adopted value
for the low $\tan\beta$ scenario is close to be excluded by Higgs boson
searches \cite{moriond98} but we keep it for reference.}
We also assume generation--diagonal, not universal, trilinear soft--breaking 
terms in order to prevent large FCNC.\footnote{For a review about FCNC effects
see for instance \cite{fcnc-cp}.}
We are left, for given values of $\tan\beta$, with the following free 
parameters: 
the gaugino mass parameter for the SU(2) sector $M_2$, the Higgs--higgsino 
mass parameter $\mu$, the mass of the pseudoscalar Higgs boson $M_A$, the 
common scalar lepton soft--breaking mass parameter $m_{\tilde l}$, the common 
scalar quark soft-breaking mass parameter $m_{\tilde q}$ and the trilinear 
soft--breaking terms $A_f$ (for $f=b,t$ or $\tau$). The conventions 
for couplings and mixing are extensively explained in the Appendices. 
In this Section all the supersymmetric parameters are taken real.
%
\subsubsection{MSSM Higgs contribution to the AWMDMs}
%
The MSSM $Zbb$ diagrams involving Higgs bosons are shown in 
Fig.~\ref{fig32a}. There $G^0$ and $G^\pm$ are the would--be--Goldstone 
bosons and $h$, $H$, $H^\pm$ and $A$ are the physical scalar and 
pseudoscalar MSSM Higgs bosons. Actually not all these diagrams contribute 
to the AWMDM: the class IV diagrams with neutral Higgs bosons 
(necessarily of opposite parity) identically vanish, also in the SM, due
to the way their couplings to fermions (\ref{c78}--\ref{c83},\ref{c85}) combine 
in ({\ref{mdm4}). 
There are only three independent parameters involved in the Higgs sector: 
$\tan\beta$, the mass of the pseudoscalar $M_A$, and the common scalar fermion mass 
parameter $m_{\tilde f}$. 
The Higgs contribution decreases with growing $M_A$ and 
$m_{\tilde f}$, consistently with the decoupling theorem 
\cite{decoupling,herrero}.

The Higgs contribution to the Re($a^w_\tau$) coming from the diagrams of 
classes III and IV are proportional to $(m_\tau/M_Z)^4\tan^2\beta$ and those 
from the diagrams of classes V and VI are proportional to $(m_\tau/M_Z)^2$. 
Due to the fourth power suppression the sum is not very sensitive to 
$\tan\beta$. We obtain 
\bea
{\rm Re} \left(a^w_\tau[{\rm Higgs}] \right) = -0.3~(-0.4) 
         \times 10^{-6} \quad \mbox{for} \quad \tan\beta = 1.6~(50), \nn
\eea
$m_{\tilde q} = 250$ GeV and $M_A=100$ GeV. 
Assuming the present lower bounds for the masses of the Higgs bosons 
\cite{higgs,moriond98} the only contribution to the Im($a^w_\tau$) can 
arise from class III diagrams and it is in general smaller than the 
SM one. 

The same considerations apply for the Higgs contribution 
to the $b$ AWMDM. The only relevant difference is the presence of a
dominant term proportional to $(m_b m_t/M_Z^2)^2$. 
The Re($a^w_b$) is: 
\bea
{\rm Re}\left(a^w_b[{\rm Higgs}]\right) = -3.8~(+0.8) 
        \times 10^{-6} \quad \mbox{ for } \quad \tan\beta = 1.6~(50),
\nonumber
\eea
$m_{\tilde q} = 250$ GeV and $M_A=100$ GeV. Again the contribution to the 
Im$(a^w_b)$ is negligible. 
%
\subsubsection{Chargino contribution to the AWMDMs}
%
There are two diagrams with charginos and scalar fermions (Fig.~\ref{fig32b}). 
In the region of MSSM parameters not ruled out by present experiments 
\cite{moriond98} charginos and 
scalar fermions cannot be pair produced in $Z$ decays. Therefore they can 
contribute only to the real part of the AWMDM. The free parameters involved 
are $\tan\beta$, $M_2$, $\mu$, $m_{\tilde f}$ (and $A_{t}$ 
in the $b$ case, since there is no mixing in the scalar neutrino sector). 
Increasing the scalar fermion common mass only produces  
decoupling of the scalar fermion from the spectrum. In the large $M_2$ and $|\mu|$ 
region also the charginos decouple.
\begin{figure}[t]
\begin{center}
\begin{tabular}{cc}
\epsfig{figure=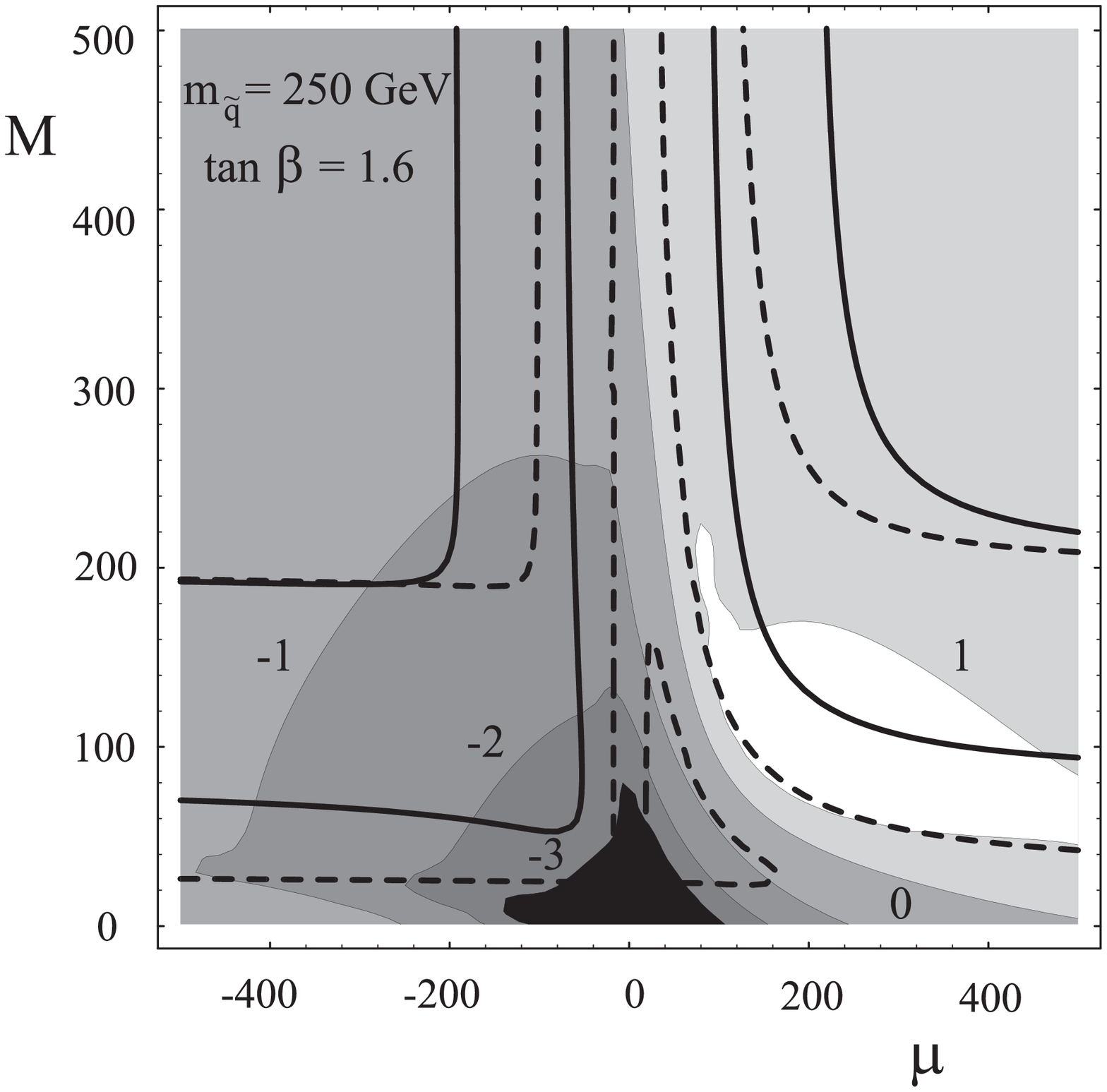,height=6.5cm,angle=0} &
\epsfig{figure=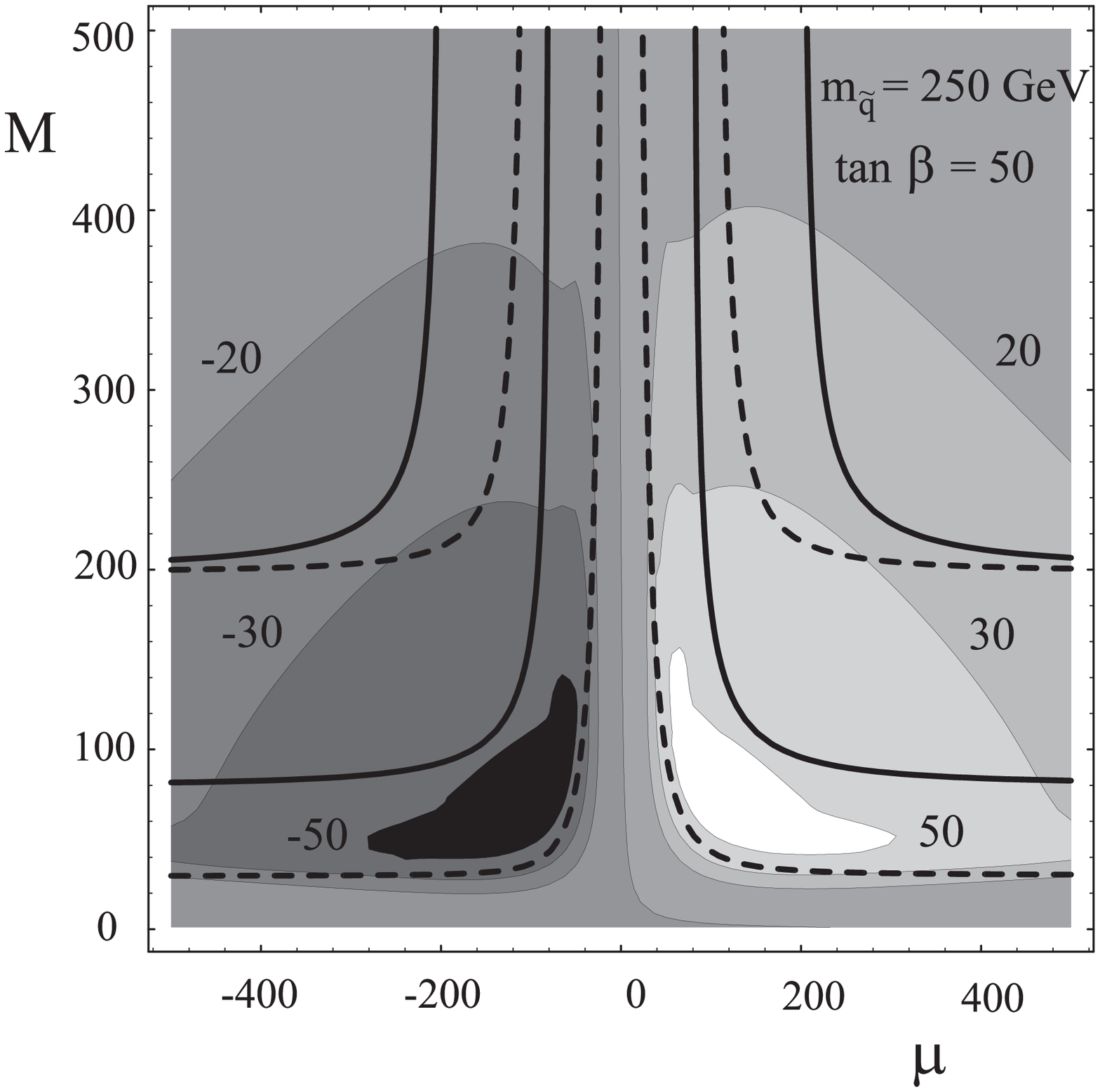,height=6.5cm,angle=0}\vspace*{-5mm}\\
(a) & (b)
\end{tabular}
\end{center}
\caption{\em Re($a^w_b$[char]) for low (a) and high (b) $\tan\beta$ in 
units of $10^{-6}$ in the plane $M_2-\mu$. The contour solid-lines 
correspond to the lightest chargino masses $m_{\tilde\chi^\pm}=90$ and $200$ 
GeV and the contour dashed-lines correspond to the lightest neutralino 
masses $m_{\tilde\chi^0}=15$ and $100$ GeV. The common scalar quark mass 
parameter is fixed to $m_{\tilde q}=250$ GeV and $m^t_{LR}=0$. From 
\cite{hirs1}.\label{fig:mdmchar}}
\vspace*{0.3cm}
\end{figure}

The chargino contribution to the real part of the $\tau$ AWMDM is:
\bea
{\rm Re} \left(a^w_\tau[{\rm char}]\right) = \pm 0.2~(\pm 7.0) \times10^{-6} 
         \quad \mbox{for} \quad \tan\beta = 1.6~(50), \nn
\eea
with $m_{\tilde q}=250$ GeV and $M_2=|\mu|=200$ GeV. The enhancement for higher 
$\tan\beta$ is due to the dependence of the chargino and neutralino 
couplings on the Higgs vacuum expectation values. 

For the real part of the $b$ AWMDM one gets:
\bea
{\rm Re} \left(a^w_b[{\rm char}]\right)= \pm 1.1~(\pm 33) \times10^{-6} 
         \quad \mbox{for} \quad \tan\beta = 1.6~(50), \nn 
\eea
with $m_{\tilde q}=250$ GeV, $M_2=|\mu|=200$ GeV and $A_t=\mu\cot\beta$.
The $\pm$ stands for ${\rm sign} (\mu)$. Varying the value 
of $A_t$ in the range $\pm \mu \cot\beta$ there is only a small effect 
in the AWMDM. We thus choose the value of $A_t$ that makes the 
off--diagonal entry of the $\tilde{t}$ scalar quarks mixing mass matrix vanish.
In such a way the scalar fermion mass eigenstates are always physical. 
Fig.~\ref{fig:mdmchar} shows the contour plot of the ${\rm Re} 
(a^w_b[{\rm char}])$ in the $M_2-\mu$ plane. The contour lines for fixed  
masses of the lightest chargino (90 and 200 GeV) and lightest neutralino 
(15 and 100 GeV) are also given for illustration. 
The contributions are enhanced by increasing 
$\tan\beta$. In the low $\tan\beta$ scenario the contribution is of order 
$10^{-6}$ (Fig.~\ref{fig:mdmchar}a), while in the high $\tan\beta$ 
scenario it becomes of order $10^{-5}$ (Fig.~\ref{fig:mdmchar}b).
The contour plots for the $\tau$ AWMDM are similar to the $b$ quark ones but 
suppressed by factors between $m_\tau/m_b$ and $(m_\tau/m_b)^2$ 
[see Eqs.~(\ref{mdm3}) and (\ref{mdm4})].
%
\subsubsection{Neutralino contribution to the AWMDMs}
%
The two diagrams involving neutralinos and scalar fermions are shown in 
Fig.~\ref{fig32b}. The scalar fermion masses are bounded by present experiments
to be heavy  but the neutralino masses can be lighter than half the mass of 
the $Z$ \cite{moriond98}. This allows 
for the possibility of a contribution to the imaginary part of the AWMDM 
through class III diagrams. A non negligible effect can arise 
only extremely near to the neutralino production threshold. 
The neutralino contribution to the 
AWMDM depends on $\tan\beta$, $M_2$, $\mu$, $m_{\tilde f}$, as well as the 
trilinear soft--breaking mass term $A_f$. No sizeable deviations due to 
changes in $A_f$ in the range $\pm \mu \tan\beta$ are observed. To avoid an
unphysical mass for the fermion we take $m^f_{LR}= 0$. 
The usual decoupling properties for scalar and fermions have been successfully 
checked.

For the real parts of the $\tau$ and $b$ AWMDM one gets:
\beq
{\rm Re} \left(a^w_\tau[{\rm neut}]\right) &=& -0.02~(-0.4) \times10^{-6} 
         ,\\
{\rm Re} \left(a^w_b[{\rm neut}] \right) &=& -0.2~(-10.5) \times10^{-6} \quad 
         \mbox{for} \quad \tan\beta = 1.6~(50). 
\eeq
and $M_2=\mu=200$ GeV. Like in the chargino contributions the result is 
enhanced with increasing $\tan\beta$.

%
\subsubsection{Gluino contribution to the $b$ AWMDM}
%
In the case of the $b$ AWMDM one more diagram must be considered: the one 
with a gluino and two $\tilde b$ scalar quarks running in the loop. Due to the 
$\alpha \raw \alpha_s$ replacement, we expect from the gluino sector a 
large contribution to the $b$ AWMDM. The present bounds on $\tilde b$ 
scalar quark masses \cite{moriond98} only allow a real contribution to the 
AWMDM. The free parameters are $\tan\beta$, $m_{\tilde q}$, $m_{\tilde g}$, 
$A_b$ and $\mu$. The last two affect the result only through the 
off--diagonal term of the $\tilde b$ scalar quark mass matrix $m^b_{LR}=A_b - 
\mu\tan\beta$. This parameter is responsible for the mixing in the $\tilde b$ 
sector and intervenes in association with the chirality flipping in the gluino 
internal line. Thus the contribution to the AWMDM is almost proportional to 
$m^b_{LR}$. We assume $\mu\tan\beta$ as typical scale for $m^b_{LR}$.

In Fig.~\ref{fig:mdmgluinos} the gluino contribution to the $b$ 
AWMDM is displayed in the $m^b_{LR}-m_{\tilde g}$ plane. In the  
low $\tan\beta$ region (Fig.~\ref{fig:mdmgluinos}a) the value of 
$m^b_{LR}$ does not affect significantly the mass of the lightest 
$\tilde b$ scalar quark (always above 200 GeV). For zero gluino mass, only the 
term proportional to the mass of the $b$ quark provides a contribution. 
As we increase the gluino mass, the term proportional to $m_{\tilde g}$ 
dominates, especially for large $m^b_{LR}$, becoming again suppressed at 
high $m_{\tilde g}$ due to the gluino decoupling. For high $\tan\beta$ 
the behavior is analogous but larger values for the AWMDM can be obtained 
as shown in Fig.~\ref{fig:mdmgluinos}b (extremely high values 
of $m^b_{LR}$ are excluded due to unphysical $m^2_{\tilde b_1}<0$).
Typical values are:
\beq
{\rm Re} \left(a^w_b[{\rm glu}] \right) &=& -1.1 \times 10^{-6} \qquad 
{\rm for}~\tan\beta = 1.6,~m^b_{LR}=200~{\rm GeV},
\label{glu1} \\
{\rm Re} \left(a^w_b[{\rm glu}]\right) &=& -18.9 \times 10^{-6} \qquad 
{\rm for}~\tan\beta = 50,~m^b_{LR}=7~{\rm TeV}, 
\label{glu2}
\eeq
with the gluino mass fixed to $m_{\tilde g} = 200$ GeV. Using the GUT
constraint (\ref{gut}) higher values for the gluino mass ($m_{\tilde g}\equiv 
M_3$) seem more likely and in this case the results in Eqs.~(\ref{glu1},
\ref{glu2}) are approximately reduced linearly with $m_{\tilde g}$.
\begin{figure}
\begin{center}
\begin{tabular}{cc}
\epsfig{figure=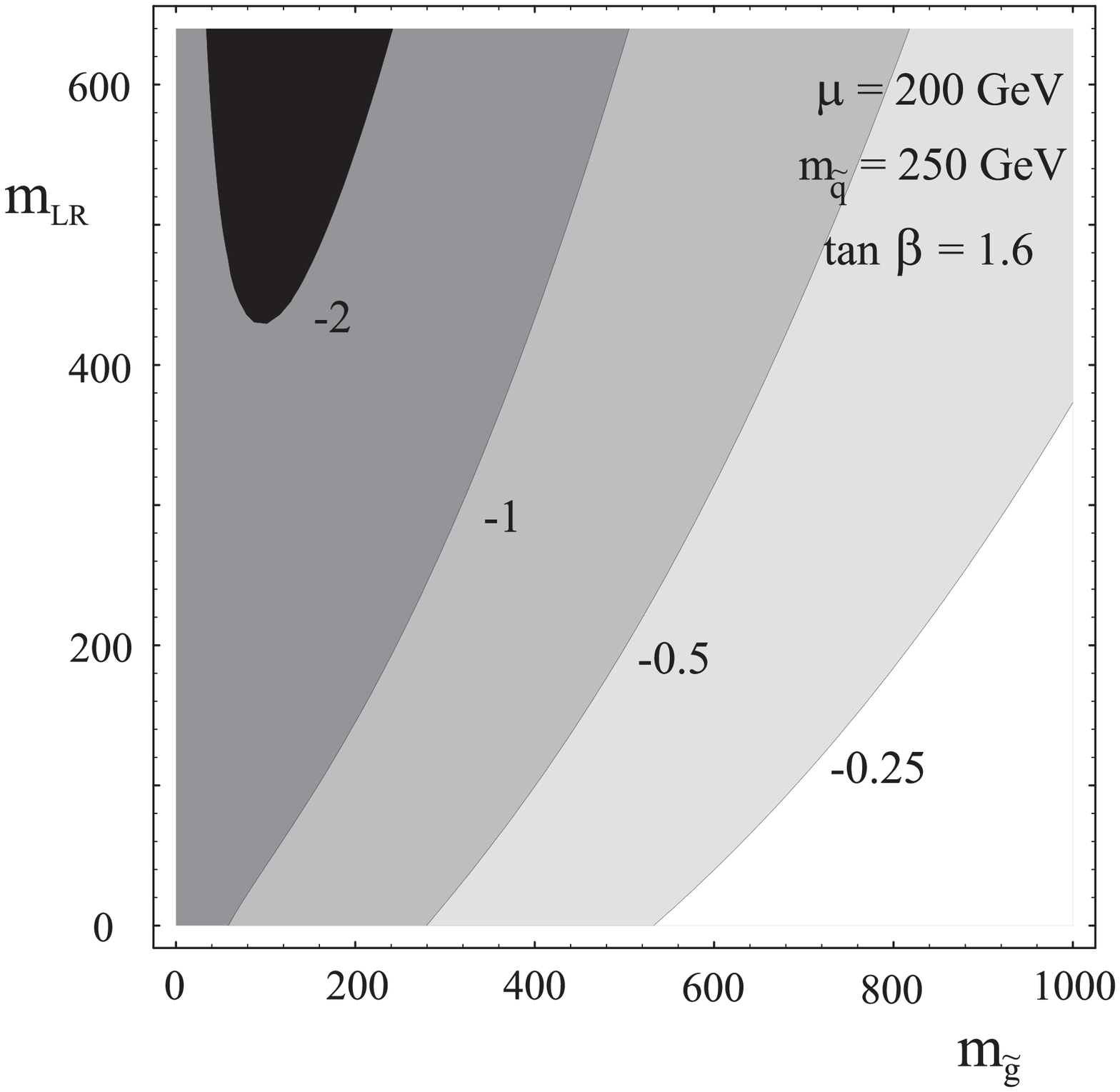,height=6.4cm,angle=0} &
\epsfig{figure=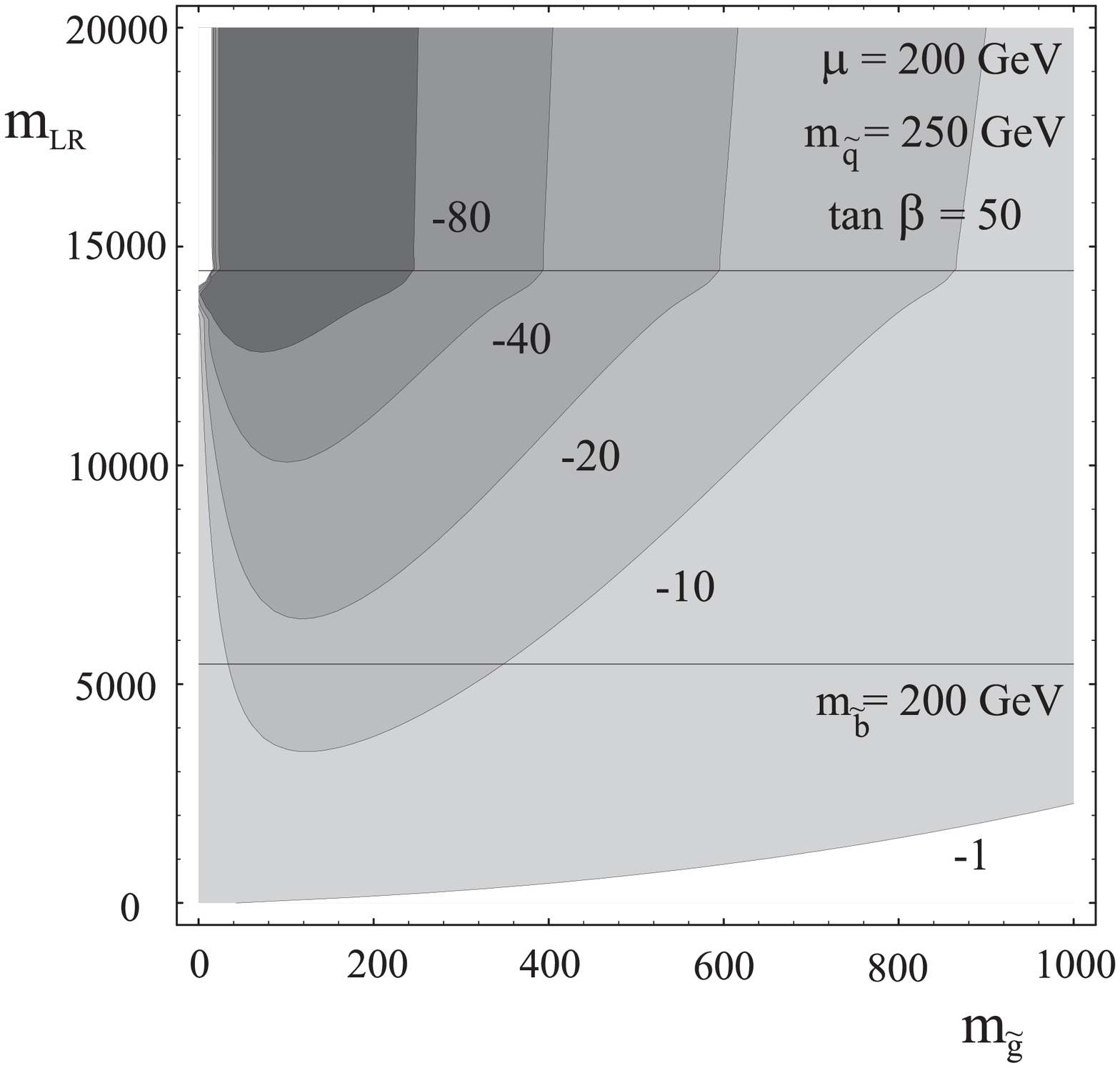,height=6.4cm,angle=0}\vspace*{-5mm}\\
(a) & (b)
\end{tabular}
\end{center}
\caption{\em Re($a^w_b$[glu]) low (a) and high (b) $\tan\beta$ in units 
of $10^{-6}$ in the plane $m^b_{LR}-m_{\tilde g}$. The common scalar quark mass 
parameter is fixed to $m_{\tilde q}=250$ GeV. The region above the upper 
horizontal line is unphysical ($m^2_{\tilde b_1}<0$). From \cite{hirs1}.
\label{fig:mdmgluinos}}
\vspace*{0.3cm}
\end{figure}
%
\subsubsection{MSSM total contribution to the AWMDMs}
%
The imaginary parts of the $\tau$ lepton and $b$ quark AWMDM are provided only 
by the diagrams with Higgs bosons and neutralinos. The total contribution in 
a reasonable region of the MSSM parameter can reach at maximum $10^{-6}$ 
and is typically lower than the corresponding SM contribution.

The real part is dominated by the chargino and the gluino sectors, when 
present. It is roughly proportional to $\tan\beta$. The total contribution 
can reach the values: 
\beq
|{\rm Re}(a^w_\tau[{\rm MSSM}])| & \sim & 0.5~(7) \times 10^{-6}, 
\label{taususywm}\\
|{\rm Re}(a^w_b[{\rm MSSM}])| & \sim & 2~(50) \times 10^{-6}
\label{bsusywm}
\eeq
for $\tan\beta=1.6~(50)$ respectively. In the low $\tan \beta$ scenario the 
MSSM contribution to the AWMDM is of the same order than in the SM
(\ref{tausmwm},\ref{bewwm}).  
More interesting is the high $\tan \beta$ scenario where the MSSM 
contribution to the real part can reach values one order of 
magnitude higher than the pure electroweak SM contribution, especially
in the $b$ quark case (\ref{bewwm}), but 
still a factor five below the standard QCD contribution (\ref{bsmwm}).
%
%
\subsection{The $\tau$ and $b$ WEDMs in the MSSM}  
%
%
Eqs.~(\ref{mdm1}--\ref{edm56}) show that for having non--vanishing 
contribution to the WEDMs one has to deal with complex couplings. 
Many of the parameters in the SUSY Lagrangian can be complex but not all
the phases originated are physical. 
Some of them can be absorbed by a redefinition of the MSSM fields
(see App.~\ref{appendix-c} for a complete discussion of the procedure). 
Our choice of CP--violating phases is:\footnote{Such a choice 
leads to a dependence on $\varphi_\mu$ of 
chargino and neutralino masses. Conversely the scalar fermion masses are 
independent of $\varphi_{\tilde{f}}$.} $\varphi_\mu\equiv{\rm arg}(\mu)$, 
$\varphi_{\tilde{f}}\equiv{\rm arg}(m^f_{\rm LR})$ ($f=\tau,\ t,\ b$) with 
$m^t_{\rm LR}\equiv A_t-\mu^*\cot\beta$ and $m^{\tau,b}_{\rm LR}\equiv 
A_{\tau,b}-\mu^*\tan\beta$. 

Since the MSSM contains a CP--conserving Higgs sector\footnote{
A one--loop non vanishing contribution to the WEDM is possible in a
general 2HDM \cite{2hdm-cp}.} 
and the SM provides contributions to the (W)EDMs beyond two loops, the 
relevant diagrams are the genuine SUSY graphs of classes III and IV 
\cite{hirs2}. 
%
\subsubsection{Chargino contribution to the WEDM}
%
\begin{figure}
\begin{center}
\begin{tabular}{cc}
\epsfig{figure=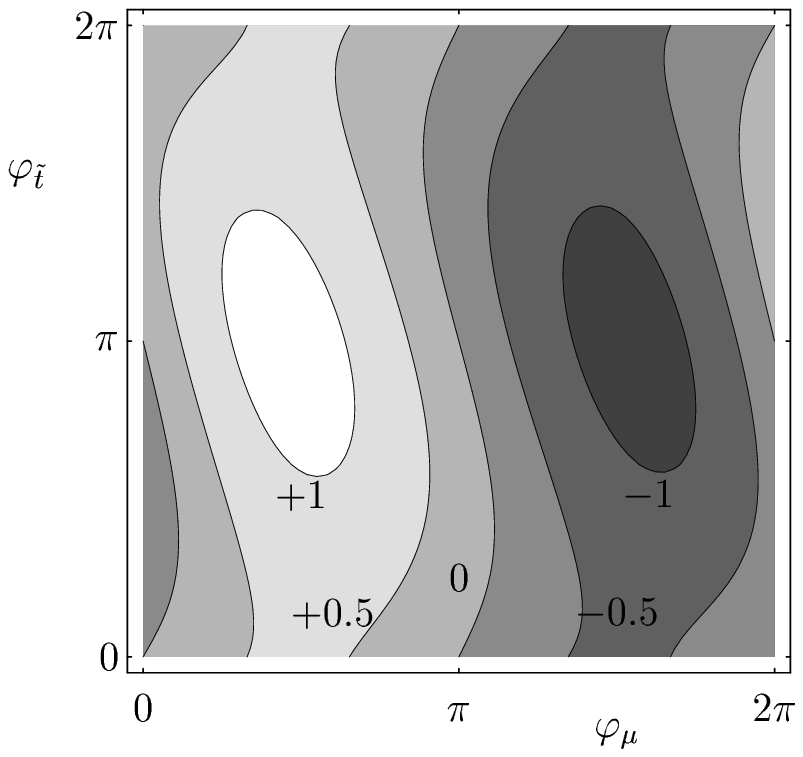,height=6.1cm,angle=0} &
\epsfig{figure=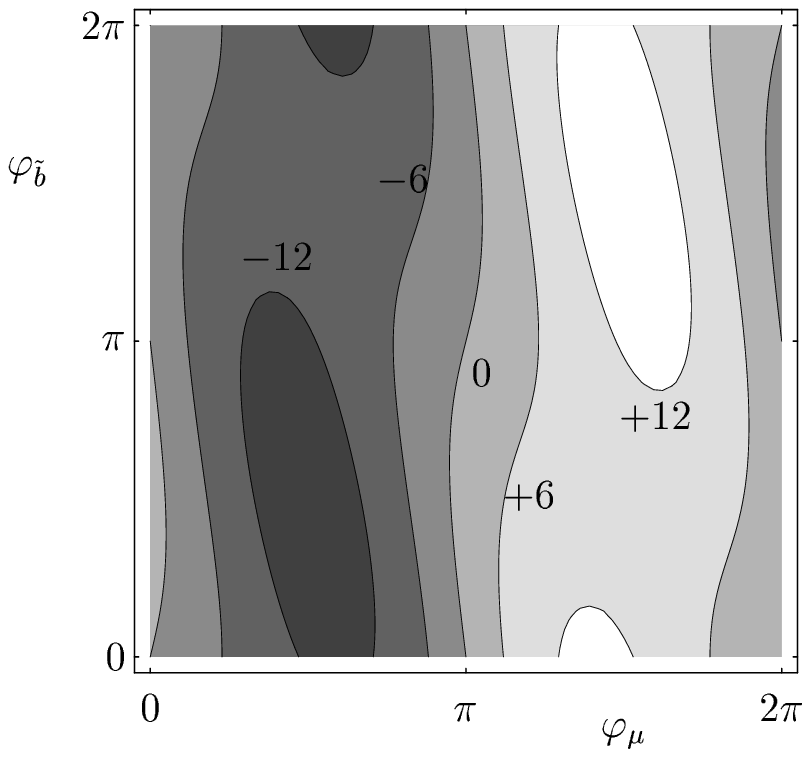,height=6.1cm,angle=0}\vspace*{-5mm}\\
(a) & (b)
\end{tabular}
\end{center}
\caption{\em Re($d^w_b$[char]) for low $\tan\beta$ (a) and Re($d^w_b$[neut]) 
for high $\tan\beta$ (b) in the plane $\varphi_{\tilde{f}}-\varphi_\mu$, in 
units of $10^{-6}\ \mu_b$ with $M_2=|\mu|=m_{\tilde q}=250$ GeV. The mixing 
$|m^t_{LR}|=|\mu|\cot\beta$\ and $|m^b_{LR}|=|\mu|\tan\beta$, respectively. 
From \cite{hirs2}.\label{fig:edmphases}}
\vspace*{0.3cm}
\end{figure}
There is only one phase, $\varphi_\mu$, involved in the chargino contribution
to the $\tau$ WEDM, as there is no mixing in the scalar neutrino sector. 
The result grows with $\tan\beta$. It also 
depends on the common scalar lepton mass (whose effect consists of diminishing 
the result through the tensor integrals) and the $|\mu|$ and $M_2$ mass 
parameters. Taking $M_2=|\mu|=m_{\tilde{l}}=250$ GeV one obtains: 
\bea
{\rm Re} \left(d^w_\tau [{\rm char}]\right) = 0.18~(5.52) 
         \times 10^{-6}~\mu_\tau \quad {\rm for} \quad \tan\beta = 1.6~(50) 
\eea
in the limit $\varphi_\mu=\pi/2$, for which the WEDM takes the largest value. 

Two CP-violating phases are involved in the $b$ case:
$\varphi_\mu$ and $\varphi_{\tilde{t}}$. In Fig.~\ref{fig:edmphases}a the 
dependence on these phases is shown, for $M_2=|\mu|=m_{\tilde{q}}=250$ GeV, 
$|m^t_{LR}|=|\mu|\cot\beta$ and both low and high $\tan\beta$ scenarios. 
The maximum effect on the WEDM is obtained for $\varphi_\mu=\pi/2$ and 
$\varphi_{\tilde{t}}=\pi$. One gets up to 
\bea
{\rm Re} \left(d^w_b [{\rm char}] \right) = 1.17~ (27.1) 
         \times 10^{-6}~ \mu_b \quad {\rm for} \quad \tan\beta = 1.6~(50).
\eea
In the high $\tan\beta$ scenario our assumed $|m^t_{LR}|$ 
takes a small value and the dependence on $\varphi_{\tilde{t}}$ tends to
disappear. To have an idea of the maximum value achievable for the
chargino contribution, we show in Fig.~\ref{fig:edmchar} the dependence on 
$M_2$ and $|\mu|$ for $\varphi_\mu=\pm\pi/2$ and $m^t_{LR}=0$ with the 
previous value for the common scalar quark mass parameter.  
Assuming the present bounds on the chargino masses \cite{moriond98}, there 
is no contribution to the imaginary part of the $\tau$ and $b$ WEDM.
\begin{figure}[t]
\begin{center}
\begin{tabular}{cc}
\epsfig{figure=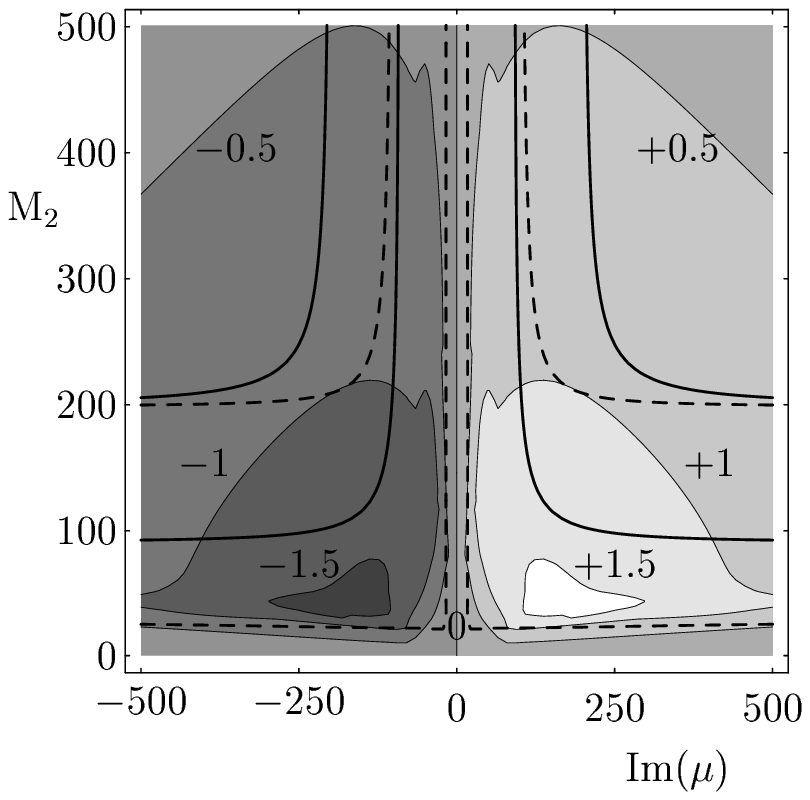,height=6.4cm,angle=0} &
\epsfig{figure=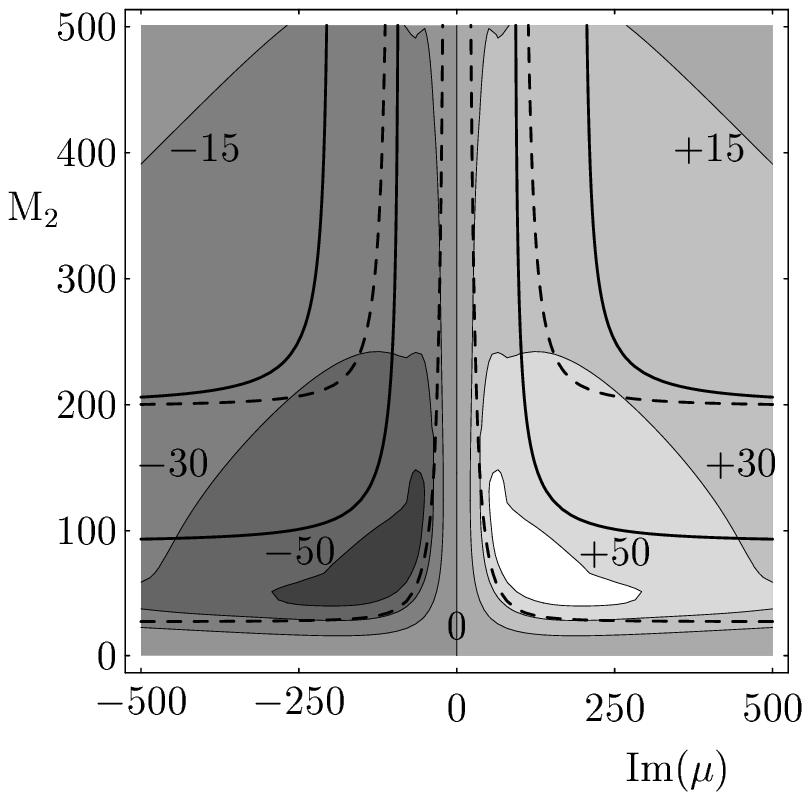,height=6.4cm,angle=0} \vspace*{-5mm}\\ 
(a) & (b)
\end{tabular}
\end{center}
\caption{\em Re($d^w_b$[char]) for low (a) and high (b) $\tan\beta$ in the 
plane $M_2-$Im$(\mu)$, in units of $10^{-6}\ \mu_b$ with $|\sin\varphi_\mu|=1$ 
and $m_{\tilde q}=250$ GeV. The mixing $|m^t_{LR}|=0$. The contour 
solid-lines correspond to the lightest chargino masses $m_{{\tilde\chi}^\pm_1}=80$ 
and $200$ GeV and the contour dashed-lines correspond to the lightest neutralino 
masses $m_{{\tilde\chi}^0_1}=15$ and $100$ GeV. From \cite{hirs2}.
\label{fig:edmchar}}
\vspace*{0.3cm}
\end{figure}
%
\subsubsection{Neutralino contribution to the WEDMs}
%
Now both $\varphi_\mu$ and $\varphi_{\tilde{\tau}}$ contribute. Assuming
that $|m^\tau_{LR}|$ is of the order of $|\mu|\tan\beta$ or below, 
there is no large influence of $\varphi_{\tilde{\tau}}$ on the neutralino
contribution,
for both low and high $\tan\beta$. This
contribution is roughly proportional to $\sin\varphi_\mu$
regardless of the value of $\varphi_{\tilde{\tau}}$. Taking $\varphi_\mu=\pi/2$,
$M_2=|\mu|=250$ GeV and $m_{\tilde{l}}=250$ GeV one finds:
\bea
{\rm Re} \left(d^w_\tau [{\rm neut}] \right) = -0.01~(-0.25) 
         \times 10^{-6}~\mu_\tau \quad {\rm for} \quad \tan\beta=1.6~(50).
\eea

The two relevant CP-violating phases for the neutralino contribution to 
the $b$ WEDM are: $\varphi_\mu$ and $\varphi_{\tilde{b}}$. As before, the 
most important effect from $\varphi_{\tilde{b}}$ arises when the off--diagonal 
term is larger, which in this case corresponds to high $\tan\beta$, as the 
trilinear soft breaking parameter is taken to be of the order of 
$|\mu|\tan\beta$. The maximum value for the neutralino contribution occurs for 
$\varphi_\mu=\varphi_{\tilde{b}}=\pi/2$ (Fig.~\ref{fig:edmphases}b). The 
total contribution increases with $\tan\beta$. Thus taking $M_2 = |\mu| = 
m_{\tilde{q}}=250$ GeV and $|m^b_{LR}|=|\mu|\tan\beta$ one gets: 
\bea
{\rm Re} \left(d^w_b [{\rm neut}] \right) = -0.29~(-12.6) 
         \times 10^{-6}~\mu_b \quad {\rm for} \quad \tan\beta=1.6~(50). 
\eea
For presently non excluded masses of the neutralinos \cite{moriond98} there 
can be a contribution to the imaginary part of the order of $10^{-6}\ \mu_f$.
%
\subsubsection{Gluino contribution to the $b$ WEDM}
%
The gluino contribution is affected only by $m^b_{LR}$, $m_{\tilde{q}}$ and 
the gaugino mass $m_{\tilde g}$. Therefore the maximum value occurs for 
$\varphi_{\tilde{b}}=\pi/2$. The mixing in the $\tilde b$ sector is determined 
by $m^b_{LR}$ and intervenes in the contribution due to chirality flipping in
the gluino internal line (the contribution proportional to $m_{\tilde g}$). 
The contribution to the WEDM is enhanced by the largest values of 
$|m^b_{LR}|$ compatible with an experimentally not excluded mass for the 
lightest $\tilde b$ scalar quark. For zero gluino mass, only the term proportional 
to the mass of the $b$ quark provides a contribution. Thus for 
$\varphi_{\tilde{b}}=\pi/2$ one gets:
\bea
{\rm Re} \left(d^w_b [{\rm gluinos}] \right) = 0.26~(9.31) 
         \times 10^{-6}~\mu_b \quad {\rm for} \quad \tan\beta = 1.6~(50), 
\eea
$|m^b_{LR}|=|\mu|\tan\beta$, $M_2=|\mu|=m_{\tilde{q}}=250$ GeV and $m_{\tilde g}
\equiv M_3$ fulfilling the GUT relation (\ref{gut}). There exists the 
possibility for a gluino contribution to the imaginary part of the $b$ WEDM 
if the hypothesis of light gluino is not discarded.
%
\subsubsection{MSSM total contribution to the WEDMs}
%
The Higgs sector does not contribute and chargino diagrams are more 
important than neutralino ones. Gluinos are also involved in the $b$ case 
and compete in importance with charginos. In the most favorable
configuration of CP-violating phases and for values of the rest of the
parameters still not excluded by experiments, these WEDMs can be as much
as twelve orders of magnitude larger than the SM estimates (\ref{smwedm}):
\beq
|{\rm Re} ( d^w_\tau )| & \lsim & 0.2~(6) \times 10^{-6}~\mu_\tau  
\label{taususywedm}\\
|{\rm Re} ( d^w_b )| &\lsim & 2~(35) \times 10^{-6}~\mu_b 
\label{bsusywedm}
\eeq
There may be a contribution to the imaginary part if the neutralinos are 
light but this contribution is at least one order of magnitude smaller 
than the real part of the $\tau$ or $b$ WEDM.

\subsection{Cancellation of the dipole moments in the supersymmetric limit} 
%
A general $Vff$ interaction is restricted to the form
(\ref{efflagm}) by Lorentz invariance. Since the Lorentz algebra is a
subalgebra of the supersymmetry algebra, this interaction is even more
constrained in a theory with unbroken supersymmetry. In fact, 
supersymmetric sum rules are derived that relate the
electric and magnetic multipole moments of any irreducible $N=1$
supermultiplet \cite{SumRules}. Applied to the $Vff$ interaction between a
vector boson coupling to a conserved current and the fermionic
component of a chiral multiplet, these sum rules force the
gyromagnetic ratio to be $g_f=2$ and forbid an electric dipole
moment:
\begin{equation}
   a^V_f=d^V_f=0.
\end{equation}
The Lagrangian of the MSSM is supersymmetric when the soft--breaking terms
are removed. For non zero value of the $\mu$ parameter the Higgs potential 
has only a trivial minimum. Therefore to keep the particles massive and 
supersymmetry preserved at the same time the choice $\mu=0$ is necessary.
Then the Higgs potential is positive semi--definite and 
it has degenerate minima corresponding to $v\equiv v_1=v_2$ ($\tan\beta=1$).
The value of $M_A=0$ follows from such a configuration.
Finally the value of the common $v$ is fixed by the phenomenology:
the muon decay constant and the gauge boson masses. 

In this supersymmetric limit the above mentioned sum rules are valid
and the magnetic and electric dipole form factors have to cancel. To verify
our expressions we checked this for the AWMDM of the $b$ quark \cite{hirs2}. 
Choosing the parameters $A_b=A_t=M_2=M_3=0$, $\mu=0$, $\tan\beta=1$ and
$M_A=0$, the SM gauge boson contribution to $a^w_b$
\cite{ber97} is indeed cancelled by the MSSM correction including
the two Higgs doublets: the gluon and gluino contribution
cancel among themselves and the neutralinos and charginos cancel the
gauge boson and Higgs contributions.

\section{SM and MSSM predictions for the top quark dipole form factors}

\subsection{SM}

The electroweak contributions to the magnetic and weak--magnetic dipole form 
factors for off--shell gauge bosons are gauge dependent. The pinch technique 
\cite{cornwall85} could be used to construct gauge--parameter independent 
magnetic dipoles in the class of $R_\xi$ gauges \cite{papa94,ber95b} but the 
prescription is not unique \cite{georg} and these quantities cannot be 
observable by themselves.
The QCD contributions (gluon exchange) to the $t$ (W)MDFF are gauge independent
and comparable in size to the electroweak predictions at $\sqrt{s}=500$ GeV 
in the 't Hooft--Feynman gauge due to the large mass of the $t$ quark
\cite{ber95b}.

There is no contribution to the electric and weak--electric dipole form factors
to one loop in the SM.

\subsection{MSSM}

The triangle diagrams with SUSY particles are gauge independent by themselves
(no gauge or Goldstone bosons involved) but the ones including Higgs scalars 
and gauge bosons in the loop are not sufficient to keep the gauge invariance in 
the case of the magnetic dipole form factor. 
The CP--violating dipole form factors (electric and weak--electric) for which 
the Higgs sector of the MSSM is irrelevant, on the other hand, can be regarded
as gauge independent quantities. These form factors have also been considered in
\cite{vienna}.\footnote{The results in \cite{vienna} after revision are 
in agreement with the ones presented here \cite{garfield}.}

We investigate the SUSY contributions to the $t$ electric and weak--electric 
dipole form factors performing a parameter scan for which a fixed value 
$\sqrt{s}=500$ GeV has been chosen.\footnote{
We use the running coupling constants evaluated at $\sqrt{s}=500$ GeV,
$\alpha_s=0.092$, $\alpha=1/126$.}
It is important to point out that the region of the supersymmetric parameter 
space that provides the maximal contributions varies with $s$ due 
to threshold effects.
We scan the mass parameters $M_2$ and $|\mu|$ in a broad range and the 
CP--violating phases: $\varphi_\mu$, $\varphi_{\tilde t}$ and 
$\varphi_{\tilde b}$ (see Figs.~\ref{fig:4.1} and \ref{fig:4.2}). 
The gluino mass is given by the GUT constraint (\ref{gut}).
We adopt a fixed value for the common scalar quark mass $m_{\tilde q}=200$ GeV: 
this is a plausible intermediate value; larger values decrease the effects. 
Finally, the moduli of the off--diagonal terms 
in the $\tilde t$ and $\tilde b$ mass matrices are also chosen at fixed values 
$|m^t_{LR}|=|m^b_{LR}|=200$ GeV, to reduce the number of free parameters.
The results are expressed in $t$ magnetons $\mu_t\equiv e/2m_t=5.64\times
10^{-17}\ e$cm.

The MSSM contributions to the top quark EDFF and WEDFF include:

\subsubsection{Neutralinos and ${\tilde t}$ scalar quarks}

\begin{figure}
\begin{tabular}{cc}
{\small Re$\{d^\gamma_t[{\tilde \chi}^0]\}$}            &
{\small Im$\{d^\gamma_t[{\tilde \chi}^0]\}$}            \\
\epsfig{file=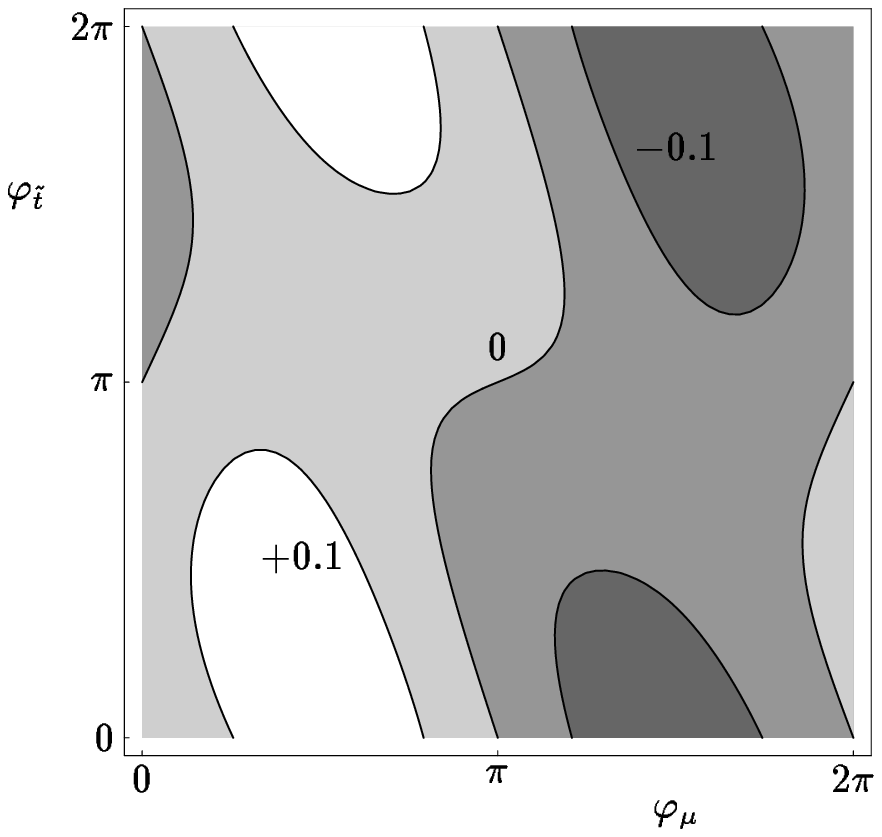,height=6.1cm}        &    
\epsfig{file=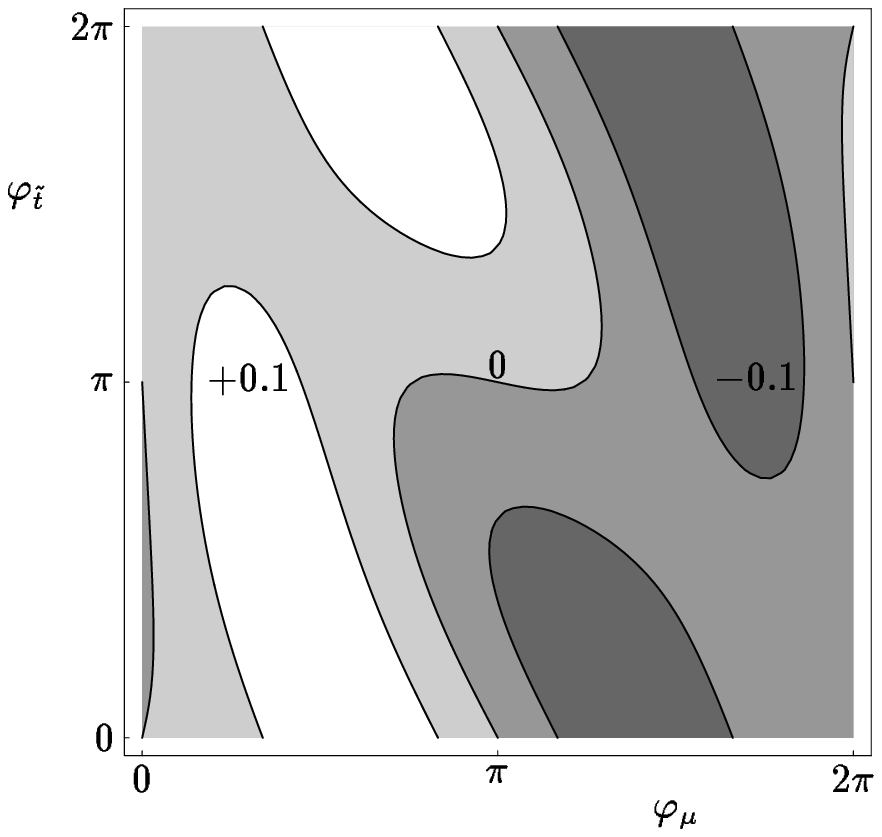,height=6.1cm}       \\
{\small Re$\{d^Z_t[{\tilde \chi}^0]\}$}                 &
{\small Im$\{d^Z_t[{\tilde \chi}^0]\}$}                 \\
\epsfig{file=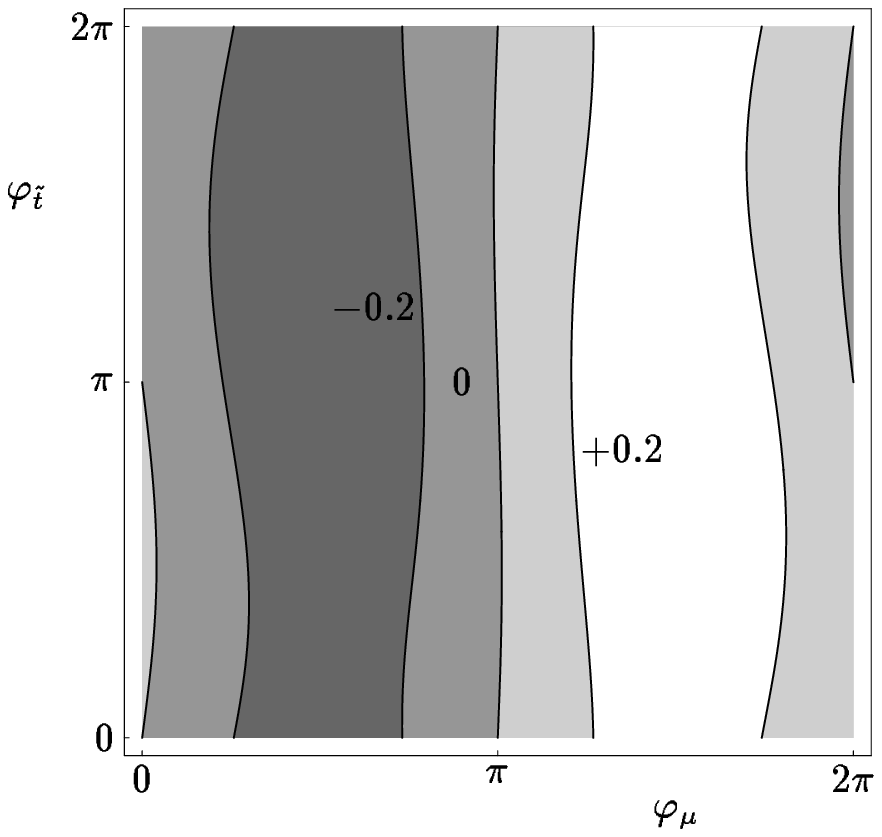,height=6.1cm}       &    
\epsfig{file=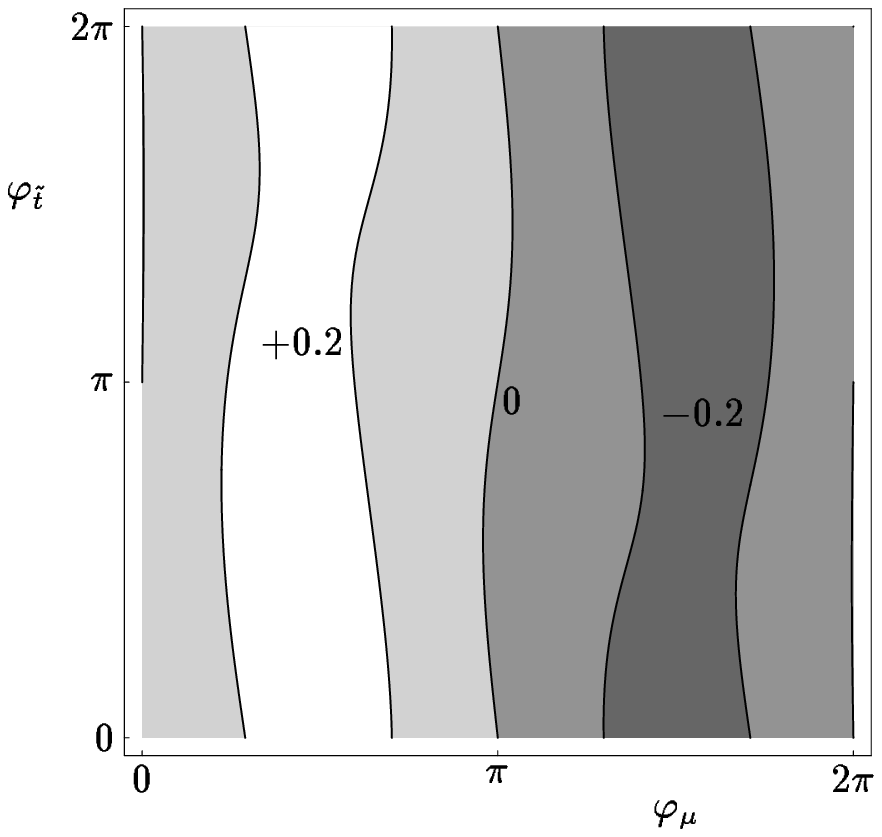,height=6.1cm}      
\end{tabular}
\caption{\em Neutralino contribution to the real and imaginary parts of the $t$ EDFF
         and WEFF [in $10^{-3}\mu_t$ units] in the plane $\varphi_{\tilde t}-\varphi_\mu$
         for $\tan\beta=1.6$ and the reference values $M_2=|\mu|=|m^t_{LR}|=
         200$ GeV at $\sqrt{s}=500$ GeV. \label{fig:4.1}}
\vspace*{0.3cm}
\end{figure}

They provide typically small contributions but quite sensitive to the
value of both the phases involved, $\varphi_\mu$ and $\varphi_{\tilde t}$
(Fig.~\ref{fig:4.1}). 

The results are larger for low $\tan\beta$ since the chirality flipping mass
terms are dominated by the $t$ quark, yielding a term proportional 
to $m_t\cot\beta$. A term proportional to the neutralino masses is also
present as well as a negligible one proportional to $m_b\tan\beta$.
The contributing diagrams belong to the classes III and IV for the $Z$ case
and only to class IV for the $\gamma$ case, as the neutralinos do not couple to
photons.
 
As a reference we take the representative values $M_2=|\mu|=200$ GeV and
$\varphi_\mu=-\varphi_{\tilde t}=\pi/2$.
For these inputs the results are
\beq
d^\gamma_t[{\tilde \chi}^0]     &=& (\ 0.080+0.081\ {\rm i})\times 10^{-3}\ \mu_t\\
d^Z_t[{\tilde \chi}^0]          &=& (-0.324+0.223\ {\rm i})\times 10^{-3} \ \mu_t      
\eeq

\subsubsection{Charginos and ${\tilde b}$ scalar quarks}

\begin{figure}
\begin{tabular}{cc}
{\small Re$\{d^\gamma_t[{\tilde \chi}^\pm]\}$}          &
{\small Im$\{d^\gamma_t[{\tilde \chi}^\pm]\}$}          \\
\epsfig{file=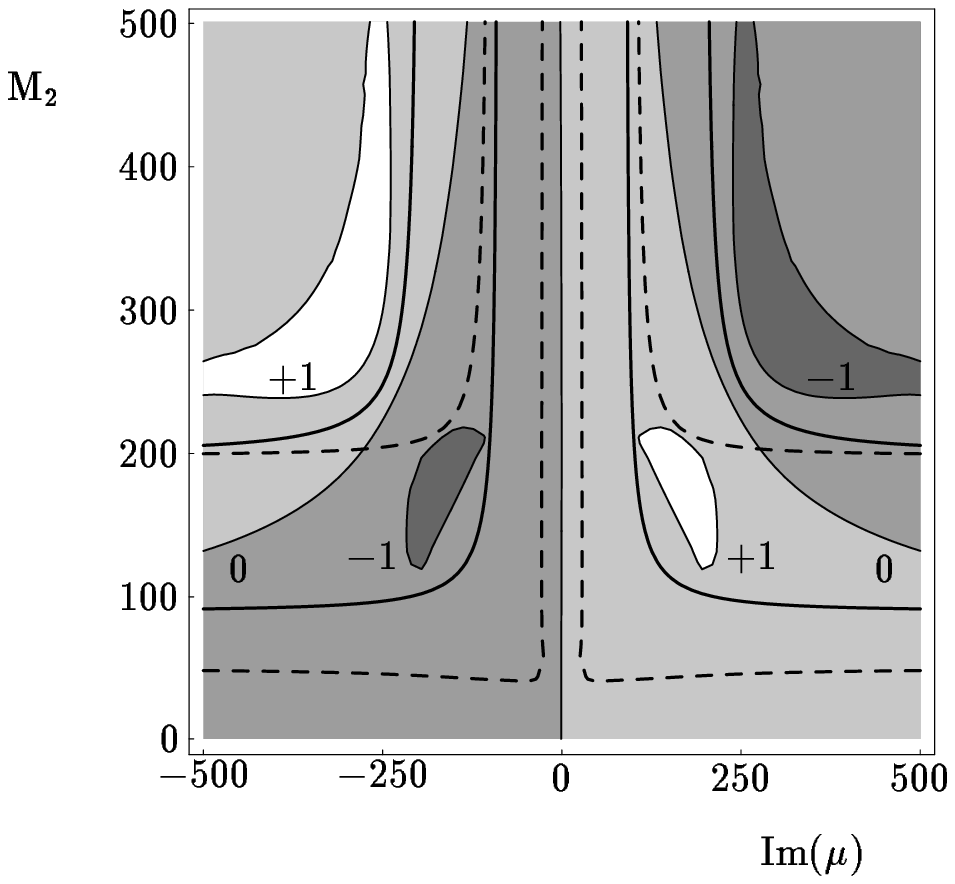,height=6.1cm}     &    
\epsfig{file=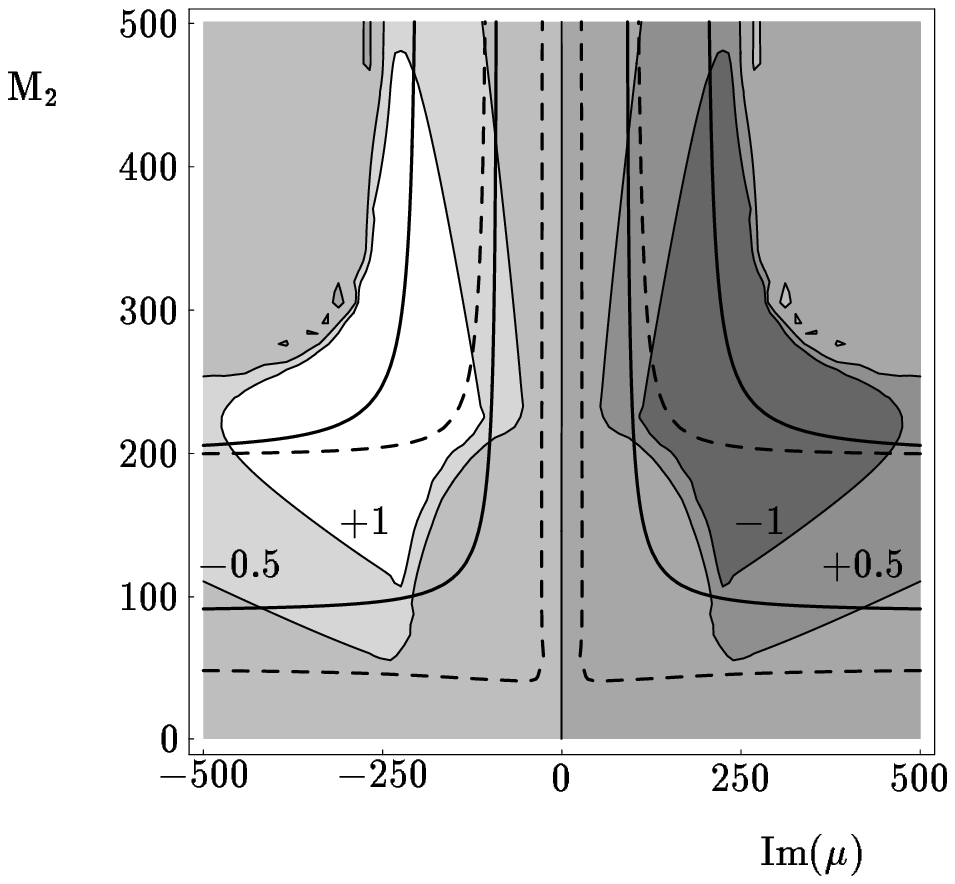,height=6.1cm}     \\
{\small Re$\{d^Z_t[{\tilde \chi}^\pm]\}$}               &
{\small Im$\{d^Z_t[{\tilde \chi}^\pm]\}$}               \\
\epsfig{file=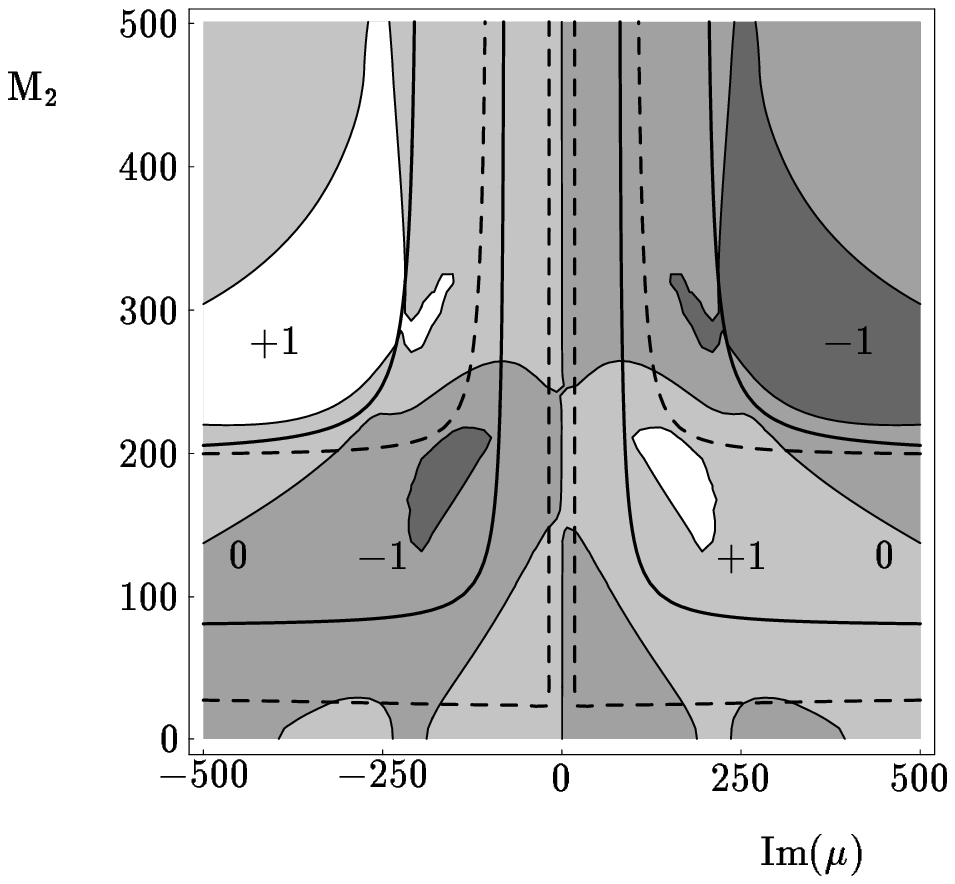,height=6.1cm}    &    
\epsfig{file=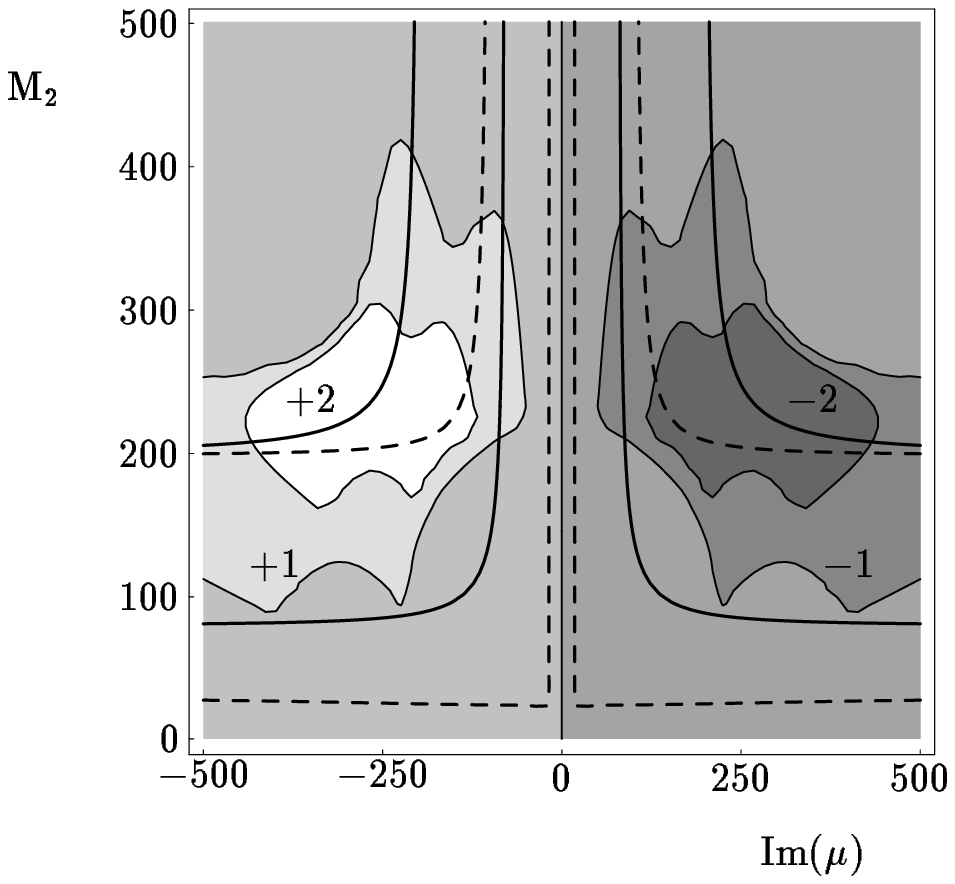,height=6.1cm}    
\end{tabular}
\caption{\em Chargino contribution to the real and imaginary parts of the $t$ EDFF 
         and WEFF [in $10^{-3}\mu_t$ units] in the plane $M_2-{\rm Im}(\mu)$ for 
         $\tan\beta=1.6$, $|m^b_{LR}|=200$ GeV and $\varphi_{\tilde b}=\pi/2$
         at $\sqrt{s}=500$ GeV. 
         The lower (upper) solid isolines correspond to $m_{\tilde{\chi}^\pm_1}
         =90\ (200)$ GeV and the dashed isolines to $m_{\tilde{\chi}^0_1}=25\ 
         (100)$ GeV. \label{fig:4.2}}
\vspace*{0.3cm}
\end{figure}

As in the case of the neutralinos, the influence of chargino diagrams is 
enhanced for low $\tan\beta$.  

The results depend very little on $\varphi_{\tilde b}$ and mainly
on $\varphi_\mu$, the maxima being close to $\varphi_\mu=\pm\pi/2$.
For increasing $\tan\beta$ the dependence on $\varphi_{\tilde b}$ grows,
as it comes with a factor proportional to $m_b\tan\beta$.

The chargino contributions are the most important ones.
In Fig.~\ref{fig:4.2} the dependence on Im$(\mu)$ and $M_2$ is displayed
for $\tan\beta=1.6$. The only relevant CP--violating phase here has been set
to the most favorable case, $\varphi_\mu=\pi/2$ (the negative values of 
Im$(\mu)$ correspond to $\varphi_\mu=-\pi/2$). The symmetry with respect to
Im$(\mu)=0$ in Fig.~\ref{fig:4.2} reflects the approximate independence of
$\varphi_{\tilde b}$, here set to $\pi/2$.\footnote{
All the contributions flip sign when the set ($\varphi_\mu,\varphi_{\tilde t},
\varphi_{\tilde b}$) is rotated by $\pi$. They vanish accordingly when all 
the phases are zero. See Fig.~\ref{fig:4.1} for illustration.} 
The same does not happen for the neutralino contributions for
a fixed value of $\varphi_{\tilde t}\ne 0,\pi$ in the plane $M_2-{\rm Im}(\mu)$.
The plots of Fig.~\ref{fig:4.2} exhibit a tendency to decoupling of the 
supersymmetric effects for increasing values of the mass parameters.
The isolines for a couple of masses of the lightest charginos and neutralinos
in the same plane are also given for orientation. The current LEP2
experimental lower limits are $m_{\tilde{\chi}^0_1}>25$ GeV and 
$m_{\tilde{\chi}^\pm_1}>90$ GeV \cite{moriond98}.

The chargino contributions for the values $M_2=|\mu|=200$ GeV and
$\varphi_\mu=\pi/2$ are

\beq
d^\gamma_t[{\tilde \chi}^\pm]   &=& (0.869-1.870\ {\rm i})\times 10^{-3}\ \mu_t \\
d^Z_t[{\tilde \chi}^\pm]        &=& (0.793-2.524\ {\rm i})\times 10^{-3}\ \mu_t
\eeq

\subsubsection{Gluinos and ${\tilde t}$ scalar quarks}

Their effect is roughly
proportional to $|m^t_{LR}|\sin\varphi_{\tilde t}$ times a chirality flipping
fermion mass, either $m_t$ or $M_3$ (the gluino mass). It is damped for
heavy gluinos circulating in the loop and also for large scalar quark masses 
(decoupling). Both terms have opposite sign to Im($m^t_{LR}$) and the one
proportional to the gluino mass dominates.

The result for $M_2=200$ GeV and $\varphi_{\tilde t}=-\pi/2$ is

\beq
d^\gamma_t[{\tilde g}]  &=& (0.457+0.170\ {\rm i})\times 10^{-3}\ \mu_t  \\
d^Z_t[{\tilde g}]       &=& (0.155+0.059\ {\rm i})\times 10^{-3}\ \mu_t
\eeq

\subsubsection{Total contribution}

In view of these results, we establish a set of SUSY parameters
and phases for which nearly all the contributions sum up constructively
at $\sqrt{s}=500$ GeV.
Our choice is
\beq
\mbox{Reference Set }\#1:& &\tan\beta=1.6 \nn \\
& &M_2=|\mu|=m_{\tilde{q}}=|m^t_{LR}|=|m^b_{LR}|=200 \mbox{ GeV} \nn \\
& &\varphi_\mu=-\varphi_{\tilde t}=-\varphi_{\tilde b}=\pi/2 ,
\label{refset1}
\eeq
for which the $t$ EDFF and WEFF reach the values:
\beq
d^\gamma_t      &=&     (1.407-1.618\ {\rm i})\times 10^{-3}\ \mu_t \\
d^Z_t           &=&     (0.624-2.242\ {\rm i})\times 10^{-3}\ \mu_t , 
\label{topff}
\eeq
at $\sqrt{s}=500$ GeV. 
To illustrate how the maximum effects appear we display in Fig.~\ref{fig:4.3}
the individual contributions as a function of $\sqrt{s}$.
The masses of the supersymmetric partners in the
loops are such that there are threshold enhancements in the vicinity of
$\sqrt{s}=500$ GeV. 

\begin{figure}
\begin{center}
\epsfig{file=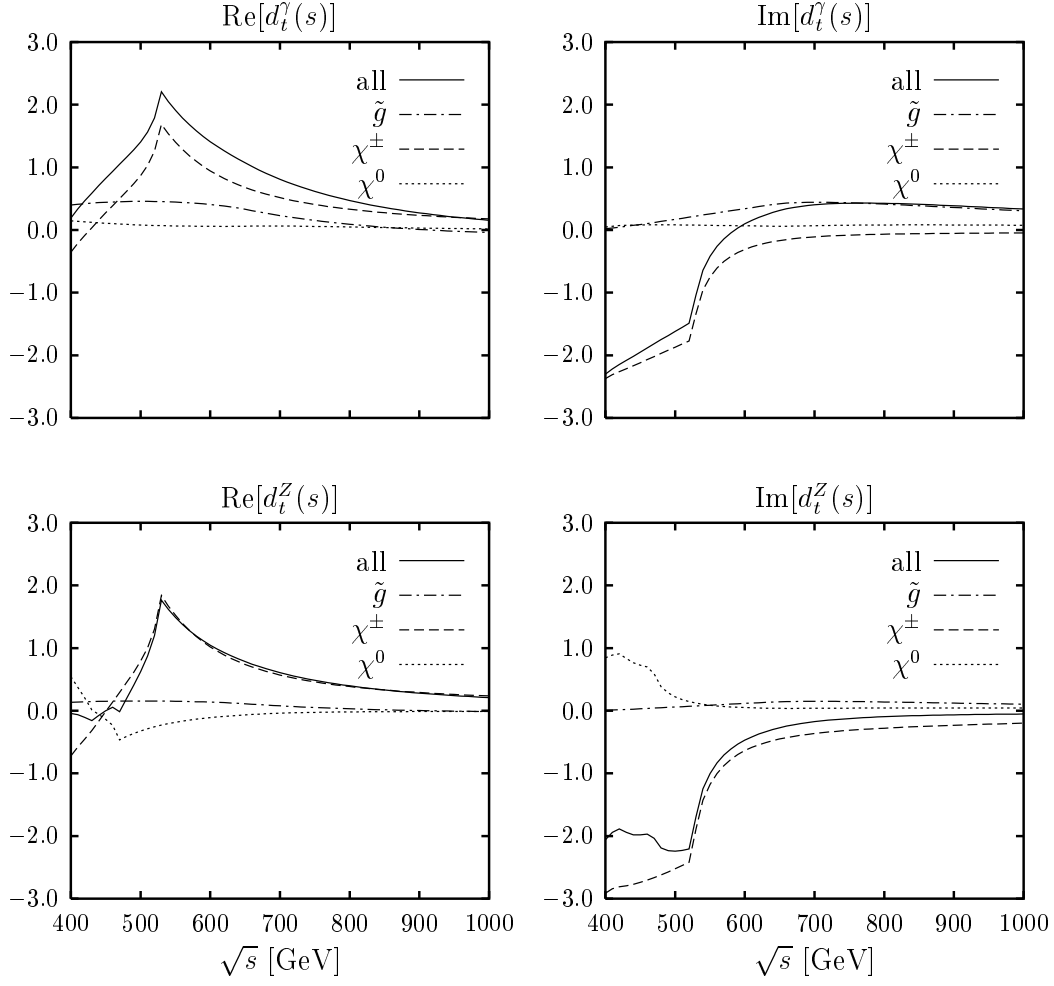,width=\linewidth}
\end{center}
\caption{\em The different contributions to the $t$ (W)EDFF [in $10^{-3}\mu_t$
units] for 
the reference set of SUSY parameters of Eq. (\ref{refset1}).\label{fig:4.3}}
\vspace*{0.3cm}
\end{figure}

\section{Observables at the \boldmath$Z$ resonance: the \boldmath$\tau$ lepton
         case}

We analyze below a set of observables sensitive to the weak dipole moments
at the $Z$ peak. In this environment the effects of electromagnetic
dipole moments to the process $e^+e^-\to f\bar{f}$ are negligible,
as the pair production proceeds through the $Z$ resonance.
The $\gamma-$exchange and box diagrams are suppressed.

  \subsection{The $Z$ width}

The $Z$ width is given in terms of generic vertex form factors by
\beq
\Gamma(Z\to
f\bar{f})&=&\frac{G_F M^3_Z}{6\sqrt{2}\pi}\sqrt{1-\frac{4m^2_f}{M^2_Z}}
\Bigg[
v^2_f\left(1+\frac{2m^2_f}{M^2_Z}\right)+
a^2_f\left(1-\frac{4m^2_f}{M^2_Z}\right)\nonumber\\
&&-6v_f s_Wc_W {\rm Re}(a^w_f)+4s_W^2c_W^2 |a^w_f|^2\left(
\frac{M^2_Z}{8m^2_f}+1\right)\nonumber\\
&&+2s_W^2c_W^2 |\hat{d}^w_f|^2\left(\frac{M^2_Z}{4m^2_f}-1\right)
\Bigg]\ ,
\eeq
where we have introduced the dimensionless parameter 
$\hat{d}^w_f\equiv 2m_f d^w_f/e$ for simplicity. 

From the previous expression one observes that the real part of the AWMDM 
contributes linearly to the total width (although it is not necessarily the
dominant term in the $\tau$ case for $a^w_\tau$ near the experimental upper 
limit, due to the small value of $v_\tau$ and the $M^2_Z/m^2_\tau$ enhancement 
factor).
The WEDM contributes only quadratically to the total $Z$ width.

One might get some upper bounds on Re$(a^w_f)$ and $|d^w_f|$
from the $Z$ partial widths measurements at LEP and SLC using the previous 
expressions. 
Actually this is not the most appropriate way to put limits to the dipoles as
much information is integrated out. Moreover the weak--magnetic and 
weak--electric contributions cannot be disentangled and some other effects
can interfere the measurements. In contrast, the differential cross section
for the production of polarized fermion pairs contains all the ingredients
needed and is the starting point for the construction of more specific 
observables.
  
  \subsection{The differential cross section {$e^-e^+\to Z \to 
                   f\bar f$}}

The differential cross section for $e^-e^+ \to f(\sone)\bar{f}(\stwo)$
with polarized final fermions 
can be written as a sum of spin independent, single--spin dependent and 
spin--spin correlation contributions as
\beq
\frac{\dd\sigma}{\dd\Omega_f}(\sone,\stwo)=
\frac{\dd\sigma^0}{\dd\Omega_f}+
\frac{\dd\sigma^S}{\dd\Omega_f}(\sone\oplus\stwo)+
\frac{\dd\sigma^C}{\dd\Omega_f}(\sone\otimes\stwo)\ ,
\eeq
where $\sone$ ($\stwo$) are the polarization vectors of 
the fermion (antifermion). We choose a reference frame in which
the $z$--axis points in the direction of the outgoing fermion and the 
$x$--$z$ plane contains both the incoming electron and the outgoing fermion. 
The polar angle $\theta_f$ is the angle between $e^-$ and the outgoing fermion. 
The azimuthal angle of the fermion can be integrated over, as the electron and 
positron are assumed unpolarized.
The masses of the electron and positron will be neglected, as well as the
quadratic terms in the anomalous couplings. The spin--correlation terms do not
contribute when the polarization of only one of the two fermions is analyzed.

In terms of the final fermion velocity $\beta$, the dilatation factor 
$\gamma=M_Z/2m_f=(1-\beta^2)^{-1/2}$ in the overall 
c.m.s. and the polarization vectors ${\bf s^*_{1,2}}$ of fermion and 
antifermion
in their rest frames, the differential cross section $e^-e^+\to Z \to f\bar f$ 
(exactly on the Z resonance) reads 
\beq
\frac{\dd\sigma^0}{\dd\cos\theta_{f}}&=&\frac{\alpha^2\beta N^f_C}{(4s_Wc_W)^4}
\frac{1}{\Gamma^2_Z}\times\Bigg\{\ \ 
(v^2_e+a^2_e)[v^2_f(2-\beta^2\sin^2\theta_{f})
\nonumber\\ & &\hspace{3.3cm}
+a^2_f\beta^2(1+\cos^2\theta_{f})]+8v_ea_ev_fa_f\beta\cos\theta_{f}
\nonumber\\ & &
-8s_Wc_W{\rm Re}(a^w_f)\ [(v_e^2+a_e^2)v_f
         +2v_ea_ea_f\beta\cos\theta_{f}]\ \Bigg\}
\ ,
\eeq
\beq
\frac{\dd\sigma^S}{\dd\cos\theta_{f}}&=&\frac{\alpha^2\beta N^f_C}{(4s_Wc_W)^4}
\frac{2s_Wc_W}{\Gamma^2_Z}\nonumber \\ 
&&\times[ \
    (\soner+\stwor)_x X_+ + (\soner+\stwor)_y Y_+ 
      + (\soner+\stwor)_z Z_+ \nonumber\\
&&\ \ + (\soner-\stwor)_x X_- + (\soner-\stwor)_y Y_- 
      + (\soner-\stwor)_z Z_- \ ]\ ,
\eeq
with
\beq
X_+&=&\Bigg\{-\frac{1}{\gamma s_Wc_W}
[2v_ea_ev^2_f+(v_e^2+a_e^2)v_fa_f\beta\cos\theta_{f}] \nonumber\\ & & \
+2\gamma{\rm Re}(a^w_f)\ [2v_ea_ev_f(2-\beta^2)+(v_e^2+a_e^2)a_f
                             \beta\cos\theta_{f}]
\Bigg\}\sin\theta_{f} \ ,\label{rea}\\
Y_+&=&-2\gamma\beta{\rm Im}(a^w_f)\ [2v_ea_ea_f+(v_e^2+a_e^2)v_f
\beta\cos\theta_{f}]\sin\theta_{f} \ , \label{ima}\\ 
Z_+&=&-\frac{1}{s_Wc_W}[(v_e^2+a_e^2)v_fa_f\beta(1+\cos^2\theta_{f})+
2v_ea_e(v^2_f+a^2_f\beta^2)\cos\theta_{f}]  \nonumber\\ & &
+2{\rm Re}(a^w_f)\ [4v_ea_ev_f\cos\theta_{f}+(v_e^2+a_e^2)a_f\beta(1+
\cos^2\theta_{f})] \ ,\label{long} \\ 
X_-&=&2\gamma\beta {\rm Im}(\hat{d}^w_f)\ 
[(v_e^2+a_e^2)v_f\cos\theta_{f}+2v_ea_ea_f\beta]\sin\theta_{f} \ , 
       \label{imd}\\
Y_-&=&2\gamma\beta {\rm Re}(\hat{d}^w_f)\ 
[2v_ea_ev_f+(v_e^2+a_e^2)a_f\beta\cos\theta_{f}]\sin\theta_{f} \ , \label{red}\\
Z_-&=&-2\beta {\rm Im}(\hat{d}^w_f)\ (v_e^2+a_e^2)v_f\sin^2\theta_{f}\ ,
\label{imdl}
\eeq
and
\beq
\frac{\dd\sigma^C}{\dd\cos\theta_{f}}&=&\frac{\alpha^2\beta N^f_C}{(4s_Wc_W)^4}
\frac{1}{\Gamma^2_Z}\ [ \
     s_{xx} C_{xx} + s_{yy} C_{yy} + s_{zz} C_{zz} \nonumber\\
 &&+(s_{xy}+s_{yx})C^+_{xy}+(s_{xz}+s_{zx})C^+_{xz}+(s_{yz}+s_{zy})C^+_{yz}
                                                                \nonumber\\
 &&+(\soner\times\stwor)_x C^-_{yz}+(\soner\times\stwor)_y C^-_{xz}
   +(\soner\times\stwor)_z C^-_{xy}
                                                                \ ]\ ,
\eeq
with
\beq
C_{xx}&=& (v^2_e+a^2_e)[v^2_f(2-\beta^2)-a^2_f\beta^2]\sin^2\theta_{f}
\nonumber\\ & & 
          -8s_Wc_W {\rm Re}(a^w_f)\ (v_e^2+a_e^2)v_f\sin^2\theta_{f}
          \ ,\\
C_{yy}&=& -(v_e^2+a_e^2)(v_f^2-a_f^2)\beta^2\sin^2\theta_{f}\ ,\\
C_{zz}&=&(v^2_e+a^2_e)[v^2_f(2+\beta^2\sin^2\theta_{f})
                      -a^2_f\beta^2(1+\cos^2\theta_{f})]
\nonumber\\ & &
+8v_ea_ev_fa_f\beta\cos\theta_{f}
\nonumber\\ & &
-2s_Wc_W{\rm Re}(a^w_f)\ [4(v_e^2+a_e^2)v_f\cos^2\theta_{f}
         +8v_ea_ea_f\beta\cos\theta_{f}]
\ ,\\
C^+_{xy}&=& 4s_Wc_W \beta{\rm Im}(a^w_f)\ (v_e^2+a_e^2)a_f\sin^2\theta_{f}
\ ,\\
C^+_{xz}&=& \Bigg\{\ \frac{1}{\gamma}
            [(v_e^2+a_e^2)v^2_f\cos\theta_{f}+2v_ea_ev_fa_f\beta]
\nonumber\\ & &
            -4s_Wc_W\gamma{\rm Re}(a^w_f)\ [(v_e^2+a_e^2)v_f(2-\beta^2)
            \cos\theta_{f}
\nonumber\\ & & \hspace{3.3cm} 
            +2v_ea_ea_f\beta]\Bigg\}\sin\theta_{f}
            ,\\
C^+_{yz}&=& 4s_Wc_W\gamma\beta{\rm Im}(a^w_f)\ [(v_e^2+a_e^2)a_f
            \cos\theta_{f}+2v_ea_ev_f\beta]\sin\theta_{f}\ ,\\
C^-_{xy}&=& 4s_Wc_W\beta{\rm Re}(\hat{d}^w_f)\
            (v_e^2+a_e^2)v_f\sin^2\theta_{f} \ ,\\
C^-_{xz}&=& 4s_Wc_W\gamma\beta{\rm Im}(\hat{d}^w_f)\ 
             [(v_e^2+a_e^2)a_f\beta\cos\theta_{f}+2v_ea_ev_f]\sin\theta_{f}\ ,\\
C^-_{yz}&=& -4s_Wc_W\gamma\beta{\rm Re}(\hat{d}^w_f)\ 
            [(v_e^2+a_e^2)v_f\cos\theta_{f}+2v_ea_ea_f\beta]\sin\theta_{f} \ ,
\eeq
where $s_{ij}\equiv \sone_i\stwo_j$, $i,j=x,\ y,\ z$ and $N^f_C=
1\ (3)$ for leptons (quarks).

Notice that, from the SM at tree level, the final fermions are longitudinally 
polarized according to Eq.~(\ref{long}). There is also a SM contribution to the 
transverse polarization (within the collision plane) but it is helicity 
suppressed (see Eq.~\ref{rea})). The transverse and the normal (to the collision
plane) single--fermion polarizations are especially sensitive to the WDMs: the 
real part of the AWMDM (WEDM) contributes to the transverse (normal) 
single--fermion polarization (Eqs.~(\ref{rea}) and (\ref{red}), respectively), 
whereas the imaginary part of the  AWMDM (WEDM) contributes to the normal 
(transverse) single--fermion polarization (Eqs.~(\ref{ima}) and (\ref{imd}), 
respectively). The comparison 
of the polarization of both $f$ and $\bar f$ isolates the CP--violating
effects: the real part of the WEDM induces a difference in the $f$ and
$\bar f$ polarizations (Eq.~(\ref{red})) orthogonal to the scattering plane 
whereas the imaginary part leads to a difference in the $f$ and $\bar f$
transverse and longitudinal polarizations (Eqs.~(\ref{imd}) and (\ref{imdl}),
respectively).
The spin--spin correlation terms also provide specific information on the 
dipoles according to their symmetry properties.

\begin{table}
\caption{\em Transformation properties of the different terms in the $Zff$ 
effective vertex.\label{tab1}}
\vspace{0.3cm}
\begin{center}
\begin{tabular}{|c|c|c|c|}
\hline
Couplings       & P             & CP            & T             \\
\hline
\hline
$v$             & $+$           & $+$           & $+$           \\
\hline
$a$             & $-$           & $+$           & $+$           \\
\hline
Re($a^w_f$)     & $+$           & $+$           & $+$           \\
\hline
Im($a^w_f$)     & $+$           & $+$           & $-$           \\
\hline
Re($d^w_f$)     & $-$           & $-$           & $-$           \\
\hline
Im($d^w_f$)     & $-$           & $-$           & $+$           \\
\hline
\end{tabular}
\end{center}
\vspace*{0.3cm}
\end{table}

\begin{table}
\caption{\em Transformation properties of single polarizations and spin--spin 
correlations. The axis are given by the direction of the fermion and the
scattering plane as shown in the text.\label{tab2}}
\vspace{0.3cm}
\begin{center}
\begin{tabular}{|l|c|c|c|}
\hline
Polarizations           & P             & CP            & T     \\
\hline
\hline
$(\sone+\stwo)_{x,z}$ & $-$             & $+$           & $+$           \\
\hline
$(\sone+\stwo)_y$         & $+$         & $+$           & $-$           \\
\hline
$(\sone-\stwo)_y$         & $+$         & $-$           & $-$           \\
\hline
$(\sone-\stwo)_{x,z}$ & $-$             & $-$           & $+$           \\
\hline
\end{tabular}
\hspace{1cm}
\begin{tabular}{|c|c|c|c|}
\hline
Correlations            & P             & CP            & T     \\
\hline
\hline
$\ba{c}
s_{xx},\ s_{yy}, \ s_{zz}, \\
(s_{xz}+s_{zx}) \ea$            & $+$           & $+$           & $+$       \\
\hline
$\ba{c}
(s_{xy}+s_{yx}), \\             
(s_{yz}+s_{zy}) \ea$            & $-$           & $+$           & $-$       \\
\hline
$(\sone\times\stwo)_{x,z}$      & $-$           & $-$           & $-$       \\
\hline
$(\sone\times\stwo)_y$  & $+$           & $-$           & $+$       \\
\hline
\end{tabular}
\end{center}
\vspace*{0.3cm}
\end{table}

\begin{table}
\caption{\em Allowed combinations of couplings and polarizations in the
cross section.\label{tab3}}
\vspace{0.3cm}
\begin{center}
\begin{tabular}{|c|c|c|c|c|c|}
\hline
 &  SM  &${\rm Re}(a^w_f)$&${\rm Im}(a^w_f)$&${\rm Re}(d^w_f)$&${\rm
 Im}(d^w_f)$\\
\hline 
\hline 
no spins     &$v^4$, $v^2a^2$, $a^4$ & $v^3$, $va^2$ & --- & --- & --- \\
\hline 
\hline 
$X_+$, $Z_+$ &$v^3a$, $va^3$ & $v^2a$, $a^3$ & --- & --- & --- \\
\hline 
$Y_+$        & --- & --- &$v^3$, $va^2$ & --- & --- \\
\hline 
$X_-$, $Z_-$ & --- & --- & --- & --- & $v^3$, $va^2$ \\
\hline 
$Y_-$        & --- & --- & --- & $v^2a$, $a^3$ &  --- \\
\hline 
\hline 
$C_{xx},\ C_{yy},\ C_{zz},\ C^+_{xz}$ 
             & $v^4$, $v^2a^2$, $a^4$ & $v^3$, $va^2$ & --- & --- & --- \\
\hline 
$C^+_{xy},\ C^+_{yz}$ & --- & --- & $v^2a$, $a^3$ & --- & --- \\
\hline 
$C^-_{xy},\ C^-_{yz}$ & --- & --- & --- & $v^3$, $va^2$ & --- \\
\hline 
$C^-_{xz}$ & --- & --- & --- & --- & $v^2a$, $a^3$ \\
\hline
\end{tabular}
\end{center}
\vspace*{0.3cm}
\end{table}
 
Table \ref{tab1} shows how the different terms of 
the effective vertex $Zff$ (identified by their couplings) transform under C, P
and T ($v$ and $a$ stand for whatever $v_i$ and $a_i$, $i=e,f$).
The properties of the transverse $({\bf s^*_i}_x)$, normal $({\bf s^*_i}_y)$ 
and longitudinal $({\bf s^*_i}_z)$ polarizations are shown in Table \ref{tab2}.
The cross section transforms as a real number under discrete transformations 
(it is C--, P-- and T--even) and thus may only contain certain combinations of 
couplings and polarizations (Table~\ref{tab3}).
All these symmetries can be easily checked in the differential cross
section. 

  \subsection{Polarization analysis}

From the previous Section one concludes that the analysis of the spin--density 
matrix of the produced fermion pair (both the single--fermion polarization
and the spin--spin correlation terms) is very important 
to disentangle the different anomalous contributions using appropriate 
observables. 
The polarization of a particle can be measured from the angular distribution
of its parity--violating weak decay products. 

In the case of colored particles the QCD interactions play a crucial
role. One needs that the quark decays before it can form hadronic 
bound states to analyze its polarization. This is not possible for light
quarks and cannot be done with reliable precision for the $b$ quark \cite{mele}.
The angular distribution of the jets in the 2--jet exclusive decays of the $Z$ 
does not carry any information on 
chirality--conserving effective couplings \cite{bernre97c} and cannot probe
CP, T or CPT invariance 
even in the case when the flavor of the jets can be tagged \cite{bernre89a}. 
CP--violating effects can be studied in 2--jet inclusive and 3--jet exclusive 
decays of the $Z$ \cite{koerner91,bernre95,abraham94}.
The $t$ quark is the only one heavy enough to decay before hadronization but it
deserves a different treatment as it cannot be pair--produced at the $Z$ peak.

The $\tau$ lepton case will be studied below as a prototype for polarization
analyses.
In the case of $\tau$ pair production the expressions are simpler 
\cite{ber95,ber94}, as 
\beq
v\equiv v_e=v_\tau=-\frac{1}{2}+2s^2_W\ , 
\ \ a\equiv a_e=a_\tau=-\frac{1}{2}\ .
\eeq
For the single polarization analysis below the direction of flight of the $\tau$ 
must be reconstructed. It has been shown \cite{kuehn93} that this is possible
if both $\tau^\pm$ decay semileptonically (at least one hadron in each decay)
and the energies and tracks of both hadrons are determined (using microvertex 
detectors). Consider the $\tau$--pair production
and their ulterior decay, in the case where at least one of them decays 
semileptonically ($\tau^-\to h^-_1 \nu_\tau$ or 
$\tau^+\to h^+_2\bar{\nu}_\tau$). The direction of emission of the final charged
hadrons $h^\pm$ in the semileptonic decays acts as a $\tau^\pm$ 
spin analyzer. 
Using the narrow width approximation \cite{tsai71}, one can write the cross 
sections for the whole process, in terms of the polar and azimuthal angles
of the charged hadrons in the $\tau^\pm$ rest frames in a very easy way:
substitute the polarization vectors $\spmr$ of
the $\tau^\pm$  by $\mp\alpha_{h^\pm}\cdot{\bf \hat{q}^*_{h^\pm}}$, 
respectively, where ${\bf \hat{q}^*_{h^\pm}}$ is the unit vector of the 
corresponding charged hadron in the $\tau^\pm$ rest frame.
The parameter $\alpha_{h}$
is a measure of the spin--analyzing power of the corresponding decay.
It is maximal, $\alpha_\pi=1$, for the decay $\tau\to\pi\nu_\tau$ and it
is given by $\alpha_\rho=(m^2_\tau-2m^2_\rho)/(m^2_\tau+2m^2_\rho)\simeq0.46$ 
for $\tau\to\rho\nu_\tau$.

Integrating over the angular distribution of the decay 
products of the $\tau$ whose polarization is not analyzed, the
spin--spin correlations between the two decay branches vanish.
In this case, one obtains:

\beq
\frac{{\rm d}\sigma}{{\rm d}\Omega_{\tau^-}{\rm d}\Omega_{h^\mp}}&=&
\frac{4}{4\pi}{\rm BR}(\tau^\mp\to h^\mp\stackrel{(-)}{\nu_\tau}){\rm BR}
(\tau^\pm\to h^\pm_2X) 
\nonumber\\ &\times& 
\Bigg[\frac{{\rm d}\sigma^0}{{\rm d}\Omega_{\tau^-}}
\pm\frac{\alpha^2\beta}{(4s_Wc_W)^4}\frac{2s_Wc_W}{\Gamma^2_Z}\alpha_{h^\mp}
\cdot
\Bigg\{(X_+\pm X_-)\sin\theta_{h^\mp}\cos\phi_{h^\mp} \nonumber \\ &&
      +(Y_+\pm Y_-)\sin\theta_{h^\mp}\sin\phi_{h^\mp} 
      +(Z_+\pm Z_-)\cos\theta_{h^\mp} \Bigg\}\Bigg]\ .
\eeq

This expression can be further simplified integrating over the
polar angle of the analyzing hadron, still keeping valuable information
on all the anomalous couplings \cite{ber95},
\beq
\frac{\dd\sigma}{\dd\cos\theta_{\tau^-}\dd\phi_{h^\mp}}&=&
{\rm BR}(\tau^\mp\to h^\mp\stackrel{(-)}{\nu_\tau}){\rm BR}(\tau^\pm\to h^\pm X)
\nonumber\\ &\times&
\Bigg[4\frac{\dd\sigma^0}{\dd\Omega_{\tau^-}}\pm
\frac{\pi\alpha^2\beta}{(4s_Wc_W)^4}\frac{2s_Wc_W}{\Gamma^2_Z}\alpha_{h^\mp}
\cdot
\Bigg\{(X_+\pm X_-)\cos\phi_{h^\mp}
\nonumber\\ & & \hspace{2cm}
      +(Y_+\pm Y_-)\sin\theta_{h^\mp}\sin\phi_{h^\mp} \Bigg\}\Bigg]\ ,
\label{hminus}
\eeq

In the single--polarization distributions above, the CP--even and CP--odd
contributions cannot be disentangled, as the CP--conserving
(CP--violating) effects represented by the real part of the AWMDM (WEDM)
may be faked by absorptive effects (unitary corrections due to the
exchange of virtual particles). For instance, as mentioned above, the
transverse polarization reads both Re$(a^w_f)$ and Im$(d^w_f)$ and the normal
polarization reads both Re$(d^w_f)$ and Im$(a^w_f)$. 
In principle, one needs to compare the polarizations of both pair--produced 
fermions to eliminate the component of unwanted CP parity.
Nevertheless, exploiting the (accidental) large suppression of the vector 
coupling of charged leptons one can still define a set of observables based on 
single polarizations which are sensitive to the dispersive and absorptive parts
of the $\tau$ weak dipole moments \cite{ber95,ber94,ber95p}. 
A {\em genuine} CP--sensitive observable must involve the two pair--produced 
fermions. This can be done comparing the two decays (at the price of half the 
statistics) or using spin--spin correlations, which manifest in angular 
correlations among their decay products. 
 
    \subsubsection{Observables from single $\tau$ polarization}

We list below the observables sensitive to the WDMs proposed by Bernab{\'e}u 
{\em et al.} \cite{ber95,ber94,ber95p} based on single $\tau$
polarization.\footnote{Here we present exact expressions in agreement
with \cite{ber95,ber94,ber95p} only for $v\ll a$.} They consist of
azimuthal asymmetries on the analyzing hadron direction and demand
the reconstruction of the $\tau$ frame. They exploit the
helicity flipping character of the dipole moments and the
symmetry properties of the anomalous terms. The present limits on the
$\tau$ AWMDM, that will be quoted later, are based on a couple of these 
asymmetries.

\paragraph{Real part of the AWMDM}

Using the distributions (\ref{hminus})  one select
events sensitive to the transverse polarization through 
\beq
\sigma^\mp_{\rm cc}(+)&\equiv&\Bigg[
\int^1_0\dd\cos\theta_{\tau^-}\int^{\pi/2}_{-\pi/2}\dd\phi_{h^\mp}+
\int^0_{-1}\dd\cos\theta_{\tau^-}\int^{3\pi/2}_{\pi/2}\dd\phi_{h^\mp}\Bigg]
\frac{\dd\sigma}{\dd\cos\theta_{\tau^-}\dd\phi_{h^\mp}}\nn \\
\sigma^\mp_{\rm cc}(-)&\equiv&\Bigg[
\int^1_0\dd\cos\theta_{\tau^-}\int^{3\pi/2}_{\pi/2}\dd\phi_{h^\mp}+
\int^0_{-1}\dd\cos\theta_{\tau^-}\int^{\pi/2}_{-\pi/2}\dd\phi_{h^\mp}\Bigg]
\frac{\dd\sigma}{\dd\cos\theta_{\tau^-}\dd\phi_{h^\mp}},\nn\\
\eeq
from which one can construct asymmetries especially sensitive to the dispersive
part of the AWMDM 
\beq
A^{\mp}_{\rm cc}&\equiv&\frac{\sigma^{\mp}_{\rm cc}(+)-\sigma^{\mp}_{\rm cc}(-)}
                      {\sigma^{\mp}_{\rm cc}(+)+\sigma^{\mp}_{\rm cc}(-)}
                      \nonumber\\
&=&
\pm\frac{\alpha_{h^\mp}}{4}s_Wc_W\frac{a}{v^2+a^2}
    \Bigg[-\frac{v}{\gamma s_Wc_W}+2\gamma{\rm Re}(a^w_\tau)
     \pm 2\gamma\frac{v}{a}{\rm Im}(\hat{d}^w_\tau)\Bigg].
\label{tote}
\eeq
The tree level contribution is helicity suppressed. A possible absorptive 
contribution from the WEDM is suppressed by $v/a$ and can be ignored.
Notice that it can otherwise be eliminated using a genuine CP--even observable
that compares both $\tau^\pm$ decaying into the same kind of hadrons 
($\alpha\equiv \alpha_{h^+}=\alpha_{h^-}$):
$A_{\rm cc}\equiv \frac{1}{2}(A^{-}_{\rm cc}-A^{+}_{\rm cc})\propto 
{\rm Re}(a^w_\tau)$.

\paragraph{Imaginary part of the AWMDM}

The following asymmetries related to the normal polarizations are sensitive 
to the absorptive part of the AWMDM: 
\beq
A^{\mp}_{\rm s}&\equiv&
\frac{\displaystyle\int^\pi_0\dd\phi_{h^{\mp}}\frac{\dd\sigma}{\dd\phi_{h^\mp}}
-\displaystyle\int^{2\pi}_\pi\dd\phi_{h^{\mp}}\frac{\dd\sigma}{\dd\phi_{h^\mp}}}
{\displaystyle\int^\pi_0\dd\phi_{h^{\mp}}\frac{\dd\sigma}{\dd\phi_{h^\mp}}
  +\displaystyle\int^{2\pi}_\pi\dd\phi_{h^{\mp}}
                                       \frac{\dd\sigma}{\dd\phi_{h^\mp}}} 
           \nonumber\\    
&=&\mp\alpha_{h^\mp}\frac{3\pi\gamma}{4}s_Wc_W\frac{va^2}{(v^2+a^2)^2} 
\Bigg[{\rm Im}(a^w_\tau)\mp \frac{v}{a}{\rm Re}(\hat{d}^w_\tau)\Bigg].
\eeq
A possible contribution of the WEDM is suppressed by $v/a$. (Again a true
CP--even observable like $A_{\rm s}\equiv \frac{1}{2}(A^{-}_{\rm s}-
A^{+}_{\rm s})$ is completely free from the WEDM dependence.)

\paragraph{Real part of the WEDM}

From events sensitive to the normal polarization of the $\tau$'s
\beq
\sigma^\mp_{\rm sc}(+)&\equiv&\Bigg[
\int^1_0\dd\cos\theta_{\tau^-}\int^{\pi}_{0}\dd\phi_{h^\mp}+
\int^0_{-1}\dd\cos\theta_{\tau^-}\int^{2\pi}_{\pi}\dd\phi_{h^\mp}\Bigg]
\frac{\dd\sigma}{\dd\cos\theta_{\tau^-}\dd\phi_{h^\mp}}\\
\sigma^\mp_{\rm sc}(-)&\equiv&\Bigg[
\int^1_0\dd\cos\theta_{\tau^-}\int^{2\pi}_{\pi}\dd\phi_{h^\mp}+
\int^0_{-1}\dd\cos\theta_{\tau^-}\int^{\pi}_{0}\dd\phi_{h^\mp}\Bigg]
\frac{\dd\sigma}{\dd\cos\theta_{\tau^-}\dd\phi_{h^\mp}}\ ,
\eeq
the following asymmetry isolates the real part of the WEDM with
a $v/a$ suppressed influence from the absorptive part of the AWMDM:
\beq
A^{\mp}_{\rm sc}\equiv\frac{\sigma^{\mp}_{\rm sc}(+)-\sigma^{\mp}_{\rm sc}(-)}
                      {\sigma^{\mp}_{\rm sc}(+)+\sigma^{\mp}_{\rm sc}(-)}
=\alpha_{h^\mp}\frac{\gamma}{2}s_Wc_W\frac{a}{v^2+a^2}
\Bigg[{\rm Re}(\hat{d}^w_\tau)\mp\frac{v}{a}{\rm Im}(a^w_\tau)\Bigg]\ . 
\eeq
The true CP--odd observable $A_{\rm CP}\equiv \frac{1}{2}(A^-_{\rm sc}+
A^+_{\rm sc})$ is completely free of the contribution from AWMDM, as expected.

\paragraph{Imaginary part of the WEDM}

A genuine CPT--odd and CP--odd observable based on single $\tau$ polarizations 
could be constructed from (\ref{tote}) although it is numerically very
suppressed:
\beq
A^{\rm CP}_{\rm cc}\equiv \frac{1}{2}(A^{-}_{\rm cc}+A^{+}_{\rm cc})=
-\frac{\alpha_{h}}{2}\gamma s_Wc_W\frac{v}{v^2+a^2}
{\rm Im}(\hat{d}^w_\tau).
\eeq

    \subsubsection{Observables from spins of both fermions}

In this Section, the polarizations of the two fermions are analyzed
simultaneously. Usually this kind of observables involve spin--spin
correlations (i.e. angular correlations of the final particles in the
decays) although in some case only single fermion polarizations might
contribute.
We restrict ourselves to observables sensitive to CP violation.
As already mentioned, only when both fermions (particle and antiparticle) 
are considered one can get genuine indications of CP violation.
The methods described below are also suitable (and extensively applied) to
search for CP violation in $e^+e^-\to f\bar f$ away from the $Z$ resonance, 
where other CP--violating form factors, besides $d_f^Z(s)$, can be relevant.
Our starting point is an initial CP--even eigenstate, in the c.m.s., as 
$e^+$ and $e^-$ beams are assumed {\em unpolarized}. Otherwise a non--zero 
expectation value of a CP--odd observable would not imply, in principle, 
CP violation. This prescription can be relaxed in some situations:

\begin{itemize}
\item
For the case where the $e^+$ and $e^-$ are transversely polarized,
opposite in direction but equal in magnitude, the initial state is
not a CP eigenstate but certain CP--odd observables still provide
a signal of CP violation \cite{bernre91c}. 

\item
Recently, also the possibility of using longitudinally polarized beams has 
been proposed. Again in this case the initial state is not
a CP eigenstate and therefore a CP--odd and CPT--even observable (sensitive to
Re($d^{\gamma,Z}_f$)) can be contaminated by CP--conserving, absorptive (T--odd) 
effects; similarly a CP--odd, CPT--odd observable (sensitive to 
Im($d^{\gamma,Z}_f$)) can be contaminated by CP--conserving, dispersive 
(T--even) effects. Nevertheless it has been argued that these contaminations 
are at the per mill level\footnote{In the SM,
at leading order in the electroweak couplings, the fermion pair production
proceeds through an intermediate CP--even eigenstate ($\gamma$ or $Z$) and
hence, to one--loop, only electroweak box contributions would be relevant
for the absorptive effects, but they are estimated to be negligible
\cite{bernre96a}. For the dispersive CP--conserving contributions, they can 
only arise in the SM from bremsstrahlung off the initial $e^-$ and $e^+$, but 
this effect is also small \cite{rindani,bernre96a}.} 
\cite{bernre96a,bernre96b} and hence far from the sensitivity of a high 
luminosity $e^+e^-$ collider: true CP--violating effects, if any, would not be 
faked.
\end{itemize}

Consider $e^+(\pplusn)+e^-(\pminusn)\to\tau^+(\kplusn,\splus)+
\tau^-(\kminusn,\sminus)$ with 
$\tau^-\to a(\qminus)+ \mbox{neutrals}$ and $\tau^+\to \bar{b}(\qplus) + 
\mbox{neutrals}$. Then the momenta and polarization vectors in the overall 
c.m.s. transform as follows:\footnote{
These vectors refer to a fixed frame. The transformation
properties of $\spm$ are not in contradiction with Table~\ref{tab2} as
$\spmr$ refer to the frame given by the scattering plane, with axes:
$\hat{z}=\kminus$, $\hat{y}=\kminus\times\pminus/|\kminus\times\pminus|$ and
$\hat{x}=\hat{y}\times\hat{z}$. The symmetries of the
different components of $\spmr$ are thus connected with the ones of
$\kminus$, $\pminus$ and $\spm$.}
\beq
 \ba{rl}
 \mbox{CP}:& \ppmn\to -\pmpn=\ppmn \\
           & \kpmn\to -\kmpn=\kpmn \\
           & \spm\to \smp \\
           & \qpmn\to -\qmpn
 \ea
&\hspace{1cm}&
 \ba{rl}     
\mbox{CPT}:& \ppmn\to \pmpn=-\ppmn \\
           & \kpmn\to \kmpn=-\kpmn \\ 
           & \spm\to -\smp \\
           & \qpmn\to \qmpn 
 \ea
\eeq

From the unit momentum of one of the $\tau$ in the c.m.s.
(e.g. $\kplus$) and both $\tau$ polarizations 
($\spm$) a basis of linearly independent {\em spin}
CP--odd tensor observables can be constructed \cite{bernre91b}. They are 
classified according to their ``$\Theta\equiv$CPT parity'' given by $(-1)^n
{\cal A}^\Theta=\eta_\Theta{\cal A}$ with $n$ being the rank of the 
observable. We show here, for later reference, the scalar observables:
\beq
{\cal A}^{(1)}&\equiv&\kplus\cdot(\splus-\sminus)
 \hspace{1cm} \mbox{[CPT--odd]} \label{eq72}\\
{\cal A}^{(2)}&\equiv&\kplus\cdot(\splus\times\sminus)
 \hspace{1cm} \mbox{[CPT--even]} \label{eq73}
\eeq 
and two vector observables:
\beq
{\cal A}^{(3)}&\equiv&\splus-\sminus 
 \hspace{1cm} \mbox{[CPT--odd]} \label{eq74} \\
{\cal A}^{(7)}&\equiv&\splus\times\sminus 
 \hspace{1cm} \mbox{[CPT--even]} \label{eq75}
\eeq 
Other spin observables are based on them. For instance \cite{bernre94}
\beq
(\pplus\times\kplus)\cdot(\splus-\sminus)&
 \hspace{1cm}& \mbox{[CPT--even]} \label{eq76}
\eeq
is just a projection of ${\cal A}^{(3)}$ sensitive to the CP--odd
combination of the transverse polarizations, $(\sminus-\splus)_y$. 
Similarly, ${\cal A}^{(1)}$ is nothing but $(\sminus-\splus)_z$
(see Table~\ref{tab2}).

The spin observables are related to more realistic (directly 
measurable) {\em momentum observables}
based on the momenta of the $\tau\to 1$ charged prong decays 
\cite{bernre89b}. The dimensionless ones are easier to measure, for instance
the scalar observables\footnote{ 
Notice that $\hat{A}_1$ is related to ${\cal A}^{(2)}$ and ${\cal A}^{(7)}$.
The observable $\hat{A}_2$ is related to ${\cal A}^{(3)}$.}
\cite{bernre92c,bernre93}:
\beq
\hat{A}_1&\equiv&\pplus\cdot\dfrac{\qplusn\times\qminusn}
                                  {|\qplusn\times\qminusn|} 
 \hspace{1cm} \mbox{[CPT--even]} 
\label{A1}\\
\hat{A}_2&\equiv&\pplus\cdot(\qplusn+\qminusn)
 \hspace{1cm} \mbox{[CPT--odd]}
\label{A2}
\eeq 
or the CP--odd traceless tensors \cite{bernre91b,bernre92c,bernre93}:
\beq
\hat{T}_{ij}&\equiv&(\qplusn-\qminusn)_i
       \frac{(\qplusn\times\qminusn)_j}{|\qplusn\times\qminusn|} + 
       (i\leftrightarrow j) \hspace{1cm} \mbox{[CPT--even]} 
\label{T33}\\
\hat{Q}_{ij}&\equiv&(\qplusn+\qminusn)_i(\qplusn-\qminusn)_j +
       (i\leftrightarrow j) \hspace{1cm} \mbox{[CPT--odd]}
\label{Q33}
\eeq
The usual decay channels are $\tau\to\ell\bar \nu_\ell\nu_\tau$ $(\ell=e,\mu)$,
$\tau\to\pi\nu_\tau$ and $\tau\to\rho\nu_\tau\to\pi\pi^0\nu_\tau$
(they amount to about 70\% of the $\tau$ decays). 
Notice that the reconstruction of the $\tau$ frame is not necessary
for the momentum observables and hence the restriction to semileptonic
$\tau$ decays is no longer mandatory. As a final comment, the observables
$\hat{A}_2$ and $\hat{Q}_{ij}$ do not involve angular correlations as
they could be measured considering separate samples of events in the
reactions $e^+e^-\to aX$ and $e^+e^-\to \bar{a}X$. Nevertheless it is 
convenient to treat them in an event--by--event basis \cite{bernre92c}.

One may obtain additional CP--odd observables from combinations
of the ones given above by multiplying them with CP--even scalar weight 
functions to maximize the sensitivity to CP--violating effects ({\em optimal
observables}). Neglecting quartic terms in the CP--violating couplings,
the differential cross section can be written in terms of the
the real and imaginary parts of the EDM and WEDM ($\lambda_i$, $i=1,...,4$),
$\dd\sigma=\dd\sigma_0+\sum_i\lambda_i\dd\sigma^i_1$. It has been shown 
\cite{optimal,optiproof} that the observables given by ${\cal O}_i=
\dd\sigma^i_1/\dd\sigma_0$ have maximal sensitivity to the CP--violating 
parameters $\lambda_i$. 

  \subsection{Experimental limits from these observables
              and comparison with theoretical predictions}

One can give a rough estimate of the sensitivity of a certain experiment to 
put limits to the form factors on a simple statistical basis. This simple
approach provides always optimistic results. More refined 
analyses require the implementation of the detector performance.

Let ${\cal A}$ be an asymmetry proportional to an anomalous form factor $F$, 
${\cal A}=c\cdot F$. The accuracy to determine this asymmetry from a 
sample of $N$ events in an experiment is given by $\Delta\langle{\cal A}
\rangle=n\sqrt{(1-{\cal A}^2)/N}\simeq n/\sqrt{N}$, as the asymmetry is expected
to be very small. $n$ is the number of standard deviations (s.d.) related to
the confidence level of the measurement. If no significant effect is observed 
in the experiment the accuracy translates into an upper bound given by 
$|F|\lsim n/(|c|\sqrt{N})$. In our case, $N=(\#$ of $Z)\cdot$BR$(Z\to\tau^-
\tau^+)\cdot$BR$(\tau^-\to A)\cdot$BR$(\tau^+\to \bar B)$ if the decay channels 
$A$ and $\bar B$ are considered as spin analyzers.

The expectation value of a CP--odd observable ${\cal O}$ is 
given by\footnote{
This relation holds only at the $Z$ peak as the on--shell amplitudes 
contain only one CP--odd form factor, the WEDM.} 
$\langle{\cal O}\rangle_{ab}\equiv\half(\langle{\cal O}
\rangle_{a\bar b}+\langle{\cal O}\rangle_{b\bar a})=c_{ab}\cdot 
\hat{d}^w_\tau$ with $\langle{\cal O}\rangle_{a\bar b}=\int 
\dd\sigma_{a\bar b}{\cal O}/\int \dd\sigma_{a\bar b}$ where the $\tau$
decay channels $a$ and $b$ are selected. Of course any phase space cut
must be CP--symmetric. The accuracy of a 
measurement based on a sample of $N$ events is 
$\Delta\langle{\cal O}\rangle_{ab}=n\sqrt{(\langle{\cal O}^2\rangle_{ab}
-\langle{\cal O}\rangle^2_{ab})/N}\simeq n\sqrt{\langle{\cal O}^2\rangle_{ab})
/N}$, as the CP--odd effects are expected to be very small. The differential
cross section is $\dd\sigma_{a\bar b}\simeq\dd\sigma^{\sc SM}_{a\bar b}+\dd
\sigma^{\sc CP}_{a\bar b}$, where the second term is CP--odd (linear in 
$\hat{d}^w_\tau$). In practice, the average $\langle{\cal O}^2\rangle_{ab}$
is given by the SM. If the measured value is (compatible with) 
zero, an upper limit given by $|\hat{d}^w_\tau|\lsim n
\sqrt{\langle{\cal O}^2\rangle_{ab})}/(|c_{ab}|\sqrt{N})$ with the confidence
level corresponding to $n$ s.d. .

It is customary to define the ratio $r\equiv\langle{\cal O}\rangle/
\sqrt{\langle{\cal O}^2\rangle}$, which is a measure of the sensitivity of the 
CP-odd observable \cite{bernre96a,bernre96b,bernre94}. 
The optimal observables are intended to maximize $|r|$. Given a 
sample of $N=2N_+=2N_-$ events, for the process and its CP image, the 
corresponding signal--to--noise ratio is then $S=
\sqrt{N}|r|$. This is nothing but the statistical significance
of the signal: $S=n$ corresponds to an $n$ s.d.
effect.

The present experimental limits are based on measurements compatible with zero 
for the AWMDM and WEDM:

\begin{itemize}

\item 
Recently a measurement by the L3 detector at LEP of the weak magnetic dipole 
moment of the $\tau$ lepton has been performed, for the first time, using the 
azimuthal asymmetries for the channels 
$\tau\to\pi\nu_\tau$ and $\tau\to\rho\nu_\tau$ \cite{eusebio}. Also a limit to
the weak--electric dipole moment was reported. The results are
\beq
|{\rm Re}(a^w_\tau)|&<&6\times 10^{-3},  
\label{retauexpwmd}\\
|{\rm Im}(a^w_\tau)|&<&2\times 10^{-2},  
\label{imtauexpwmd}\\
|{\rm Re}(d^w_\tau)|&<&4\times 10^{-17}\ e{\rm cm} =  7.2\times10^{-3}\ \mu_\tau
\eeq
(95 $\%$ C.L.), an order of magnitude less stringent than the expectations in
\cite{ber95p}, where all the semileptonic decay channels (52$\%$ of the total 
decay rate) are considered. The $d^w_\tau$ result is not competitive with that
obtained by the CP--odd tensor methods, given below.

\item
For the WEDM the present limits were obtained by the OPAL collaboration 
using optimal observables \cite{opal97} with 95$\%$ C.L.: 
\beq
|{\rm Re}(d^w_\tau)|&<&5.6\times 10^{-18}\ e{\rm cm} = 10^{-3}\ \mu_\tau,  
\label{retauexpwed}\\
|{\rm Im}(d^w_\tau)|&<&1.5\times 10^{-17}\ e{\rm cm} = 2.7\times10^{-3}\
\mu_\tau.
\label{imtauexpwed}
\eeq

\end{itemize}
 
These experimental limits are not in conflict with the theoretical expectations
of the SM and the MSSM. They both predict a real (imaginary) part for the $\tau$
AWMDM at least three (four) orders of magnitude smaller (\ref{tausmwm},
\ref{taususywm}) than the present experimental sensitivity (\ref{retauexpwmd},
\ref{imtauexpwmd}). Closer are the $\tau$ WEDM expectations in the MSSM: the 
real part can reach values of the order of $3\times 10^{-20}\ e$cm in the large 
$\tan\beta$ scenario (\ref{taususywedm}) roughly two orders of magnitude below 
the experimental reach (\ref{retauexpwed},\ref{imtauexpwed}).

\section{Observables off the \boldmath$Z$ resonance: the $t$ quark case}

  
One needs to go beyond the $Z$ peak to produce $t$ quark pairs and this 
implies to account for several new features:

\begin{enumerate}

\item
The photon--exchange diagram is no longer suppressed in this regime and 
therefore one needs to separate the contributions of electromagnetic and 
weak form factors. It is then enough to employ as many independent
observables as necessary. For instance, looking for CP--violating effects,
one may use $\hat{T}_{ij}$ and $\hat{A}_1$ to isolate the real parts and
$\hat{Q}_{ij}$ and $\hat{A}_2$ to get the imaginary parts of $d^\gamma_f(s)$
and $d^Z_f(s)$ (cf. Ref.~\cite{bernre93} for the $\tau$ and 
Ref.~\cite{bernre92c} for the $t$ quark).

\item
Not only vertex-- but also box--diagrams correct the tree 
level process to one loop. The former corrections are parameterized by the 
electromagnetic and the weak vertex form factors but the latter demand the 
introduction of new form factors according to the more general topology of the 
process. 

\item
In general, the dipole form factors $a^V_f(s)$ and $d^V_f(s)$ are not gauge 
invariant to one loop for an off--shell $V$. Exceptions are the
CP--violating ones because they do not receive contributions from vertex 
corrections in which gauge or Goldstone bosons are involved. This at least 
guarantees the gauge--parameter independence for the class of $R_\xi$ gauges.

\end{enumerate}

In any case, for a realistic theory, one expects that any CP--odd observable 
will depend not only on the CP--violating effects due to
vertex corrections parameterized by the dipole form
factors but also on possible CP--violating box contributions. 
We concentrate on supersymmetric CP violating effects in $t$--pair
production at $e^+e^-$ colliders. 
CP conserving MSSM one--loop contributions to $e^+e^-\to f\bar f$ are discussed
in \cite{hs}.
To our knowledge, only the electric and the 
weak--electric form factors have been considered so far to parameterize these 
effects.\footnote{
A similar analysis for hadron colliders has been recently presented in 
Ref.~\cite{zhouhad}.}
Our purpose is:

\begin{itemize}

\item
To briefly review a specific set of CP--odd observables in $t$--pair production.

\item
To evaluate the expectation value of several observables in the 
context of the MSSM with complex parameters to one loop.
  
\item
To compare the contribution from the electric and weak--electric form
factors to these observables with the CP--violation effects coming from
the box contributions. In this way we illustrate, with the MSSM as a concrete 
example, the importance of the box effects.

\end{itemize}

  \subsection{CP--odd observables}

The spin effects can be analyzed through the angular correlation of the weak 
decay products, both in the nonleptonic and in the semileptonic channels
\beq 
&&t(\kplusn)\to b({\bf q_b}) X_{\rm had}({\bf q_{X}})\ , \\  
&&t(\kplusn)\to b({\bf q_b})\ell^+(\qplus)\nu_\ell\quad (\ell=e,\mu,\tau)
\eeq
and the charged conjugated ones.\footnote{The leading QCD corrections to
$e^+e^-\to t\bar{t}$, that include a gluon emission, have a very small
effect on the $t$ spin orientation \cite{kodaira98}.} 
As $m_t>M_W+m_b$, the $t$ quark decays proceed
predominantly through $Wb$. Within the SM the angular distribution of the
charged lepton is a much better spin analyzer of the $t$ quark than that
of the $b$ quark or the $W$ boson arising from semileptonic or nonleptonic 
$t$ decays \cite{analtopspin}. Usual CP--odd observables are:

\begin{enumerate}

\item
The analogous to the ones for the $\tau^+\tau^-$ case \cite{bernre92c}
(Eqs.~(\ref{eq72}--\ref{Q33})), including the optimal observables
\cite{optimal,zhounlc}, obtained from their multiplication
with appropriate CP--even scalar weight functions depending on the energies or
momenta of the final state particles that maximize the
sensitivity to CP--violating effects.
\item
As the $t$ quark is a heavy fermion, the CP conjugate modes $t_L\bar t_L$ and
$t_R\bar t_R$ are produced with a sizeable rate. This allows to construct
the following CP--odd asymmetry \cite{peskin,ber95p}
\beq
A=\frac{\#(t_L\bar t_L)-\#(t_R\bar t_R)}{\#(t_L\bar t_L)+\#(t_R\bar t_R)}
\hspace{1cm} \mbox{[CPT--odd]}
\label{LR}
\eeq
This asymmetry is related to the one that can be
measured through the energy spectra of prompt leptons in the decays 
$t\to W^+b\to\ell^+\nu_\ell b$ and its conjugate.
The $W^+$ is predominantly longitudinally polarized and, assuming the standard 
$V$--$A$ interaction, the $b$ quark is preferentially
left--handed. The $W^+$ is mostly collinear with the $t$ polarization and
so is the $\ell^+$ anti--lepton. Above the $t\bar t$ threshold a
$\ell^+$ coming from a $t_R$ has more energy, due to the Lorentz boost, than
one produced in a $t_L$ decay. The same happens for the conjugate channel,
the  $\ell^-$ from a $\bar{t}_L$ is in average more energetic than the one
from the $\bar{t}_R$. Therefore, in the decay of the $t_R\bar{t}_R$ the 
anti--lepton from $t_R$ has a higher energy $E_+$, while in the decay of the 
pair $t_L\bar t_L$ the lepton from $\bar t_L$ has a higher energy $E_-$. 
Thus the asymmetry $A$ is sensitive to the energy asymmetry of the leptons
\cite{bernre89a,bernre92a},
\beq
\theta(E_+-E_-)-\theta(E_--E_+)
\hspace{1cm} \mbox{[CPT--odd]}
\eeq
where $\theta(x)$ is the step function. Using the $b$ quark as spin
analyzer, a similar asymmetry based on $b$ and
$\bar b$ energies has also been proposed \cite{garfieldval}:
\beq
A^E_{\rm CP}&\equiv&A^E_b-A^E_{\bar b} 
\hspace{1cm} \mbox{[CPT--odd]} \nonumber \\
\mbox{with}& &A^E_{b(\bar b)} \equiv
\frac{\#_{b(\bar b)}(E_{b(\bar b)}>E_0)-\#_{b(\bar b)}(E_{b(\bar b)}<E_0)}
     {\#_{b(\bar b)}(E_{b(\bar b)}>E_0)+\#_{b(\bar b)}(E_{b(\bar b)}<E_0)}
\eeq
where $E_0=\sqrt{s}(m^2_t-M^2_W)/4m^2_t$ is the average energy of the
$b$ or $\bar b$ quarks.
\item
The T--odd {triple product correlations}
\cite{bernre92c,bernre94,bernre96a,bernre96b,vienna}: 
\beq
{\cal A}^{\rm Re}_{\rm CP}&\equiv&\langle{\cal T}_+\rangle-
\langle{\cal T}_-\rangle 
\hspace{1cm} \mbox{[CPT--even]} \nonumber\\
\mbox{with}& & 
{\cal T}_+\equiv({\bf q_1 q_2 q_3})\equiv({\bf q_1}\times{\bf q_2\cdot q_3})
\ \mbox{and}\ {\cal T}_-\equiv{\rm CP}({\cal T}_+)
\eeq
and T--even correlations \cite{bernre94,garfieldval}:
\beq
{\cal A}^{\rm Im}_{\rm CP}&\equiv&\langle{\cal D}_+\rangle-
\langle{\cal D}_-\rangle
\hspace{1cm} \mbox{[CPT--odd]} \nonumber\\
\mbox{with}& & 
{\cal D}_+\equiv {\bf q_1\cdot q_2}
\ \mbox{and}\ {\cal D}_-\equiv{\rm CP}({\cal D}_+)
\eeq
and their dimensionless versions, where ${\bf q_{123}}$ can be any of the 
3--momenta in $e^+e^-\to t\bar{t}$ or the $t$ ($\bar{t}$) decay products
($\hat{A}_1$ and $\hat{A}_2$ are just two of them). 
These asymmetries have been proven suitable to
investigate whether the source of CP violation resides in the $t\bar t$
production or in their decays. In particular, the ones based on 
$(\qplusn {\bf \hat{q}_{\bar X}} \pplus)$ and ${\bf \hat{q}_{\bar X}}\cdot
\qplusn$ are sensitive to CP violation in $t\bar t$ production while the
asymmetry from $(\qplusn {\bf \hat{q}_b} \pplus)$ (close to threshold) traces 
CP violation in semileptonic $t$ and $\bar t$ decays \cite{bernre94}.

\end{enumerate}

We will avoid the possible CP violation in the $t$ or $\bar t$ decays
by considering (ideal) spin observables. In this way we face
the evaluation of the influence on the CP--odd observables of the vertex and 
box diagrams without any interference of other CP--violating effects in the 
$t$ decays. The expectation value of some of the realistic momentum observables
given above will also be presented (assuming SM top decays) for comparison 
with experimental capabilities.

  \subsection{MSSM full contribution to CP--odd observables}

\begin{figure}
\begin{center}
\epsfig{file=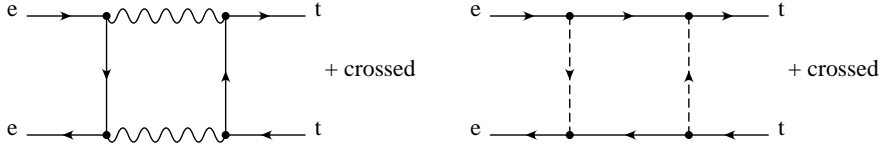,width=12cm}
\end{center}
\caption{\em Generic box graphs for the MSSM at one loop.
\label{boxgraphs}}
\vspace*{0.3cm}
\end{figure}

The one--loop differential cross section for polarized $t$--pair production
in the MSSM \cite{hs} involves the box diagrams indicated in 
Fig.~\ref{boxgraphs} besides the vertex graphs of Fig.~\ref{fig1}. The class 
with vector boson exchange contains only SM contributions ($[eZZt]$, $[\nu_e 
WWb]$) whereas the other one is purely supersymmetric ($[\tilde e\tilde\chi^0
\tilde\chi^0\tilde t]$, $[\tilde\nu_e\tilde\chi^\pm\tilde\chi^\pm\tilde b]$) and
provides CP violating effects.
The box diagrams with Higgs boson exchange are proportional to the
electron mass and can be neglected in the whole calculation (they are CP even 
in any case). When we refer to one--loop calculation in the following, the 
QED as well as the standard QCD corrections to the tree level process are 
excluded: they need real photonic and gluonic corrections to render an 
infrared finite result and constitute an unnecessary complication as they are 
CP--even and do not affect qualitatively our conclusions.

\subsubsection{Spin Observables}

\begin{table}
\caption{\em CP--odd spin observables and the coefficients for the expectation 
value of their integrated version
at $\sqrt{s}=500$ GeV, where only the CP--violating dipole form factors
are taken into account.
\label{tab6}}
\vspace{0.3cm}
\begin{center}
\begin{tabular}{|c|l|c|c|c|r|r|r|r|}
\hline
$i$ &CPT & ${\cal O}_i$ & {\bf a}&{\bf b} & $c_1$ & $c_2$ & $c_3$ 
                                     & $c_4$ \\
\hline \hline
1 & even& $(\soner-\stwor)_y$     & N$\uparrow$&N$\downarrow$
        & 0.526 & 0     & 1.517 & 0     \\
2 & even& $(\soner\times\stwor)_x$& N$\uparrow$&L$\uparrow$     
        &$-0.465$& 0    &$-0.061$& 0    \\
3 & even& $(\soner\times\stwor)_z$& N$\uparrow$&T$\uparrow$     
        & 0.708 & 0     & 0.144 & 0     \\
4 & odd & $(\soner-\stwor)_x$     & T$\uparrow$&T$\downarrow$
        & 0     & 0.930 & 0     & 0.123 \\
5 & odd & $(\soner-\stwor)_z$     & L$\uparrow$&L$\downarrow$
        & 0     &$-1.417$& 0    &$-0.287$\\
6 & odd & $(\soner\times\stwor)_y$& L$\uparrow$&T$\downarrow$
        & 0     & 0.263 & 0     & 0.758 \\
\hline
\end{tabular}
\end{center}
\vspace*{0.3cm}
\end{table}
           
Consider the process $e^+({\bf p})+e^-(-{\bf p})\to t({\bf k},\sone)+
\bar{t} (-{\bf k},\stwo)$ (pair--production of polarized $t$ quarks). 
A list of CP--odd spin observables\footnote{
Notice that ${\cal O}_{1,3,5}$ are equivalent to (\ref{eq76},\ref{eq72},
\ref{eq73}) respectively and the rest are some other projections of 
(\ref{eq74}) and (\ref{eq75}). ${\bf s^*_{1,2}}$ are the $t$ and $\bar t$ 
polarizations defined in their rest frames.}  
classified according to their CPT properties is given in Table~\ref{tab6}.
Their expectation values as a function of $s$ and the scattering angle of the
$t$ quark in the overall c.m. frame are given by
\beq
\langle{\cal O}\rangle_{\mathbf{ab}} &=& \frac{1}{2\dd\sigma}\left[
 \sum_{\soner,\stwor=\pm \mathbf{a},\pm \mathbf{b}}
+\sum_{\soner,\stwor=\pm \mathbf{b},\pm \mathbf{a}}\right]
 \dd\sigma(\soner,\stwor)\ {\cal O}\ ,
\label{obsaverage} \\
\dd\sigma&=&\sum_{\pm\soner,\pm\stwor}\dd\sigma(\soner,\stwor)\ .
\label{spinaverage}
\eeq
The directions of polarization of $t$ and $\bar t$ ({\bf a}, {\bf b}) are 
taken normal to the scattering plane (N), transversal (T) or 
longitudinal (L). They can be either parallel ($\uparrow$) or antiparallel 
($\downarrow$) to the axes defined by $\hat{z}={\bf k}$, $\hat{y}={\bf k}\times
{\bf p}/|{\bf k}\times{\bf p}|$ and $\hat{x}=\hat{y}\times\hat{z}$. Notice that
\beq
\langle{\cal O}^2\rangle_{\mathbf{ab}} &=& \frac{1}{\dd\sigma}
\sum_{\soner,\stwor=\pm \mathbf{a},\pm \mathbf{b}}
\dd\sigma(\soner,\stwor)\ {\cal O}^2
\eeq
and the same for the average of any CP--even quantity. If the information
of the $t$ scattering angle is not available one can consider integrated
observables 
\beq
\langle{\cal O}\rangle_{\mathbf{ab}} &=& \frac{1}{2\sigma}\left[
 \sum_{\soner,\stwor=\pm \mathbf{a},\pm \mathbf{b}}
+\sum_{\soner,\stwor=\pm \mathbf{b},\pm \mathbf{a}}\right]
 \sigma(\soner,\stwor)\ {\cal O}\ .
\eeq

The contributions to the CP--odd observables are linear in the $t$ EDFF and 
WEDFF and in the CP--violating parts of the one--loop box graphs. 
The shape of the different dipole contributions to these observables 
is depicted in Fig.~\ref{fig6.1}. 
The coefficients of the dipoles in the linear expansion of the
integrated spin observables,
\beq
\langle{\cal O}\rangle_{\mathbf{ab}}\equiv\frac{2m_t}{e}\left(
  c_1\ {\rm Re}[d^\gamma_t(s)]
+ c_2\ {\rm Im}[d^\gamma_t(s)]
+ c_3\ {\rm Re}[d^Z_t(s)]
+ c_4\ {\rm Im}[d^Z_t(s)]\right) \nonumber \\
\eeq
are shown in 
Table~\ref{tab6} for their integrated version at $\sqrt{s}=500$ GeV.
They are model independent.

\begin{figure}
\begin{center}
\epsfig{file=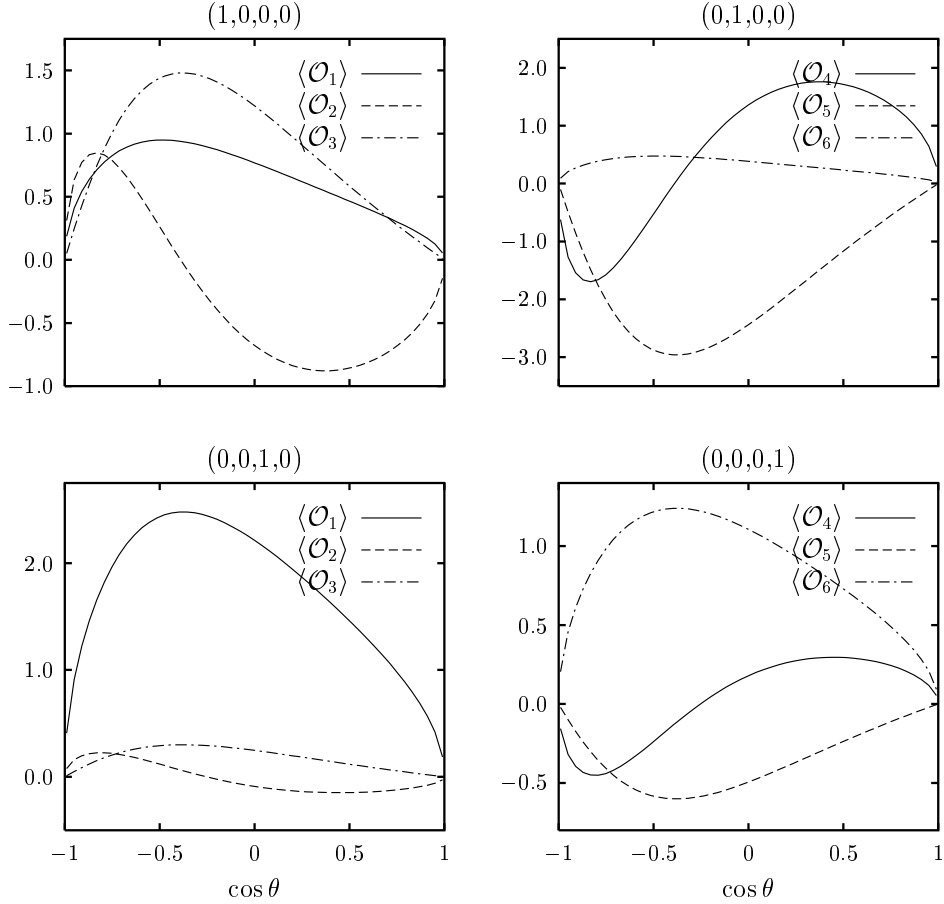,width=0.9\linewidth}
\end{center}
\caption{\em Dipole contributions to the expectation values of the spin 
observables at $\sqrt{s}=500$ GeV for different values of $({\rm Re}[d^\gamma_t], 
\ {\rm Im}[d^\gamma_t],\ {\rm Re}[d^Z_t],\ {\rm Im}[d^Z_t])$ in $\mu_t$ units. 
\label{fig6.1}}
\vspace*{0.3cm}
\end{figure}

\begin{figure}
\begin{center}
Reference Set $\#1$ \\ ~ \\
\epsfig{file=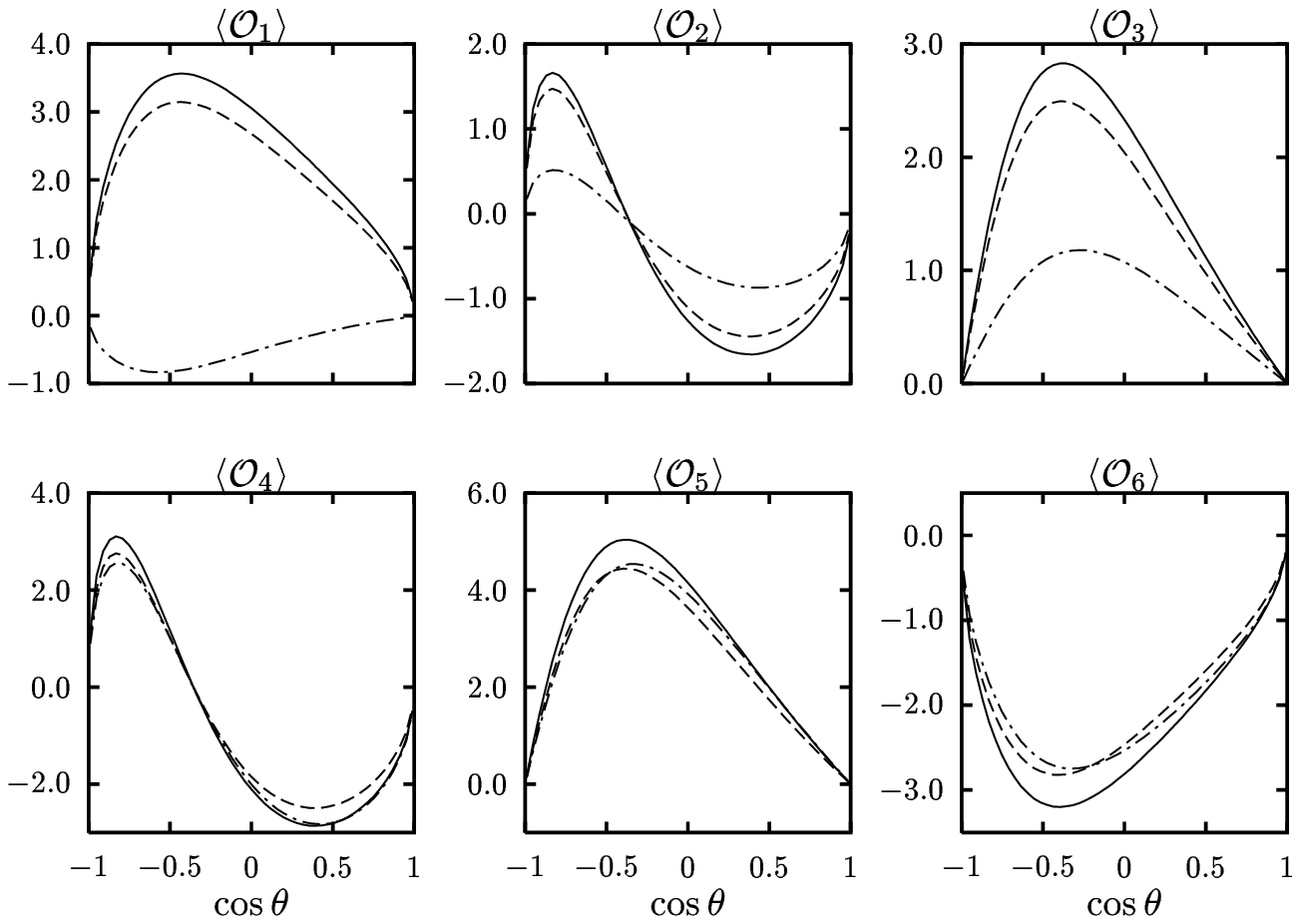,width=0.9\linewidth}
\\ ~ \\
Reference Set $\#2$ \\ ~ \\
\epsfig{file=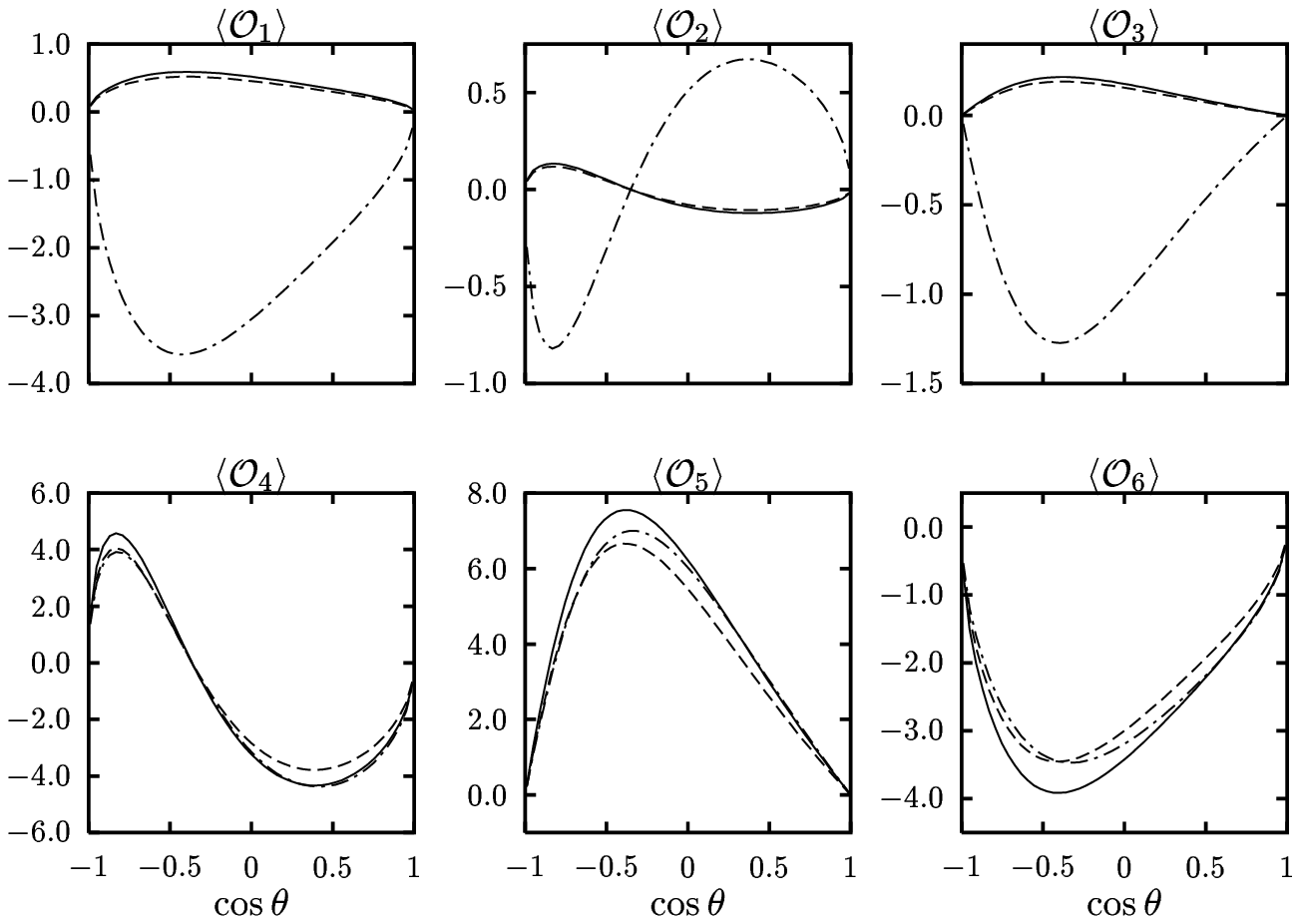,width=0.9\linewidth}
\end{center}
\caption{\em Expectation value of the spin observables [in $10^{-3}$ units] for 
the two reference sets of SUSY parameters, assuming for the cross section: 
  (i) tree level plus contribution from (W)EDFFs only (solid line);  
 (ii) one loop including all the vertex corrections and the self--energies
      (dashed line);
(iii) complete one loop (dot--dashed line).
\label{fig2}}
\vspace*{0.3cm}
\end{figure}

In Fig.~\ref{fig2} we compare the contributions of dipoles and boxes to
the spin observables, for two reference sets of parameters: the first
was given in (\ref{refset1}) and is the one that has been shown to enhance 
the dipole effects; the second is given by
\beq
\mbox{Reference Set }\#2:& &\tan\beta=1.6 \nn \\
& &M_2=|\mu|=m_{\tilde{q}}=|m^t_{LR}|=|m^b_{LR}|=200 \mbox{ GeV} \nn \\
& &\varphi_\mu=\varphi_{\tilde t}=\varphi_{\tilde b}=\pi/2\ ,
\label{refset2}
\eeq
where, compared to (\ref{refset1}), only the CP--violating phases have been 
changed.
The shape of the solid and dashed curves is the same in all cases, as expected. 
Their different size is due to the contributions to the normalization factors 
coming from self--energies, A(W)MFFs and other CP--even corrections.
The plots show that for the Set $\#1$ the MSSM box graphs contribute 
in general to CP violation in the process $e^+e^-\to t \bar t$ by roughly the 
same amount and with a different profile than the EDFFs and WEDFF of the MSSM. 
For nearly all cases this eventually results in a large reduction of the value 
of any CP--odd observable with respect to the expectations from the dipoles
alone, as it will be shown below. Other sets of SUSY 
parameters, that do not enhance the dipole contributions, can provide
instead {\em smaller dipole form factors but larger observable effects}. This
is the case of the Set $\#2$, that yields much smaller values for the 
real parts of the dipole form factors at $\sqrt{s}=500$ GeV than we had before 
in Eq.~(\ref{topff}),
\beq
d^\gamma_t      &=&     (0.547-1.889\ {\rm i})\times 10^{-3}\ \mu_t , \\
d^Z_t           &=&     (0.362-2.319\ {\rm i})\times 10^{-3}\ \mu_t . 
\label{topff2}
\eeq
As a consequence, the CPT--even observables receive their contributions
mainly from the box diagrams (Fig.~\ref{fig2}). 

\begin{table}
\caption{\em Ratio $r=\langle{\cal O}\rangle/\sqrt{\langle{\cal O}^2\rangle}$
 [in $10^{-3}$ units] of the integrated spin observables at $\sqrt{s}=500$ GeV 
 for the two reference sets of SUSY parameters. The left column excludes the box 
 corrections and the right one comes from the complete one--loop cross section 
 for $e^+e^-\to t\bar t$.\label{tab7}}
\vspace{0.3cm}
\begin{center}
\begin{tabular}{|c|l|c|c|c|r|r|r|r|}
\hline
$i$ &CPT & ${\cal O}_i$ & {\bf a}&{\bf b} & 
\multicolumn{2}{|c|}{Set $\#1$} &
\multicolumn{2}{|c|}{Set $\#2$} \\
\hline \hline
1 & even&  $(\soner-\stwor)_y$ & N$\uparrow$&N$\downarrow$
	& 1.216 &$-0.231$& 0.207 &$-1.394$      \\
2 & even&  $(\soner\times\stwor)_x$& N$\uparrow$&L$\uparrow$	
	&$-0.755$&$-0.489$&$-0.053$&$0.318$       \\
3 & even&  $(\soner\times\stwor)_z$& N$\uparrow$&T$\uparrow$	
	& 1.184	&$0.625$&0.090	&$-0.598$       \\
4 & odd &  $(\soner-\stwor)_x$     & T$\uparrow$&T$\downarrow$
	&$-1.230$&$-1.421$&$-1.888$&$-2.217$ \\
5 & odd &  $(\soner-\stwor)_z$     & L$\uparrow$&L$\downarrow$
	& 2.550 &$2.739$ & 3.823 &4.216   \\
6 & odd &  $(\soner\times\stwor)_y$& L$\uparrow$&T$\downarrow$
	&$-1.683$&$-1.751$&$-2.050$&$-2.216$ \\
\hline
\end{tabular}
\end{center}
\vspace*{0.3cm}
\end{table}

The previous arguments are reflected in Table~\ref{tab7}.
There we show the ratio $r=\langle{\cal O}\rangle/\sqrt{\langle{\cal O}^2
\rangle}$ for the integrated spin observables.
We compare the result when only the self energies and vertex corrections
are included (left column) with the complete one--loop calculation (right 
column). The shape of the observables as a function of the $t$
polar angle (Fig.~\ref{fig2}) is such that the signal for CP--violation $r$ is 
still not very large for a couple of CPT--even observables with the Set $\#2$ 
but it is much better for all the others. This illustrates that the dipole form
factors of the $t$ quark are not sufficient to parameterize observable 
CP--violating effects and the predictions can be wrong by far.

As final comments, notice that ${\cal O}_5$ is nothing but the asymmetry of 
Eq.~(\ref{LR}),
\beq
\langle{\cal O}_5\rangle
=\frac{\#(t_L\bar t_L)-\#(t_R\bar t_R)}{\#(t_L\bar t_L)+\#(t_R\bar t_R)}\ .
\eeq
Attending to Table~\ref{tab6} it happens to be the most sensitive observable
to the imaginary part of the EDFF and also the best one to test CP violation
for our choice of SUSY parameters (Table~\ref{tab7}).
The observables $(\soner-\stwor)$ can still give information on CP violation 
when the polarization of one of the $t$ quarks is not analyzed.\footnote{
Notice that anyway a comparison between two samples, one with polarized
$t$ and the other with polarized $\bar t$, is necessary to build the
genuine CP--odd observable of Eq.~(\ref{spinaverage}).} The sensitivity
of the single--spin polarization to CP--violation is indeed worse: 
$\langle{\cal(\soner-\stwor)}_{x,y,z}\rangle_{\bf ab}
=2\langle{\cal(\soner)}_{x,y,z}\rangle_{\bf a}$ for {\bf a}({\bf b})=
N$\uparrow$($\downarrow$),T$\uparrow$($\downarrow$),L$\uparrow$($\downarrow$).

\subsubsection{Momentum Observables}

Consider now the decay channels labeled by $a$ and $c$ acting as spin analyzers 
in $t + \bar t\to a(\qplus) + \bar c(\qminus)+X$. 
The expectation value of a CP--odd observable is given by the average over the 
phase space of the final state particles,
\beq
\langle{\cal O}\rangle_{ac} = \frac{1}{2}\left[ 
\langle{\cal O}\rangle_{a\bar c} + \langle{\cal O}\rangle_{c\bar a} \right] =
\frac{1}{2\sigma_{ac}}\int
\left[ \dd\sigma_{a\bar c} + \dd\sigma_{c\bar a} \right]\ {\cal O}\ ,
\eeq
where both the process ($a\bar c$) and its CP conjugate ($c\bar a$) are 
included and 
\beq
\sigma_{ac}=\int\dd\sigma_{a\bar c}=\int\dd\sigma_{c\bar a} \ ,
\eeq
in full analogy with Eqs.~(\ref{obsaverage},\ref{spinaverage}).
The differential cross section for $t$--pair production and decay is
evaluated for every channel using the narrow width approximation.

\begin{table}
\caption{\em Ratio $r$ [in $10^{-3}$ units] of the momentum observables 
(\ref{A1}--\ref{Q33}) and their ratio $r$  
at $\sqrt{s}=500$ GeV for three different channels: $t + \bar t\to a(\qplus) + 
\bar c(\qminus)+X$, given the Set $\#2$ of SUSY parameters 
(\ref{refset2}). The left column includes only the $t$ (W)EDFF corrections  
and the right one comes from the complete one--loop cross section for 
$e^+e^-\to t\bar t$. \label{tab9}}
\vspace{0.3cm}
\begin{center}
\begin{tabular}{|l|c|r|r|r|r|r|r|}
\hline
CPT     & ${\cal O}$ & \multicolumn{2}{|c|}{$b-b$} 
        & \multicolumn{2}{|c|}{$\ell-b$} 
        & \multicolumn{2}{|c|}{$\ell-\ell$} \\ 
\hline \hline
even& $\hat{A}_1$   &
$-0.036$	&0.242 		&
0.030		&$-0.202$ 	& 	
0.068	&$-0.467$\\
odd & $\hat{A}_2$ 	&  
0.270		&0.304 	&
$-0.180$	&$-0.204$& 	
$-0.501$	&$-0.812$\\
even& $\hat{T}_{33}$&
$-0.006$	&$0.042$&
0.021	&$-0.140$& 	
$-0.037$	&$0.248$\\
odd & $\hat{Q}_{33}$&
0.486		&0.542 	& 
$-0.335$	&$-0.374$& 	
$-1.274$	&$-1.420$\\
\hline
\end{tabular}
\end{center}
\vspace*{0.3cm}
\end{table} 

In Table~\ref{tab9} the ratio $r$ is shown for three different decay channels 
at $\sqrt{s}=500$ GeV and some {\em realistic} CP--odd observables involving 
the momenta of the decay products analyzing $t$ and $\bar t$ polarizations 
in the $e^+e^-$ c.m.s. (laboratory frame). As expected the leptonic
decay channels are the best $t$ spin analyzers but the number of leptonic
events is also smaller. The Reference Set $\#2$ has 
been chosen. As expected, the dipole contributions (left columns) to the CPT 
even observables are very small for this choice of SUSY parameters 
but the full expectation values (right columns) are larger.

Concerning the experimental reach of this analysis,
the statistical significance of the signal of CP violation is given
by $S=|r|\sqrt{N}$ with $N=\epsilon{\cal L}\sigma_{t\bar t}\mbox{BR}
(t\to a)\mbox{BR}(\bar t\to\bar c)$ where $\epsilon$ is the detection
efficiency and ${\cal L}$ the integrated luminosity of the collider.
The branching ratios of the $t$ decays are BR $\simeq 1$ for the $b$ channel
and BR $\simeq 0.22$ for the leptonic channels ($\ell=e,\mu$).
At $\sqrt{s}=500$ GeV the total cross section for $t$--pair production is
$\sigma_{t\bar t}\simeq 0.5$ pb and the NLC integrated luminosity ${\cal L}
\simeq 50$ fb$^{-1}$ \cite{nlc}. Assuming a perfect detection efficiency, one
gets $\sqrt{N}\simeq 160, 75, 35$ for the channels $b-b$, $\ell-b$,
$\ell-\ell$, respectively. With these statistics, values of $|r|\sim 10^{-2}$
would be necessary to achieve a 1 s.d. effect, which does not seem to
be at hand in the context of the MSSM as Table~\ref{tab9} shows. 
Nevertheless, the ratio $r$ should improve considerably ($\sim$30\%) using 
optimal observables (suitable to enhance the dipole effects) and also with 
(longitudinally) polarized beams \cite{bernre96a}. 

\section{Summary and conclusions}

The one--loop expressions of the dipole form factors of fermions in terms 
of arbitrary complex couplings in a general renormalizable theory for the 
't Hooft--Feynman gauge have been given. The CP--violating (--conserving)
dipole form factors depend explicitly on the imaginary (real) part of 
combinations of the couplings. 

As an application, the weak--magnetic and weak--electric dipole moments of
the $\tau$ lepton and the $b$ quark (defined at $s=M^2_Z$) and the 
electric and the weak--electric dipole form factors of the $t$ quark
(for $s>4m^2_t$) have been evaluated for the MSSM with preserved R--parity
and non--universal soft--breaking terms. A version with complex couplings
of the MSSM has been implemented in order to get one--loop CP--violating 
effects. The quark dipoles depend on three physical CP phases and the lepton
ones only on two since there is no mixing in the scalar neutrino sector.
The supersymmetric parameter space has been scanned in search for the
largest contributions to the dipoles. 
Both the $\tau$ and $b$ dipole moments are enhanced for large $\tan\beta$. 
There the AWMDMs can be one order of magnitude larger than in the electroweak 
SM (in particular the real parts, ${\cal O}(10^{-5})$) but still a factor 
five below the standard QCD contribution for the $b$ case. The WEDMs can be 
instead twelve orders of magnitude larger than the tiny three--loop prediction 
by the SM and reach ${\cal O}(10^{-20})\ e$cm for the $\tau$ and 
${\cal O}(10^{-19})\ e$cm for the $b$.
Conversely, the $t$ dipole form factors are larger in the low $\tan\beta$ 
scenario. The values depend strongly on the interplay between the energy at 
which they are evaluated and the set of supersymmetric parameters used as
inputs. The real and imaginary parts can reach values of similar size: 
at $\sqrt{s}=500$ GeV, the $t$ EDFF and WEDFF are of ${\cal O}(10^{-19})\ e$cm.

At the $Z$ resonance the electromagnetic dipole form factors are irrelevant
and the weak dipole form factors (weak dipole moments of the pair--produced
fermions) are gauge invariant objects that can also be related to physical
observables. Such observables are based on the polarization analysis of
the fermions, equivalently on the angular and energy distributions of their 
decay products. The cross section for polarized fermion production in $e^+e^-$
collisions at the $Z$ peak has been given in terms of the AWMDM and WEDM,
as well as a review on observables able to discriminate the dipole moments for 
the $\tau$ case (the polarization analysis does not appear feasible for the $b$
case). The present experimental limits based on them have been compared to the 
SM and the MSSM expectations. 
The experimental sensitivity to the weak dipole moments of the $\tau$
are far from the MSSM predictions, at least three (two) orders of magnitude 
for the weak--magnetic (weak--electric) dipoles.

Away from the $Z$ peak both the electromagnetic and the weak dipole form
factors are equally relevant but not enough to parameterize all the physical
effects (in particular CP--violation). The case of $t$--pair production
in high energy $e^+e^-$ colliders has been considered to illustrate this fact. 
Taking several CP--odd spin--observables based on the $t$ and $\bar t$
polarization vectors, the different contributions from vertex and box
corrections have been evaluated. There is no one loop contribution from the
SM sector. It has been shown that, for the set of supersymmetric parameters 
that provides sizeable values of the $t$ CP--violating dipoles, the SUSY box 
contributions happen to contribute with opposite sign and in a similar 
magnitude, yielding altogether a much smaller CP--violating observable effect.
Another configuration has been shown for which the dipoles are
smaller but the combined effect is larger. The same analysis has been performed
using instead realistic observables based on the momenta of $t$ and $\bar t$ 
decay products, with similar results. The channels in which a $bb$, $b\ell$ 
and $\ell\ell$ act as spin analyzers have been used under the assumption of 
standard CP--conserving decays and using the narrow width approximation. As 
expected, the leptons are the best $t$ spin analyzers yielding the maximal 
values for the signal of CP--violation, $r\simeq 0.5\times 10^{-3}$. 
Nevertheless the statistics for such events is smaller.

\section*{Acknowledgments}

We wish to thank T. Hahn for very valuable assistance in the optimization of 
the computer codes and the preparation of the figures. 
Useful discussions with J. Bernab{\'e}u and A. Masiero are also gratefully 
acknowledged. 
One of us (S.R.) would like to thank the INFN Theoretical Group of Padua for 
the nice and fruitful hospitality enjoyed during the preparation of part 
of this paper.  
J.I.I. has been supported by the Fundaci{\'o}n Ram{\'o}n Areces and partially by 
the Spanish CICYT under contract AEN96-1672 and S.R. by the EC under contract 
ERBFMBICT972474.

\appendix

\section{The one-loop tensor integrals \label{appendix-a}}

\begin{figure}[htb]
\begin{center}
\begin{picture}(15,4)
\epsfxsize=12cm
\put(1.5,0){\epsfbox{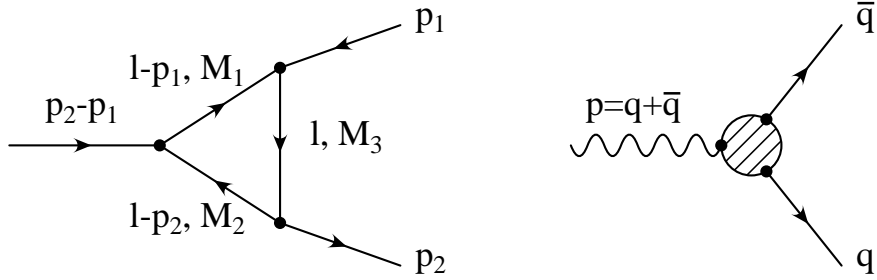}}
\end{picture}
\end{center}
\caption{\em Momentum convention for three-point-tensor integrals.
\label{fig:3pf}}
\vspace*{0.3cm}
\end{figure}
The systematic way of dealing with one-loop integrals consists of
reducing the tensor integrals to scalar ones. We employ the following 
set of tensor integrals (see Fig.~\ref{fig:3pf}):
\beq
{C}_{\{0,\ \mu,\ \mu\nu\}}(p_1,p_2,M_1,M_2,M_3)\equiv\frac{16\pi^2}{i}
\mu^{4-D}\int\frac{\dd^D l}{(2\pi)^D}\frac{\{1,\ l_\mu,\ l_\mu l_\nu\}}
{{\cal D}_1{\cal D}_2{\cal D}_3}\ ,
\label{c}
\eeq
with
\beq
{\cal D}_1&=&(l-p_1)^2-M^2_1+i\epsilon\ , \nonumber\\
{\cal D}_2&=&(l-p_2)^2-M^2_2+i\epsilon\ , \nonumber\\
{\cal D}_3&=&l^2-M^2_3+i\epsilon\ .
\eeq
and the orthogonal reduction ($k_{\pm}=p_1\pm p_2$) \cite{topol},
\beq
{C}_\mu&\equiv&C^+_1 k_{+\mu}+C^-_1 k_{-\mu} \ , \\
{C}_{\mu\nu}&\equiv&C^+_2 k_{+\mu}k_{+\nu}+C^-_2 k_{-\mu}k_{-\nu} 
        +C^{+-}_2 [k_{+\mu}k_{-\nu}+k_{+\nu}k_{-\mu}]+C^0_2 g_{\mu\nu}\ .
\eeq

For us these definitions are very convenient, because the integral
(\ref{c}) is invariant under the combined replacements 
$p_1\leftrightarrow p_2$ and $M_1\leftrightarrow M_2$, and we are 
dealing with equal mass final-state particles $p_1^2=p_2^2=m^2$.
Therefore $C^-_1$ and $C^{+-}_2$ are antisymmetric under these
replacements while the other scalar integrals are symmetric, and the
case $M_1=M_2$ automatically yields $C^-_1=C^{+-}_2=0$.

For comparison we also list the tensor decomposition defined in
\cite{Denner93}. 
\beq
{C}^{(D)}_{\{0,\ \mu,\ \mu\nu\}}(p_1,p_2,M_3,M_1,M_2)\equiv\frac{16\pi^2}{i}
\mu^{4-D}\int\frac{\dd^D l}{(2\pi)^D}\frac{\{1,\ l_\mu,\ l_\mu l_\nu\}}
{{\cal D}_1{\cal D}_2{\cal D}_3}\ ,
\eeq
with
\beq
{\cal D}_1&=&l^2-M^2_3+i\epsilon\ , \nonumber\\
{\cal D}_2&=&(l+p_1)^2-M^2_1+i\epsilon\ , \nonumber\\
{\cal D}_3&=&(l+p_2)^2-M^2_2+i\epsilon\ .
\eeq
and the tensor decomposition
\beq
{C}^{(D)}_\mu&\equiv&{p_1}_\mu C^{(D)}_1 + {p_2}_\mu C^{(D)}_2\ ,\\
{C}^{(D)}_{\mu\nu}&\equiv&g_{\mu\nu} C^{(D)}_{00}
               + {p_1}_\mu{p_1}_\nu C^{(D)}_{11}
               + {p_2}_\mu{p_2}_\nu C^{(D)}_{22}
               + ({p_1}_\mu{p_2}_\nu + {p_1}_\nu{p_2}_\mu)
               C^{(D)}_{12}\ .\nonumber\\&&
\eeq
These scalar integrals are related to the ones obtained by orthogonal
reduction in the following way:
\begin{eqnarray}
C^{(D)}_1 & = & -C_1^+ -C_1^-,
\\
C^{(D)}_2 & = & -C_1^+ +C_1^-,
\\
C^{(D)}_{00} & = & C_2^0,
\\
C^{(D)}_{11} & = & C_2^+ +C_2^- +2C_2^{+-},
\\
C^{(D)}_{22} & = & C_2^+ +C_2^- -2C_2^{+-},
\\
C^{(D)}_{12} & = & C_2^+-C_2^-,
\end{eqnarray}
with the arguments
$$
C_i^{(D)} = C_i^{(D)}(p_1,p_2,M_3,M_1,M_2),\quad
C_i^j = C_i^j(-p_1,-p_2,M_1,M_2,M_3).
$$

\section{Conventions for fields and couplings in the SM and the MSSM
         \label{appendix-c}}

  \subsection{The conventions}

We use the conventions of Ref.~\cite{mssm,HHG}. The covariant 
derivative acting on a SU(2)$_L$ weak doublet field with hypercharge $Y$ is 
given by
\beq
D_\mu=\partial_\mu+{\rm i}g\frac{\tau^a}{2}W^a_\mu+{\rm i}g'\frac{Y}{2}B_\mu \ ,
\label{covder}
\eeq
where $\tau^a$ are the usual Pauli matrices and the electric charge
operator is $Q_f=I^f_3+Y/2$, with $I^f_3=\tau^3/2$. The $Z$ and
photon fields are defined by
\beq
Z_\mu&=&W^3_\mu\cos\theta_W-B_\mu\sin\theta_W \ ,\\
A_\mu&=&W^3_\mu\sin\theta_W+B_\mu\cos\theta_W \ .
\eeq
The charged weak boson field is $W^\pm_\mu=\frac{1}{\sqrt{2}}
(W^1_\mu\mp {\rm i} W^2_\mu)$.\s

In the Standard Model (SM), there is only one Higgs doublet ${\bf H}$ with 
hypercharge $Y=1$. After spontaneous symmetry breaking (SSB), the physical 
Higgs field $H^0$ and the would--be--Goldstone bosons $\chi$ and $\phi^\pm$ are
given by
\beq
{\bf H}=\left(\ba{c} \phi^+ \\ \frac{1}{\sqrt{2}}[v+H^0+{\rm i}\chi] 
\ea\right)\ .
\eeq\s

In the Minimal Supersymmetric Standard Model (MSSM) there are the same matter
and gauge field multiplets as in the SM, each supplemented
by superpartner fields to make up complete supersymmetry
multiplets. The matter and gauge fields have the same quantum number
assignments as in the SM. But in the MSSM there are two
Higgs doublets with opposite hypercharges. Each forms a chiral 
supersymmetry multiplet and an SU(2) doublet. 
The Lagrangian we use is defined in Ref.~\cite{Haber93}, especially in
Eq.~(1.45) and Eq.~(1.54). We do not impose any reality constraint
onto the parameters except for the reality of the
Lagrangian.

Spontaneous breakdown of the SU(2)$\times$U(1) gauge symmetry leads to the 
existence of five physical Higgs particles: two CP-even Higgs bosons $h$ and 
$H$, a CP-odd or pseudoscalar Higgs boson $A$, and two charged Higgs particles 
$H^\pm$. They are grouped in two doublets,
${\bf H}_1\equiv i\tau^2{\bf\Phi}_1^*$ and ${\bf H}_2\equiv{\bf\Phi}_2$,
with opposite  hypercharges ($Y=\mp 1$), where  
\beq
{\bf\Phi}_1=\left(\ba{c} \phi^+_1\\ \phi^0_1 \ea\right)\ , \ \
{\bf\Phi}_2=\left(\ba{c} \phi^+_2 \\ \phi^0_2 \ea\right)\ . 
\eeq
After SSB these are expressed in terms of the physical fields $h$, $H$, $A$, 
$H^\pm$ and the would-be-Goldstone bosons $G^0$, $G^\pm$ by
\beq
\phi^+_1&=&-H^+\sin\beta+G^+\cos\beta \ , \\
\phi^0_1&=&\frac{1}{\sqrt{2}}\left\{v_1+[(-h\sin\alpha+H\cos\alpha)+
                        {\rm i}(-A\sin\beta+G^0\cos\beta)]\right\} \ , \\
\phi^+_2&=& H^+\cos\beta+G^+\sin\beta \ , \\
\phi^0_2&=&\frac{1}{\sqrt{2}}\left\{v_2+[( h\cos\alpha+H\sin\alpha)+
                        {\rm i}( A\cos\beta+G^0\sin\beta)]\right\} \ . 
\eeq

Besides the four masses, two additional parameters are needed to describe the 
Higgs sector at tree-level: $\tan\beta=v_2/v_1$, the ratio of the two 
vacuum expectation values, and a mixing angle $\alpha$ in the CP-even sector. 
However, only two of these parameters are independent. Using $M_A$ and 
$\tan\beta$ as input parameters, the masses and the mixing angle $\alpha$ in 
the $H,h$ sector read \cite{HHG} 
\begin{eqnarray}
M_{h,H}^2& = &\frac{1}{2} (M_A^2+M_Z^2+\epsilon) \nonumber\\
 & \times&\Bigg[\,  1 \mp 
\sqrt{1 - 4 \frac{ M_A^2 M_Z^2 \cos^22\beta + \epsilon(M_A^2 
\sin^2\beta + M_Z^2 \cos^2\beta)}{(M_A^2+M_Z^2+\epsilon)^2} }\, 
\Bigg] \ , \\
M_{H^\pm} &=& M_A \left[ 1+ \frac{M_W^2}{M_A^2} \right]^{1/2} \ , \\
\tan 2 \alpha& =& \tan 2 \beta \frac{ M_A^2+M_Z^2} {M_A^2-M_Z^2+ 
\epsilon/\cos{2\beta}} \ ; \ \ -\frac{\pi}{2}<\alpha<0 \ ,
\end{eqnarray}
which include the leading radiative correction
\beq
\epsilon = \frac{3 G_{F}}{\sqrt{2}\pi^2} \frac{m_t^4}{\sin^2\beta}
\log\left( 1+\frac{m_{\tilde{q}}^2}{m_t^2} \right) \ .
\eeq
Here $G_F$ is the Fermi constant and $m_{\tilde{q}}$ is the common
mass scale for the squarks.\s

The mass terms for the neutral gauginos and Higgsinos come out of the
bilinear Higgs part of the superpotential (the $\mu$ term) and the 
soft--SUSY--breaking gaugino mass terms with masses
$M_2$ and $M_1$ for the SU(2) and U(1) gauginos $\lambda^a$ and
$\lambda^\prime$, respectively. The gaugino mass parameters are
constrained by the GUT relation
\beq M_1 = \frac{5}{3} \frac{s_W^2}{c_W^2} M_2\ .
\eeq
Mixing terms arise from the minimal
coupling terms between Higgs, Higgsino and gaugino fields, where the
Higgs fields have been replaced by their vacuum expectation values.
In terms of two--component spinors the mass terms add up to
\beq
{\cal L}^{\tilde{\chi}^0}_m = -\frac{1}{2}
(-{\rm i}\lambda^\prime,-{\rm i}\lambda^3,\psi^0_{H_1},\psi^0_{H_2}) Y
(-{\rm i}\lambda^\prime,-{\rm i}\lambda^3,\psi^0_{H_1},\psi^0_{H_2})^T + h.c.\ ,
\eeq
with the symmetric mass matrix
\beq
Y = \left(\begin{array}{rrrr}
     M_1 &  0 & .  &   . \\
     0 & M _2& . & . \\
     -M_Z s_W \cos \beta & M_Z c_W \cos \beta &  0 &  -\mu \\
     M_Z s_W \sin \beta & -M_Z c_W \sin \beta & -\mu & 0
\end{array}\right)\ .
\eeq
We define the unitary diagonalization matrix $N$ and the matrix
$N^\prime$, which is often more convenient, by the equations
\beq
N^*YN^{-1} = N_{diag},
\ N^\prime = N \cdot \left(\begin{array}{rrrr} c_W&-s_W&0&0 \\
                                         s_W&c_W&0&0  \\
                                         0&0&1&0      \\
                                         0&0&0&1
                     \end{array}\right)\ .
\eeq
The matrix $N$ is not unique. We could impose the requirement that
$N_{diag}$ has only non--negative entries (which would force $N$ to
have an imaginary part if $Y$ is not positive definite), but at least
if the parameters in $Y$ are real
it is also possible to choose $N$ real (which may lead to some negative
entries of $N_{diag}$). However our formulas do not depend on this
choice. In general the $\mu$ parameter may be complex and then the
neutralino masses depend also on the phase of $\mu$.
The neutralino mass eigenstates are four
Majorana spinors $\tilde{\chi}_j^0$ given by
\begin{eqnarray}
{\tilde{\chi}}_j^0 & \equiv &
(P_L N_{jk} + P_R N^*_{jk}) \tilde{\Psi}^0_k\ , \\
{\tilde{\Psi}}_k^0 & \equiv &
\left({{-{\rm i}\lambda^\prime}\choose {{\rm i}\bar{\lambda}^\prime}}\ , 
{{-{\rm i}\lambda^3}\choose {{\rm i}\bar{\lambda}^3}}\ ,
{{\psi_{H_1}^0} \choose {\bar{\psi}_{H_1}^0}}\ ,
{{\psi_{H_2}^0} \choose {\bar{\psi}_{H_2}^0}} \right)\ .
\end{eqnarray}\s

The mass terms for the charged gaugino and Higgsino have the same
origin and, in terms of two-component spinors, read 
\begin{eqnarray}
{\cal L}^{\tilde{\chi}^\pm}_m & = & 
-\frac{1}{2} ( \psi^+\ , \ \ \psi^-)
\left(\begin{array}{cc} 0&X^T \\ X & 0 \end{array}\right)
\left(\begin{array}{r} \psi^+ \\ \psi^- \end{array}\right) , \nonumber\\
X & \equiv & \left(\begin{array}{cc} M_2&M_W \sqrt{2}\sin \beta\\ 
             M_W \sqrt{2} \cos \beta& \mu\end{array}\right)\ ,
\end{eqnarray}
with the abbreviations
\beq
\psi^+_j = (-{\rm i}\lambda^+, \psi_{H_2}^+), \quad \psi^-_j = 
(-{\rm i}\lambda^-, \psi_{H_1}^-)\ .
\eeq
We now define unitary diagonalization matrices $U, V$ by the equation
\beq
U^* X V^{-1} = M_{diag}.
\eeq
Again these matrices can be chosen to be real if $X$ is real. Taking in
general a complex $\mu$ parameter, the masses of the charginos are given by
\beq
m^2_{\tilde{\chi}^{\pm}_{1,2}}=\frac{1}{2}\Big\{M^2_2+|\mu|^2+2M^2_W\mp
[(M^2_2-|\mu|^2)^2+4M^2_W\cos^22\beta\\
+4M^2_W(M^2_2+|\mu|^2+2M_2{\rm Re}(\mu)\sin 2\beta)]^{1/2}\Big\}.
\eeq

The chargino mass eigenstates are two Dirac spinors $\tilde{\chi}^+_j
$ given by
\begin{eqnarray}
\tilde{\chi}^+_j &\equiv& (P_L V_{jk} + P_R U_{jk}^*)\tilde{\Psi}_k \ ,\\
\tilde{\Psi}_k & \equiv &
 \left({{-{\rm i}\lambda^+} \choose {{\rm i}\bar{\lambda}^-}}\ , 
       {{\psi_{H_2}^+} \choose {\bar{\psi}_{H_1}^-}} \right).
\end{eqnarray}
We abbreviate the charge conjugated fields as
\beq \tilde{\chi}^-_i \equiv \tilde{\chi}^{+ C}_i\ .
\eeq\s

The mass terms for the scalar quarks come from the Yukawa couplings
to the Higgs fields (which yield the corresponding quark
masses) and the soft--SUSY--breaking squark mass terms and squark-Higgs
interactions parameterized by $m_{\tilde{q}_{L/R}}$ and $A$, 
respectively.
Moreover there are mass and mixing terms from auxiliary field terms
involving one or two Higgs bosons and two squark fields.
The resulting mass matrix for the two squarks of  a given flavor is
(the dependence of the parameters $m_{\tilde{q}_{L/R}}$ and $A$ on the
generation/flavor has been dropped):
\beq
{\cal M}^2_{\tilde{q}}=\left(\begin{array}{rr}
       m_L^2+m_q^2 & m^*_{LR} m_q \\ 
       m_{LR} m_q  & m_R^2+m_q^2 \end{array}\right)\ ,
\eeq
where
\begin{eqnarray}
m_L^2 & = & m_{\tilde{q}_L}^2 + \cos 2 \beta\ (I^f_3-Q_fs_W^2) M_Z^2\ , 
\label{b28}\\
m_R^2 & = & m_{\tilde{q}_R}^2 + \cos 2 \beta\ (Q_fs_W^2) M_Z^2\ ,
\end{eqnarray}
and, for up-type squarks:
\beq
m_{LR} = A - \mu^*\cot\beta\ ,
\eeq
while for down-type squarks:
\beq
m_{LR} = A - \mu^*\tan\beta \ .
\eeq
This hermitian mass matrix is diagonalized by a unitary matrix $S$, so
we can write the squark mass eigenstates of flavor $i$ as
\begin{eqnarray}
\tilde{q}_{i1} & = & S^i_{11}\ \tilde{q}_{iL} + S^i_{12}\ 
\tilde{q}_{iR}\ ,\nonumber\\
\tilde{q}_{i2} & = & S^i_{21}\ \tilde{q}_{iL} + S^i_{22}\ 
\tilde{q}_{iR}\ .
\end{eqnarray}
In general $m_{LR} = |m_{LR}|e^{i\phi}$ is a complex number and $S$
may be written as
\beq S^{i} = 
 \left(\begin{array}{rr} 
  e^{{\rm i}\phi/2}\cos\theta_i &  e^{-{\rm i}\phi/2}\sin\theta_i \\
 -e^{{\rm i}\phi/2}\sin\theta_i &  e^{-{\rm i}\phi/2}\cos\theta_i
                        \end{array}\right)\ .
\eeq\s

The mass terms of the sleptons arise in the same way as the
squark masses. The main difference appears for the sneutrinos: there
is only one sneutrino for every generation, $\tilde{\nu}_l$, and hence there
is no mixing. Moreover, we cannot add trilinear soft--breaking terms
to shift the masses of the sneutrinos, whose value is given just by
$m^2_{\tilde{\nu}}=m^2_{\tilde{l}_L}+1/2\cos2\beta M^2_Z$. (See
Eq.~(\ref{b28})).

In general, the physical sfermion masses are given by
\beq
m^2_{\tilde{f}_{1,2}}=m^2_f+\frac{1}{2}\left\{(m^2_R+m^2_L)\mp
\left[(m^2_R-m^2_L)^2+4m^2_f|m_{LR}|^2\right]^{1/2}\right\},
\eeq
independent of the phase of $m_{LR}$.

\subsection{Physical Phases in the MSSM with complex couplings}

As we do not constrain the parameters of the MSSM to be real, the
following parameters may take complex values: the
Yukawa couplings, $\mu$, the soft parameters $m_{12}^2$, $M_1$, $M_2$,
$M_3$ and the $A$ parameters.
But not all combinations of phases in these parameters lead to
different physical results because several phases can be absorbed by
redefinitions of the fields. We now describe a procedure to eliminate
the unphysical phases. We assume the GUT relation between the $M_i$,
so they have one common phase. The only remaining phases will be
chosen to be those of $\mu$, the $A$ parameters and, for three
generations, one phase for all the Yukawa couplings (the $\delta_{\rm CKM}$).

Analogous to the Standard Model case, the Yukawa couplings can be
changed by redefinitions of the quark superfields in such a way that
there remains only one phase for three generations and only real
couplings for less than three generations.

\begin{table}[htb]
\caption{\em The charges $n_i$ for three U(1) symmetries that leave
  ${\cal L}_{\rm MSSM}$ invariant.}
\label{PQCharges}
\vspace{0.3cm}
\center
\begin{tabular}{|c|r|r|r|r|r|r|r|r|r|c|}
\hline
U(1) & $M_i$ & $A$ & $m_{12}^2$ & $\mu$\ & $H_1$ & $H_2$ & $QU$\ 
     & $QD$\ & $LE$\ & $\theta$\\
\hline
\hline 
$PQ$  & 0 & 0 & $-1$ & $-1$ & 1/2 & 1/2 & $-1/2$ & $-1/2$ 
      & $-1/2$ & 0 \\
$R_1$ & $-1$ & $-1$ & 0 & 1 & 0 & 0 & 1 & 1 & 1 & $-1/2$ \\
\hline 
\end{tabular}
\vspace*{0.3cm}
\end{table}

Following \cite{relax} we consider two U(1) transformations $PQ$
and $R_1$ that do not only transform the fields but also the
parameters of the MSSM. In table \ref{PQCharges} the charges $n_i$ of
the various quantities are displayed with which the Lagrangian ${\cal
  L}_{\rm MSSM}$ is invariant under the combined multiplications with
$e^{i\alpha n_i}$. The first transformation is a Peccei-Quinn
symmetry and $R_1$ is an R symmetry that also transforms the $\theta$
variables appearing in the arguments of the superfields in ${\cal
  L}_{\rm MSSM}$. 

Since ${\cal L}_{\rm MSSM}$ is invariant under these transformations,
so are the physical predictions of the MSSM. If the parameters $M_i$, $A$,
$m_{12}^2$ and $\mu$  have the phases $\varphi_M$, $\varphi_A$,
$\varphi_{m_{12}^2}$ and $\varphi_\mu$, we first apply $R_1$ with the
angle $\varphi_M$, then $PQ$ with the angle $\varphi_{m_{12}^2}$ and
obtain for an arbitrary observable:
\beq
\begin{array}{clcccc}
&\sigma(|\mu|,|A|,|M_i|,|m_{12}^2|,&\varphi_\mu,&\varphi_A,
&\varphi_M,&\varphi_{m_{12}^2})
\\ = & 
\sigma(|\mu|,|A|,|M_i|,|m_{12}^2|,&\varphi_\mu+\varphi_M,
&\varphi_A-\varphi_M,&0,&\varphi_{m_{12}^2})
\\ = & 
\sigma(|\mu|,|A|,|M_i|,|m_{12}^2|,&\varphi_\mu+\varphi_M-\varphi_{m_{12}^2},
&\varphi_A-\varphi_M,&0,&0).
\end{array}
\eeq
So the physical predictions only depend on the absolute
values of the parameters and the phases
\beq
\phi_A\equiv {\rm arg}(A M_i^*), \quad \phi_B\equiv {\rm arg}(\mu A m_{12}^{2*})
\ .
\eeq
One can replace these phases by another set where only the $\mu$ and the $A$ 
parameters are complex. In our choice the phase of $A$ is traded for
the phase of the off--diagonal term in the corresponding sfermion mixing matrix:
$m^f_{LR}=A_f-\mu^*\{\cot,\tan\}\beta$ (C.31,C.32). Relaxing universality
for the soft--breaking terms, every $A_f$ contains a different
CP--violating phase. 

\subsection{Vertex factors}

\begin{figure}[htb]
\begin{center}
\begin{picture}(15,14)
\epsfxsize=18cm
\put(1,0){\epsfbox{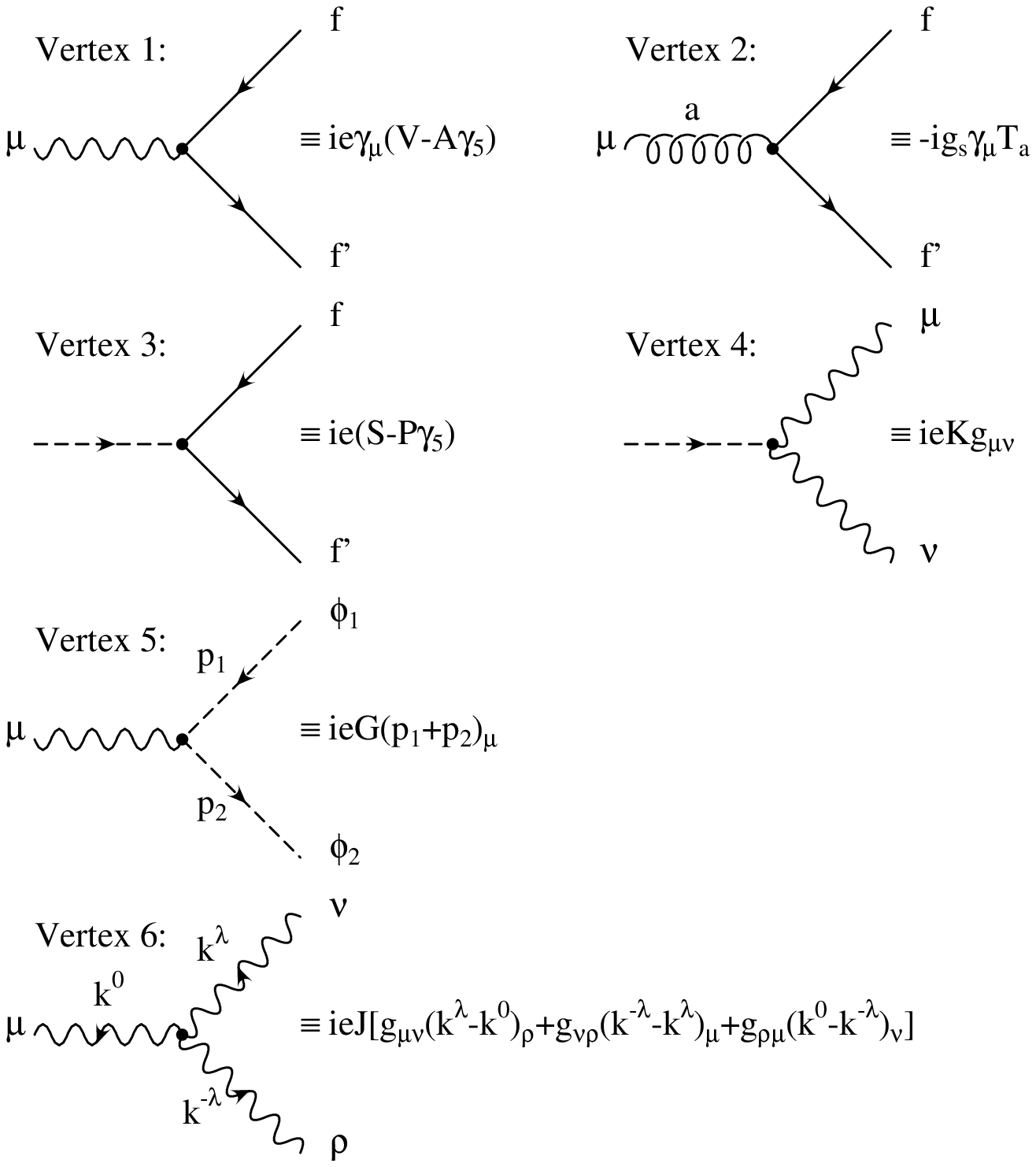}}
\end{picture}
\end{center}
\caption{\em Generic Vertices.\label{fig:ver}}
\vspace*{0.3cm}
\end{figure}

We employ the notation and conventions of the previous sections. 
The list of generic vertices is shown in Fig.~\ref{fig:ver}.

    \subsubsection{Couplings in the SM}

\begin{description}

\item{\bf{Vertex 1:}} Coupling of electroweak gauge bosons and 
fermions: ${\rm i}e\gamma_\mu (V-A\gamma_5)$.
\beq
\gamma \bar{f}f: & V&=-Q_f \ , \ \ A=0 \ ,  \\
Z\bar{f}f:       & V&=-\frac{v_f}{2s_Wc_W} \ ,  \ \ A=-\frac{a_f}{2s_Wc_W} \ ,
\\
W^\pm \bar{f}'f: & V&=A=-\frac{1}{2\sqrt{2}s_W}\ ,
\eeq 
with $v_f\equiv(I^f_3-2s^2_WQ_f)$, $a_f\equiv I^f_3$.

\item{\bf{Vertex 2:}} Coupling of gluons and quarks: 
$-{\rm i}eg_s\gamma_\mu T_a$.

\item{\bf{Vertex 3:}} Coupling of two fermions and one Higgs boson
(${\cal H}\bar{f}'f$): ${\rm i}e (S-P\gamma_5)$.
\beq
H^0\bar{f}f:          & S=-\mu_f \ ,  & P=0 \ , \\
\chi \bar{f}f:        & S=0 \ ,       & P=-2{\rm i}I_3^f\mu_f \ , \\
\phi^\pm \bar{f}'f:   & S= \sqrt{2}I_3^f[\mu_f-\mu_{f'}] \ ,  
                      & P=-\sqrt{2}I_3^f[\mu_f+\mu_{f'}]\ ,
\eeq
with $\mu_f\equiv\displaystyle\frac{m_f}{2s_WM_W}=
\displaystyle\frac{m_f}{2s_Wc_WM_Z}$.

\item{\bf{Vertex 4:}} Coupling of one Higgs boson and two gauge 
bosons: ${\rm i}e K g_{\mu\nu}$.
\beq
H^0ZZ:                  & K=&\displaystyle\frac{M_Z}{s_Wc_W} \ , \\
H^0W^\pm W^\mp:         & K=&\displaystyle\frac{M_W}{s_W} \ , \\
\phi^\pm W^\mp\gamma:   & K=&M_W \ ,  \\
\phi^\pm W^\mp Z:       & K=&-M_Zs_W \ .  
\eeq 
The rest of the couplings are $K=0$.

\item{\bf{Vertex 5:}} Coupling of one gauge boson and 
two Higgs bosons ($V\phi_1\phi_2$): \\${\rm i}e G (p_1+p_2)_\mu$.

\beq
\gamma\phi^\lambda\phi^{-\lambda}:    &G=&-\lambda \ , \\
Z\phi^\lambda\phi^{-\lambda}:&G=&-\lambda\displaystyle\frac{\cos2\theta_W}
                             {2s_Wc_W}    \ , \\
Z\chi H^0:                   &G=&\displaystyle\frac{{\rm i}}{2s_Wc_W} \ , \\
W^\lambda\phi^{-\lambda}H^0: &G=&\displaystyle\frac{\lambda}{2s_W} \ , \\
W^\lambda\phi^{-\lambda}\chi:&G=&\displaystyle\frac{{\rm i}}{2s_W} \ .
\eeq 
Interchanging the two Higgs bosons causes the coupling constant to change sign.

\item{\bf{Vertex 6:}} Coupling of three gauge bosons 
(outgoing momenta): 
$${\rm i}e J \{g_{\mu\nu}(k^\lambda-k^0)_\rho
       +g_{\nu\rho}(k^{-\lambda}-k^\lambda)_\mu
       +g_{\rho\mu}(k^0-k^{-\lambda})_\nu \} \ .$$
\beq
\gamma W^\lambda W^{-\lambda}: &J=&-\lambda \ , \\
Z W^\lambda W^{-\lambda}:      &J=&-\lambda\displaystyle\frac{c_W}{s_W} \ .
\eeq 

\end{description}

    \subsubsection{Couplings in the MSSM}

\begin{description}

\item{\bf{Vertex 1:}} Coupling of two neutralinos or charginos and
one gauge boson: ${\rm i}e\gamma_\mu (V-A\gamma_5)$.

The fermion flow direction in our Feynman graphs is fixed by the
outgoing fermion antifermion pair. The relation between the gauge
boson vertex factors for the two different fermion flow directions is
obtained by substituting spinors by charge conjugated spinors in the
interaction term:
\beq
V_\mu \bar{\tilde{\chi}}^+ \gamma^\mu (P_L g_L + P_R g_R)
\tilde{\chi}^+
=
V_\mu \bar{\tilde{\chi}}^- \gamma^\mu (-P_L g_R - P_R g_L)
\tilde{\chi}^-\ . 
\eeq
Taking into account also the symmetry factor $S=2$ for the
neutralino coupling and $S=1$ for the charginos, the boson vertex
factors $V\equiv(g_L+g_R)/2$ and $A\equiv(g_L-g_R)/2$ are given by:
\beq
Z\bar{\tilde{\chi}}^0_j\tilde{\chi}^0_k: & 
g_L=&\displaystyle\frac{1}{2s_Wc_W}
(N^\prime_{k4}N^{\prime *}_{j4}-N^\prime_{k3}N^{\prime *}_{j3})\ , \\ &  
g_R=&\displaystyle\frac{1}{2s_Wc_W}
(N^{\prime *}_{k3}N^\prime_{j3}-N^{\prime *}_{k4}N^\prime_{j4})\ , \\ 
Z\bar{\tilde{\chi}}^+_k {\tilde{\chi}}^+_j
: & 
g_L=&\displaystyle - \frac{1}{s_Wc_W}
\left[\left(\frac{1}{2} - s_W^2\right) V^*_{k2}V_{j2}+c_W^2 
V^*_{k1}V_{j1}\right]
\ , \\ &
g_R=&\displaystyle - \frac{1}{s_Wc_W}
\left[\left(\frac{1}{2} - s_W^2\right) U_{k2}U^*_{j2}+c_W^2 
U_{k1}U^*_{j1}\right]
\ , \\  
Z\bar{\tilde{\chi}}^-_j\tilde{\chi}^-_k: & 
g_L=&\displaystyle\frac{1}{s_Wc_W}
\left[\left(\frac{1}{2} - s_W^2\right) U_{k2}U^*_{j2}+c_W^2 
U_{k1}U^*_{j1}\right]
\ , \\ & 
g_R=&\displaystyle\frac{1}{s_Wc_W}
\left[\left(\frac{1}{2} - s_W^2\right) V^*_{k2}V_{j2}+c_W^2 
V^*_{k1}V_{j1}\right]\ ,
\eeq
and
\beq
\gamma\bar{\tilde{\chi}}^+_k {\tilde{\chi}}^+_j
: & 
g_L=g_R=&-\delta_{jk}\ , \\  
\gamma\bar{\tilde{\chi}}^-_j\tilde{\chi}^-_k: & 
g_L=g_R=&\delta_{jk}\ .
\eeq

\item{\bf{Vertex 2:}} There is no genuine supersymmetric vertex of
this kind.

\item{\bf{Vertex 3:}} There are three couplings of the kind
${\rm i}e (S-P\gamma_5)$ :

(Some of the couplings are more easily written in terms of
$g_{L,R}\equiv S\pm P$).

\noindent\hspace{-0.33cm}$\bullet$  
Coupling of one neutralino or chargino to fermions and scalar fermions. 

The couplings of neutralinos and charginos to quarks and scalar quarks are
given by
\begin{eqnarray}
{\cal L}_{\tilde{\chi} q \tilde{q}} & = & e\tilde{q}^\dagger
\bar{\tilde{\chi}}(P_L g_L + P_R g_R) q 
+ e\bar{q} (P_R g_L^* + P_L g_R^*) \tilde{\chi} \tilde{q}\ .
%
\end{eqnarray}


\underline{down-type quarks} ($i=1,3,5$):
\begin{eqnarray}
\tilde{q}_{i,k}^\dagger \bar{\tilde{\chi}}^0_j q_i :&
g_L=& 
-\sqrt{2} \left[
      \left(Q_i  N^{\prime *}_{j1}
      + (I^i_3-Q_i s_W^2)  \frac{1}{s_Wc_W} N^{\prime *}_{j2} \right)
      S_{k1}^i  \right.\nonumber\\
&&\left.
      + \frac{ m_{q_i}}{2 M_W s_W\cos\beta}N^{\prime *}_{j3}
      S_{k2}^i
      \right]\ , 
\\
&g_R=& 
-\sqrt{2} \left[
      - \left(Q_i     N^\prime_{j1}
      + (-Q_i s_W^2)  \frac{1}{s_Wc_W} N^\prime_{j2} \right)
      S_{k2}^i \right.\nonumber\\
&&\left.
     + \frac{ m_{q_i}}{2 M_W s_W\cos\beta} N^\prime_{j3}
     S_{k1}^i \right]\ ,
\\
\tilde{q}_{(i+1),k}^\dagger \bar{\tilde{\chi}}^-_j q_i :&
g_L=& 
\frac{1}{s_W} \left[  - V^*_{j1} S_{k1}^{i+1}
          + \frac{m_{q_{i+1}}}{\sqrt{2} M_W \sin\beta}
            V^*_{j2} S_{k2}^{i+1} \right]\ ,
\\
&g_R=& 
\frac{1}{s_W} \left[   \frac{m_{q_i}}{\sqrt{2} M_W \cos\beta}
           U_{j2}  S_{k1}^{i+1}
\right]\ .
\end{eqnarray}

\underline{up-type quarks} ($i=2,4,6$):
\begin{eqnarray}
\tilde{q}_{i,k}^\dagger \bar{\tilde{\chi}}^0_j q_i :&
g_L=& 
-\sqrt{2} \left[
       \left(Q_i   N^{\prime *}_{j1}
       + (I^i_3 - Q_i s_W^2)  \frac{1}{s_Wc_W} N^{\prime *}_{j2} \right)
       S_{k1}^i  \right.\nonumber\\
&&\left.
       + \frac{ m_{q_i}}{2 M_W s_W\sin\beta} N^{\prime *}_{j4}
       S_{k2}^i \right]\ ,
\\
&g_R=& 
-\sqrt{2} \left[
      - \left(Q_i   N^\prime_{j1}
         + (-Q_i s_W^2)  \frac{1}{s_Wc_W} N^\prime_{j2} \right)
         S_{k2}^i \right.\nonumber\\
&&\left.
         + \frac{m_{q_i}}{2 M_W s_W\sin\beta} N^\prime_{j4}
         S_{k1}^i \right]\ ,
\\
\tilde{q}_{(i-1),k}^\dagger \bar{\tilde{\chi}}^+_j q_i :&
g_L=& 
\frac{1}{s_W} \left[ - U^*_{j1} S_{k1}^{i-1}
         + \frac{m_{q_{i-1}}}{\sqrt{2} M_W \cos\beta}
           U^*_{j2} S_{k2}^{i-1} \right]\ ,
\\
&g_R=& 
\frac{1}{s_W} \left[ \frac{m_{q_i}}{\sqrt{2} M_W \sin\beta}
           V_{j2} S_{k1}^{i-1} \right]\ .
\end{eqnarray}

The couplings of neutralinos and charginos to leptons and scalar leptons are
given analogously, performing the following substitutions:
\beq
i=1,3,5&:&\tilde{q}_{i1}\to\tilde{l}_1
\ , \ \ S^i_{jk}\to S^l_{jk}\ , \ \ Q_i=-1\ , \ \ I^i_3=-\frac{1}{2}\ ,
\ \ m_{q_i}=m_l \ ,\nonumber\\
 & & \tilde{q}_{i2}\to\tilde{l}_2\ ,\nonumber\\
i=2,4,6&:&\tilde{q}_{i1}\to\tilde{\nu}_l
\ , \ \ S^i_{jk}\to \delta_{jk}\ , \ \ Q_i=0\ , \ \ I^i_3=\frac{1}{2}\ , 
\ \ m_{q_i}=0 \ ,\nonumber\\
 & & \tilde{q}_{i2}\ \mbox{does not exist} \ . \nonumber
\eeq

\noindent\hspace{-0.33cm}$\bullet$  
Coupling of one gluino to a quark and a scalar quark.

The interaction between gluinos, quarks and squarks is described by
the terms
\begin{eqnarray}
{\cal L} & = & \tilde{q}^\dagger
e\overline{\tilde{g}}^a (P_L g_L + P_R g_R) \frac{\lambda^a}{2} q 
+ e\overline{q} (P_R g_L^* + P_L g_R^*) \frac{\lambda^a}{2}
\tilde{g}^a \tilde{q}\ ,
\end{eqnarray}
yielding the vertex factors
\beq
{\rm i}e(P_L g_L + P_R g_R)\frac{\lambda^a}{2},\quad
{\rm i}e(P_R g_L^* + P_L g_R^*)\frac{\lambda^a}{2}.
\eeq
In our calculations the Gell-Mann matrices appear only in the
combination 
\beq
\sum_{a=1}^8 \left(\frac{\lambda^a \lambda^a}{4} \right)_{AB} =
C_2(F)\delta_{AB} = \frac{4}{3}\delta_{AB} .
\eeq
The couplings are
\begin{eqnarray}
\tilde{q}_{i}^\dagger \overline{\tilde{g}} q 
:
&eg_L =& -\sqrt{2} \displaystyle{g_s}  S_{i1} \ , \\
&eg_R =& +\sqrt{2} \displaystyle{g_s}  S_{i2} \ .
\end{eqnarray}

\noindent\hspace{-0.33cm}$\bullet$  
Coupling of two fermions and one Higgs boson (${\cal H}\bar{f}'f$).
\begin{eqnarray}
H\bar{u}u: & S=
-\mu_u\sin\alpha/\sin\beta\ , & P=0 \ , \label{c78} 
\\
H\bar{d}d: & S=
-\mu_d\cos\alpha/\cos\beta\ , & P=0\ , 
\\
h\bar{u}u: & S=
-\mu_u\cos\alpha/\sin\beta\ , & P=0\ , 
\\
h\bar{d}d: & S=
\mu_d\sin\alpha/\cos\beta\ , & P=0\ , 
\\
A\bar{u}u: &S=0 \ , & 
P=-{\rm i}\mu_u\cot\beta\ ,  
\\
A\bar{d}d: &S=0  \ , & 
P=-{\rm i}\mu_d\tan\beta \ , \label{c83}
\\
H^+\bar{u}d: & S=
\displaystyle\frac{1}{\sqrt{2}}(\mu_u\cot\beta +\mu_d\tan\beta) \ , &
\nonumber\\
&
P=\frac{1}{\sqrt{2}}(\mu_u\cot\beta-\mu_d\tan\beta) \ , &
\\
G^0\bar{f}f: & S=0 \ ,        & P=-2{\rm i}I_3^f\mu_f \ , \label{c85}\\
G^\pm \bar{f}'f: & S= \sqrt{2}I_3^f[\mu_f-\mu_{f'}] \ ,  
           & P=-\sqrt{2}I_3^f[\mu_f+\mu_{f'}]\ .
\end{eqnarray}
with $\mu_f\equiv m_f/2s_WM_W=m_f/2s_Wc_WM_Z$. For the vertices corresponding 
to the hermitian conjugated fields we have to replace ($S,P$) by ($S^*,-P^*$).

\item{\bf{Vertex 4:}} Coupling of one Higgs boson and two gauge bosons:
${\rm i}e K g_{\mu\nu}$.

The couplings of a neutral Higgs to two $Z$ bosons are:
\begin{eqnarray}
hZZ: & K=&\frac{M_Z}{s_Wc_W} \sin(\beta-\alpha)\ , \\
HZZ: & K=&\frac{M_Z}{s_Wc_W} \cos(\beta-\alpha)\ . 
\end{eqnarray}
And the only nonzero couplings of a charged Higgs to neutral gauge and $W$ 
bosons are:
\begin{eqnarray}
G^\pm W^\mp \gamma:     & K=& M_W \ , \\
G^\pm W^\mp Z:          & K=& -M_Z s_W \ .
\end{eqnarray}
The rest of the couplings are $K=0$.

\item{\bf{Vertex 5:}} There are two types of couplings $Z\phi_1\phi_2$
of the kind ${\rm i}e G (p_1+p_2)_\mu$\ . 

\noindent\hspace{-0.33cm}$\bullet$  
Coupling of two Higgs bosons and one neutral gauge boson.

\begin{eqnarray}
ZAH: &G=&
-\frac{{\rm i}\sin(\beta-\alpha)}{2 s_Wc_W} \ ,
\\
ZAh: &G=&
\frac{{\rm i}\cos(\beta-\alpha)}{2 s_Wc_W} \ ,
\\
ZG^0H: &G=&
\frac{{\rm i}\cos(\beta-\alpha)}{2 s_Wc_W} \ ,
\\
ZG^0h: &G=&
\frac{{\rm i}\sin(\beta-\alpha)}{2 s_Wc_W} \ ,
\\
ZH^\lambda H^{-\lambda}: &G=&
-\lambda \frac{\cos2\theta_W}{2s_Wc_W} \ ,  
\\
Z G^\lambda G^{-\lambda}: &G=&
-\lambda \frac{\cos2\theta_W}{2s_Wc_W} \ ,
\\
\gamma H^\lambda H^{-\lambda}: &G=&
-\lambda  \ ,  
\\
\gamma G^\lambda G^{-\lambda}: &G=&
-\lambda  \ .    
\end{eqnarray}
Interchanging the $\phi_1$ and $\phi_2$ causes the coupling 
constant to change sign.

\noindent\hspace{-0.33cm}$\bullet$  
Coupling of two scalar fermions and one neutral gauge boson.
\beq
\gamma\tilde{q}_{i,j}\tilde{q}^\dagger_{i,k}: & G= &
-Q_i \delta_{jk} \ , 
\\
Z\tilde{q}_{i,j}\tilde{q}^\dagger_{i,k}: & G= &
-\frac{1}{s_Wc_W} 
\left[(I^i_3 - Q_i s_W^2) S^{i*}_{j1} S^{i}_{k1}
- Q_i s_W^2 S^{i*}_{j2} S^{i}_{k2} \right] \ .
\eeq

\item{\bf{Vertex 6:}} There is no genuine supersymmetric vertex of
this kind.

\end{description}



\begin{thebibliography}{99}

\bibitem{gelectron}
{J. Schwinger}, \prd{73}{48}{416}.

\bibitem{databook}
Particle Data Group, {\em Eur. Phys. J.} {\bf C3} (1998) 1.

\bibitem{smemdm}
        {T. Kinoshita}, 
        \prl{75}{95}{4728}; \\
        {A. Czarnecki, B. Krause, W.J. Marciano}, 
        \prd{52}{95}{2619}, \prl{76}{96}{3267}; \\
        {F. Jegerlehner},  
        {\em Nucl. Phys. Proc. Suppl.} {\bf 51C} (1996) 131; \\
        {S. Laporta, E. Remiddi}, 
        {\em Phys. Lett.} {\bf B379} (1996) 283; \\ 
        {A. Czarnecki, B. Krause},
        {\em Phys. Rev. Lett.} {\bf 78} (1997) 4339;\\
        {B. Krause}, \plb{390}{97}{392}; \\
        {M. Hayakawa, T. Kinoshita}, \prd{57}{98}{465}; \\
        {M. Davier, A. H\"ocker}, \hepph{9801361};\\
        {G. Degrassi, G.F. Giudice}, \hepph{9803384}.

\bibitem{brook}
V. Huges, in {\em Frontiers of High Energy Spin Physics}, 
Proceedings of the 10th International Symposium, Nagoya, 
Japan, edited by T. Hasegawa et al., Universal Academy Press, 
Tokyo, 1992.

\bibitem{munpemdm} 
{G. Couture, H. K\"onig}, \prd{53}{96}{555}; \\
{U. Chattopadhyay, P. Nath}, {\em  Phys. Rev.} {\bf D53} (1996) 1648; \\
{T. Moroi}, {\em Phys. Rev.} {\bf D53} (1996) 6565; \\
{M. Carena, G.F. Giudice, C.E.M. Wagner},
{\em Phys. Lett.} {\bf B390} (1997) 234; \\ 
{M. Krawczyk, J. Zochowski}, \prd{55}{97}{6968}.

\bibitem{ckm}
{N. Cabibbo}, \prl{10}{63}{531};\\
{M. Kobayashi, T. Maskawa}, \ptp{49}{73}{652}.

\bibitem{cpreview}
For recent reviews on CP violation see e.g.: \\
K. Gronau, D. London, \prd{55}{97}{2845}; \\
Y. Grossman, Y. Nir, R. Rattazzi, \hepph{9701231}; \\
Y. Nir, \hepph{9709301}; \\
X.-G. He, \hepph{9710551}.

\bibitem{b-asym}
G.R. Farrar, M.E. Shaposhnikov, \prd{50}{94}{774};\\
M.B. Gavela et al., \npb{430}{94}{382};\\
P. Huet, E. Sather, \prd{51}{95}{379}.

\bibitem{susycp}
W. Buchm\"uller, D. Wyler, \plb{121}{83}{321};\\
J. Polchinski, M.B. Wise, \plb{125}{83}{393};\\
F. del Aguila, M.B. Gavela, J.A. Grifols, A. M\'endez,
\plb{126}{83}{71} [E: {\em Phys. Lett.} {\bf B129} (1983) 473].

\bibitem{dugan}
M. Dugan, B. Grinstein, L.J. Hall, \npb{255}{85}{413}.

\bibitem{gavela}
J.M. Frere, M.B. Gavela, \plb{132}{83}{107}.

\bibitem{relax}
S. Dimopoulus, S. Thomas, \npb{465}{96}{23}.

\bibitem{b-susyasym}
N. Turok, J. Zadrozny, \npb{369}{92}{729};\\
M. Dine, P. Huet, R. Singelton Jr., \npb{375}{92}{625}; \\
A. Cohen, A. Nelson, \plb{297}{92}{111};\\
D. Comelli, M. Pietroni, A. Riotto, \plb{343}{95}{343};\\
P. Huet, A.E. Nelson, \prd{53}{96}{4578}; \\
M. Aoki, N. Oshimo, A. Sugamoto,
\hepph{9612225}; \hepph{9706287}; \hepph{9706500};\\
M. Carena, M. Quir\'os, A. Riotto, I. Vilja, C.E.M. Wagner,
\hepph{9702409};\\
G.M. Cline, M. Joyce, K. Kaimulaine, \hepph{9708393}.

\bibitem{dipoles}
X.-G. He, B.H.J. Mc Kellar, S. Pakvasa, \ijmp{A4}{89}{5011};\\
W. Bernreuther, M. Suzuki, \rmp{63}{91}{313};\\
Y. Kizukuri, N. Oshimo, \prd{45}{92}{1806}, {\em Phys. Rev.} {\bf D46} (1992) 
3025;\\  
S. Bertolini, F. Vissani, \plb{324}{94}{164}.

\bibitem{smedm}
J.F. Donoghue, \prd{18}{78}{1632};\\
E.P. Shabalin, \sjnp{28}{78}{75};\\
A. Czarnecki, B. Krause, \prl{78}{97}{4339}.

\bibitem{edm-n}
N.F. Ramsey, \arnps{40}{90}{1};\\
I.S. Altarev et al., \plb{276}{92}{242}.

\bibitem{edm-l}
J. Bailey et al., {\em J. Phys.} {\bf C4} (1978) 345, 
			\npb{150}{79}{1};\\
Y. Semertzidis et al., E821 Collaboration at BNL, 
{\em AGS Expression of Interest: Search for an Electric Dipole 
Moment of Muon}, September 1996.

\bibitem{nath1}
P. Nath, \prl{66}{91}{2565}.

\bibitem{phase0}
R. Kuchimanchi, \prl{76}{96}{3486};\\
R. Mohapatra, A. Rasin, \prl{76}{96}{3490};\\
Y. Nir, A. Rattazzi, \plb{382}{96}{363}.

\bibitem{cptop}
E. Christova, M. Fabbrichesi, \plb{315}{93}{113}, \plb{315}{93}{338}, 
\plb{320}{94}{299};\\
E. Grzadkowski, W.Y. Keung, \plb{316}{93}{137};\\
M. Aoki, N. Oshimo, \hepph{9801294}, \hepph{9808217}. 

\bibitem{nlc}
ECFA/DESY LC Physics Working Group, \hepph{9705442}.

\bibitem{garisto}
R. Garisto, J.D. Wells,
\prd{55}{97}{1611}.

\bibitem{nath2}
T. Ibrahim, P. Nath, \plb{418}{98}{98}, \prd{57}{98}{478}, \hepph{9807501}.

\bibitem{falkolive}
T. Falk, K.A. Olive, \hepph{9806236}.

\bibitem{abel}
S.A. Abel,  \plb{410}{97}{173}.

\bibitem{fcnc-cp}
F. Gabbiani, A. Masiero, \npb{322}{89}{235};\\
G.S. Hagelin, S. Kelley, T. Tanaka, \npb{415}{94}{293};\\
F. Gabbiani, E. Gabrielli, A. Masiero, L. Silvestrini,
{\em Nucl. Phys.} {\bf B477} (1996) 321.

\bibitem{ber95}
J. Bernab\'eu, G.A. Gonz\'alez-Sprinberg, M. Tung, J. Vidal,
{\em Nucl. Phys.} {\bf B436} (1995) 474.        
		
\bibitem{ber97}
J. Bernab\'eu, G.A. Gonz\'alez-Sprinberg, J. Vidal,
\plb{397}{97}{255}.        

\bibitem{ber94}
J. Bernab\'eu, G.A. Gonz\'alez-Sprinberg, J. Vidal,
\plb{326}{94}{168}.

\bibitem{ber95p}
J. Bernab\'eu, G.A. Gonz\'alez-Sprinberg, J. Vidal,
in {\em Proceedings of the Ringberg Workshop on Perspectives for electroweak
interactions in $e^+e^-$ collisions}, ed. B.A. Kniehl, World Scientific 1995, 
p. 329.

\bibitem{mele}
B. Mele, G. Altarelli, \plb{299}{93}{345}; \\
B. Mele, \mpl{A49}{94}{1239}.

\bibitem{ber95b}
	{J. Bernab\'eu, D. Comelli, L. Lavoura, J. P. Silva},
	{\em Phys. Rev.} {\bf D53} (1996) 5222.

\bibitem{hirs1}
W. Hollik, J.I. Illana, S. Rigolin, D. St\"ockinger, 
\plb{416}{98}{345};\\
B. de Carlos, J.M. Moreno,
\hepph{9707487}.

\bibitem{bernre97b}
W. Bernreuther, A. Brandenburg, P. Overmann,
\plb{391}{97}{413}.

\bibitem{hirs2}
W. Hollik, J.I. Illana, S. Rigolin, D. St\"ockinger, 
\plb{425}{98}{322}.

\bibitem{vienna}
A. Bartl, E. Christova, W. Majerotto,
\npb{460}{96}{235} [E: \npb{465}{96}365];\\
A. Bartl, E. Christova, T. Gajdosik, W. Majerotto,
\hepph{9705245}.

\bibitem{iz}
C. Itzykson, J. Zuber, {\em Quantum Field Theory}, McGraw--Hill, 1985.

\bibitem{decoupling}
	{T. Appelquist, J. Carrazone},
	{\em Phys. Rev.} {\bf D11} (1975) 2856.

\bibitem{herrero}
A. Dobado, M.J. Herrero, S. Pe\~naranda, \hepph{9710313}, \hepph{9806488}.

\bibitem{higgs}
{OPAL Collaboration}, \zpc{73}{97}{189};\\
{ALEPH Collaboration}, CERN PPE/97-071.

\bibitem{moriond98}
S. De Jong, talk at the XXXIIIrd Rencontres de Moriond; \\
M. Maggi, {\em ibid}; \\
S. Katsanevas, talk at SUSY98 Conference; \\
D. Treille, plenary talk at ICHEP98 Vancouver. 

\bibitem{2hdm-cp}
G.C. Branco, M.N. Rebelo, \plb{160}{85}{11};\\
S. Weinberg, \prd{42}{90}{860}; \\
W. Bernreuther, T. Schr\"oder, T.N. Pham, \plb{279}{92}{389}.

\bibitem{SumRules} 
        S. Ferrara, E. Remiddi,
        \plb{53}{74}{347};\\
	I. Giannakis, J.T. Liu, M. Porrati,
        {\bf hep-th}/{9803073}.

\bibitem{cornwall85}
J.M. Cornwall,
\prd{26}{82}{1453}.

\bibitem{papa94}
J. Papavassiliou, C. Parrinello, \prd{50}{94}{3059}.

\bibitem{georg}
A. Denner, S. Dittmaier, G. Weiglein, 
\plb{333}{94}{420}, \npb{440}{95}{95}.

\bibitem{garfield}
T. Gajdosik, private communication.

\bibitem{bernre97c}
W. Bernreuther, 
talk at the 20th Johns Hopkins Workshop, Heidelberg, 1996,
\hepph{9701357}.

\bibitem{bernre89a}
W. Bernreuther, U. L\"ow, J.P. Ma, O. Nachtmann,
\zpc{43}{89}{117}.

\bibitem{koerner91}
J. K\"orner, J.P. Ma, R. M\"unch, O. Nachtmann, R. Sch\"opf,
{\em Z. Phys.} {\bf C49} (1991) 447.

\bibitem{bernre95}
W. Bernreuther, G.W. Botz, D. Bruss, P. Haberl, O. Nachtmann,
\zpc{68}{95}{73}.

\bibitem{abraham94}
K.J. Abraham, B. Lampe,
\plb{326}{94}{175}.

\bibitem{kuehn93}
J.H. K\"uhn,
\plb{313}{93}{458}.

\bibitem{tsai71}
Y.S. Tsai,
\prd{4}{71}{2821} [E: \prd{13}{76}{771}].

\bibitem{bernre91c}
W. Bernreuther, O. Nachtmann,
\plb{268}{91}{424}.

\bibitem{bernre96a}
W. Bernreuther, P. Overmann,
\zpc{72}{96}{461}.

\bibitem{rindani}
B. Ananthanarayan, S.D. Rindani, \prd{52}{95}{2684}.

\bibitem{bernre96b}
W. Bernreuther, A. Brandenburg, P. Overmann,
in Proc. Workshop on $e^+e^-$ Linear Colliders; Annecy, Assergi,
Hamburg, 1995, \hepph{9602273}.

\bibitem{bernre91b}
W. Bernreuther, G.W. Botz, O. Nachtmann, P. Overmann,
{\em Z. Phys.} {\bf C52} (1991) 567.

\bibitem{bernre94}
W. Bernreuther, P. Overmann,
\zpc{61}{94}{599}.

\bibitem{bernre89b}
W. Bernreuther, O. Nachtmann,
\prl{63}{89}{2787} [E: {\em Phys. Rev. Lett.} {\bf 64} (1990) 1072].

\bibitem{bernre92c}
W. Bernreuther, O. Nachtmann, P. Overmann, T. Schr\"oder,
{\em Nucl. Phys.} {\bf B388} (1992) 53 [E: \npb{406}{93}{516}].

\bibitem{bernre93}
W. Bernreuther, O. Nachtmann, P. Overmann,
\prd{48}{93}{78}.

\bibitem{optimal}
D. Atwood, A. Soni, \prd{45}{92}{2405}.

\bibitem{optiproof}
M. Diehl, O. Nachtmann, \zpc{62}{94}{397}.

\bibitem{eusebio}
E. S\'anchez \'Alvaro,
talk at the Fourth International Workshop on Tau Lepton Physics (TAU96) 
Estes Park, Colorado, September 1996;
PhD thesis, Universidad Complutense de Madrid, 1997.

\bibitem{opal97}
The OPAL Collaboration, \zpc{74}{97}{403}.

\bibitem{hs}
W. Hollik, C. Schappacher, \hepph{9807427}.

\bibitem{zhouhad}
H.-Y Zhou, \hepph{9805358}, \hepph{9806323}.

\bibitem{kodaira98}
J. Kodaira, T. Nasuno, S. Parke,
\hepph{9807209}.

\bibitem{analtopspin}
A. Czarnecki, M. Jezabek, J.H. K\"uhn, \npb{351}{91}{70}.

\bibitem{zhounlc}
H.-Y Zhou, \hepph{9806239}.

\bibitem{peskin}
C.R. Schmidt, M.E. Peskin, \prl{69}{92}{410};\\
C.R. Schmidt, \plb{293}{92}{111};\\
D. Chang, W.-Y. Keung, I. Phillips,
\npb{408}{93}{286} [E: \npb{429}{94}{255}].

\bibitem{bernre92a}
W. Bernreuther, T. Schr\"oder,
\plb{279}{92}{389}.

\bibitem{garfieldval}
A. Bartl, E. Christova, T. Gajdosik, W. Majerotto,
\hepph{9712380}.

\bibitem{topol}
W. Beenakker, S.C. van der Marck, W. Hollik, 
\npb{365}{91}{24}.

\bibitem{Denner93} A. Denner, \fp {\bf 41} {93} 307.

\bibitem{mssm}
H.E. Haber, G.L. Kane, \prep{117}{85}{75};\\
J.F. Gunion, H.E. Haber,
\npb{272}{86}{1} [E: {\em Nucl. Phys.} {\bf B402} (1993) 567].

\bibitem{HHG} 
        J.F. Gunion, H.E. Haber, G. Kane, S. Dawson, 
        {\sl The Higgs Hunter's Guide}, Addison-Wesley 1990.
                
\bibitem{Haber93}
        H.E. Haber,
        in {\sl Proc. of the 1992 Theoretical Advanced Study Institute
        in Particle Physics}, ed. J. Harvey and J. Polchinski (World
        Scientific, Singapore, 1993), p. 583.

\end{thebibliography}
\end{document}